\documentclass  [usegraphicx] {mn2e}

\newcommand\kms{$\rm{km\,s^{-1}}$}
\newcommand\HII{H\,{\sc ii}}

\title[OH polarization spectra, longitudes 350$^\circ$ to 
41$^\circ$]
{Parkes full polarization spectra of OH masers - I. Galactic longitudes 
350$^\circ$ through the Galactic Centre to 41$^\circ$}
\author[J.L.~Caswell, J.A.~Green \& C.J.~Phillips]
        {J.L.~Caswell\thanks{james.caswell@csiro.au}, 
J.A.~Green,
and C.J.~Phillips
\\CSIRO Astronomy and Space Science, Australia Telescope National 
Facility, PO Box 76, Epping, NSW, Australia 2121 \\}

\date{Accepted   2013 Feb6
      Received	 2013 Jan8}

\pagerange{\pageref{firstpage}--\pageref{lastpage}}
\pubyear{2013}

\begin{document}

\maketitle

\label{firstpage}

\begin{abstract}

Full polarization measurements of 1665 and 1667-MHz OH masers at 
sites of massive star formation have been made with the Parkes 64-m radio 
telescope.  Here we present the resulting spectra for 104 
northerly sources, from Galactic longitude 350$^\circ$ through the 
Galactic Centre to 41$^\circ$.  
Some maser positions were previously uncertain by many arcseconds, and 
thus for more than 20 masers we made new measurements with the 
ATCA (which also revealed several hitherto unreported 
masers), in most cases yielding arcsecond  precision to match the 
majority of sites. 
Position improvements have assisted us  in distinguishing OH 
masers with accompanying methanol masers from those without (thought to 
be at a later stage of evolution).  
There was no existing linear polarization information at many sites, 
and spectral resolution was sometimes poor, or velocity 
coverage incomplete.  
These inadequacies are addressed by the present Parkes spectra. 
The whole OH maser sample exhibit the well-known predominance of highly 
circularly polarized features.  We find that linear polarization is also 
common, but usually much weaker, and we highlight the rare cases of very 
pronounced linear polarization that can extend to 100 per cent.    
Unusually large velocity ranges of at least 25 \kms\ are present at 
seven sites.  
Our spectra measurements for most sources are at two epochs 
spaced by nearly one year, and reveal high stability at most sites, and 
marked variability 
(more than factors of two in the strongest feature) at only 
five sites.  The spectra also provide a valuable reference for 
longer term variability, with  high stability evident over the 
past decades at 10 sites and  marked variability for four of the sample.  
Future systematic monitoring of these variables may uncover further 
examples of periodicity, a phenomenon so far recognised in only one 
source.

\end{abstract}


\section{Introduction}

Our objective has been to obtain full polarization spectra, at high
velocity resolution, of masers at the 1665- and 1667-MHz transitions of 
OH  accessible to the Parkes telescope. 
We restricted our sample to sites in star formation regions 
(SFRs), thus excluding masers around late-type AGB stars which 
represent a quite different field of study. Often associated with OH in 
SFRs are methanol masers.  
Blind maser surveys are a major tool for pinpointing sites of 
massive star formation and, for such surveys, methanol at 6668 MHz 
has recently become the maser of choice (e.g. Green et al. 2009, 2010; 
Caswell et al. 2010b) since it is exclusively associated with high mass 
star formation and is commonly stronger than OH.  
But OH masers, none the less, retain a special role in star formation 
studies by virtue of their remarkable polarization properties, which are 
clear signatures of the strong magnetic fields where the masers 
originate.

Early measurements of OH masers, despite the low spatial resolution of 
most single dishes used more than 40 years ago, revealed remarkable 
circular and linear polarization, sometimes approaching 100 per cent.  
Reliable measurement of linear polarization, as well as circular 
(e.g. Robinson et al. 1970), was performed for a handful of sources, 
but was observationally intensive in view of intrumental limitations.  
As a consequence, polarization studies of OH masers thereafter have  
mostly been confined to the simpler measurement of circular 
polarization.  
Catalogued southern OH masers with precise positions now exceed  several 
hundred (Caswell 1998), but some lack published spectra, especially 
sources discovered with the Australia Telescope Compact Array 
(ATCA), observed only with low spectral resolution, and restricted to 
total intensity information.  
For many other sources, published spectra are available from 
Parkes observations, but date from 25 years ago, and limited to 
circular polarization.

Notable amongst the rare studies where full polarization has been 
measured are those using long baseline interferometer arrays such as 
the VLBA (e.g. Fish et al. 2005) and MERLIN (e.g. Hutawarakorn \& Cohen 
1999).  Such high spatial resolution might be thought essential to 
minimise depolarization from blending of differently polarized features;  
but fortunately, the high resolution studies have shown that, at any 
single velocity, there is often a single dominant feature, with only 
minor contributions from spatially nearby features, and thus 
depolarization in observations with low spatial resolution is not 
excessive.  
In view of this, and the slow progress in long baseline interferometry 
observations of most masers, a practical and faster means to advance 
polarization and other studies is to first acquire high quality 
single dish spectra of all the known OH masers. 

Even for single dish observations, it is only in the past decade that 
instrumentation has allowed optimum spectral line observations for 
studying OH masers, i.e. with full polarimetry, high spectral 
resolution, adequate velocity coverage, and preferably simultaneous 
coverage of more than one transition.  Further 
instrumental development extending such capabilities to interferometer 
arrays is even more recent, as provided for the ATCA by CABB (the 
Compact Array Broadband Backend (Wilson et al. 2011;  Caswell \& Green 
2011)).  Here we report the results of an extensive polarimetric survey 
of OH masers conducted in 2004 and 2005 with the Parkes telescope.  

Positions for most targets have arcsecond accuracy and were primarily 
taken from Caswell (1998). A few southern sources discovered since 1998 
were added to this sample, and some more northerly sources were also  
added, mostly those with positions 
reported by Forster \& Caswell (1989, 1999) and by Argon, 
Reid, \& Menten (2000).  Where required, additional new positions were 
obtained with the ATCA, as described in the next section.  

Contemporaneously with our observations, a similar survey targeting 
nearly 100 northern sources was conducted by Szymczak \& Gerard (2009) 
with the Nancay radio telescope (hereafter the NRT as abbreviated by the 
authors) and this has allowed us to make valuable 
comparisons with the sources in common.  In the region of sky 
overlapping the NRT observations, our measurements of about 100 sources 
include 50 in common with the NRT sample.  

The present paper (Paper I) reports our results for sources in the 
Galactic longitude range 350$^\circ$ through the Galactic Centre to 
40$^\circ$, thus capturing the full overlap with the NRT sample, and any 
VLBA or MERLIN targets.
In total, we present spectra for 104 distinct maser sites, of which 
23 have improved positions reported for the the first time in the 
present paper.  A subsequent paper (Paper II)  will present results for 
the remainder of the sample, a somewhat larger number of more southerly 
sites from Galactic longitude 240$^\circ$ to 350$^\circ$, for which 
there have been extremely few other observations.

The present large sample of full polarization spectra allows a good  
assessment of the incidence of linear polarization, and reveals 
some remarkable examples whose further study may elucidate reasons for 
the occasional occurrence of extremely high linear polarization.  
\\
\\
The full interpretation of the polarization spectra is beyond the 
scope of the present paper, but indications of the outcomes expected 
are shown in the analysis by Wright et al. (2004a,b)  of the much  
studied source W3(OH).  The study of masers 
around this young ultra-compact \HII\ region (uc\HII) allows uniquely    
detailed characterisation of the molecular material around the uc\HII\ 
region, revealing the orientation of a slowly expanding, 
rotating, torus, and its magnetic field distribution, in addition to the 
physical properties of density, OH abundance and temperature implied by 
the masing of the OH molecules.  In other studies, Hutawarakorn \& Cohen 
(1999) link magnetic field orientations to outflow phenomena, and 
Fish \& Reid (2006) argue persuasively for magnetic field direction 
preservation during collapse to a star;  the high magnetic energy 
density implied from the maser polarization measurements suggest that 
magnetic fields may be a controlling influence in the collapse to form 
the star (e.g. Asanok et al. 2010, Caswell, Kramer \& Reynolds 
2011a).  Results of the present study reveal candidates that are 
especially worthy of further exploration with the expectation of 
fruitful interpretation similar to the above.

\section[]{Observations and data reduction}

\subsection{Position determinations with the ATCA}

Improved position determinations for more than 20 sources were made in 
two periods with the ATCA.  
The first session (within project c906f) 2005 March 28, was conducted in 
a 6-km configuration, 
yielding typical position accuracy of 1 arcsec.  The second session  
(project c1386, 2005 May 24), specifically targeted sources close to 
declination zero, and used the hybrid configuration H168 which  has some 
baselines NS as well 
as EW, but extending only to slightly beyond 168m.  For such short 
baselines, the 
synthesised beam is approximately 2 x 3 arcmin, and rms position 
uncertainties, evaluated individually for each source, ranged  
from several arcseconds for strong sources to greater than 10 arcseconds 
for the weakest sources, but in all cases usefully improving 
upon existing information.

\subsection{Parkes spectral line observations}

The Parkes 64-m radio telescope was used in two observing periods, 2004
November 23-27 and  
2005 October 26-30 (project p484).   
The receiver accepted two orthogonal linear polarizations, followed by  
digital filters that restricted the processed signal to a 4-MHz 
bandpass before entering a correlator.  
Several correlators are available at Parkes, and the one selected as 
most suitable was the `Parkes multibeam correlator', with a new 
configuration.  The 32 blocks of 1024 spectral channels were 
concatenated to provide single-beam full polarimetry yielding 
four polarization products, each with 8192 channels.  These outputs can 
be manipulated in the software processing to provide the four Stokes 
parameters, I, Q, U and V, and any other equivalent representation such 
as RHCP and LHCP (right-hand and left-hand circular polarization 
respectively) for I and V; and linearly polarized flux density and 
its polarization position angle for Q and U.  
A similarly novel correlator configuration with concatenated blocks 
of the multibeam correlator was used in earlier Parkes observations to 
provide  16384 channels for each of two input polarizations when only 
the autocorrelations were required (Caswell 2004b).

In the present work, since two orthogonal linear polarizations are 
sampled, the autocorrelations are summed to provide total intensity 
(I), and differenced to produce Q (in the case where the Position Angle 
of one 
feed is set at zero).  With IAU conventions as summarized by Hamaker \& 
Bregman (1996), the cross-correlation provides U from the `in-phase' or 
`real' part, and V from the `quadrature' or `imaginary' part, 
provided that there is zero phase path difference between the inputs.  
During the hardware set-up, the phase path difference was adjusted to 
zero, with error no worse than 5$^\circ$, so as to minimise leakage 
between the in-phase and in-quadrature outputs.  

For each target, an observation of 10 min was made in most cases, 
but reduced to 4 min for some strong sources, and increased to 20 
min for weak sources.  
During each observation, in order to maintain a constant Receiver 
Position Angle on the sky, we slowly rotated the receiver at the rate  
needed to compensate for the parallactic angle change that occurs with 
an altitude-azimuth mounted telescope.  No reference spectra 
were taken, since the digital filters provide an inherently flat 
baseline and we found that a linear sloping baseline subsequently proved 
adequate for the final processing in most cases.  
Man-made interference at these frequencies is rare at the Parkes 
observatory.  Internally generated interference affected a small amount 
of 1667-MHz data but with negligible effect on the final results.  The 
few instances with interference on the displayed profiles are at
1667 MHz near velocity +110 \kms\ in the spectra of 
12.026--0.032, 24.494--0.039, 29.862--0.040, 29.956--0.015 and 
30.820--0.060.  

In addition to real time display of the correlator output as it 
accumulated, data quality was monitored at completion of each source 
observation, using the ATNF spectral line reduction program 
{\sc spc} for rapid preliminary display of RHCP, LHCP, and Q and U.   


During the 2005 sessions, the Position Angle for one of the probes 
was kept at zero (by maintaining a Receiver Position Angle of 
+45$^\circ$, or alternatively at --45$^\circ$ for locations where 
rotation limitations of the cabling prevented +45$^\circ$).  
At this epoch, Q is therefore the difference of the auto-correlations, 
and U is the real part of the cross-correlations.  
During the 2004 sessions, the Position Angle for one of the probes 
was kept at 45$^\circ$;   at this epoch, U is therefore the difference 
of the autocorrelations (and Q is the real part of the 
cross-correlations).  
The different observing procedure at the two epochs was intentional, 
since it gave the opportunity of comparing the errors arising from 
differencing and cross-correlation, which are quite distinct.

\subsection{Alternative polarization observing strategies}

Since full polarization spectroscopy from a dual-channel 
receiver on a single dish can be achieved by a variety of methods, 
we briefly review their respective advantages and disadvantages.  

The Position Angle of the probes on the sky needs to be known, although 
not necessarily set to zero or 45$^\circ$;  but if allowed to be at some 
intermediate angle, then each linear polarization measurement is a 
combination of Q and U that must later be disentangled; furthermore, if 
the Position Angle is changing, this must be 
corrected for within each short integration before summing over the 
typical 10-minute observation.   Our procedure removes this requirement.

We note that, in an alternative strategy, if the two circular 
polarizations are sampled, then autocorrelations provide RHCP and LHCP, 
with their sum yielding total intensity, I, and their difference,  V; 
cross-correlation  (in phase and in quadrature), produces U and Q.   In 
this observing mode, it is the circular polarization 
that is crucially dependent on the precise amplitude scaling of the 
two channels.  The issue of determining the polarization position angle 
remains the same as for the linear probe case.  

For comparison with the NRT observations (Szymczak \& Gerard 2009) 
we note that their signal is derived from two orthogonal feeds, with 
receiver orientation maintained at a constant position angle.  Cross 
correlations are not measured directly, but in  
parallel to the processing of the signal from each feed, the signals are 
fed into an RF hybrid so as to electronically generate RHCP and LHCP 
(and hence V as the difference in Stokes parameter terminology).  
The direct processing provides I as the sum, and Q as the difference;  a 
separate observation is made after feed rotation by 45$^\circ$ which 
provides U as the difference and a repeat measurement of I and V.  

Strategies with interferometric arrays are less varied.  For 
the most recent capabilities on the ATCA with CABB, 
we note that linearly polarized feeds are used and the correlations 
(between antennas) of the parallel feeds are recorded, equivalent to 
single dish auto-correlations, and also the correlations (between 
antennas) of the orthogonal feeds, equivalent to the 
single dish cross-correlations.    
However, no physical receiver rotation is employed, with the 
corresponding equivalent rotation performed in software at the 
reduction stage.  
VLA and VLBA procedures are similar to the ATCA except that the `native' 
polarization sampling is performed with circularly polarized feeds.

\subsection{Final spectral line reduction in {\sc asap}}

The ATNF program {\sc asap} (ATNF Spectral line Analysis Package) 
was used for 
final reduction of data.  Amplitude scaling was applied so that the 
final intensity calibration is relative to the source 1934-638, for 
which a total intensity at 1666 MHz of 14.16 Jy has been adopted.    
This is equivalent to a flux density of 36.4 Jy for Hydra A, and thus 
essentially the same as used in earlier work (Caswell \& Haynes 
1987 and references therein) where Hydra A was used as a calibrator 
with assumed flux density 36 Jy.  

The recorded data comprised 8192-channel spectra across 4 MHz for each of 
the autocorrelations from the orthogonal linear probes, XX, and YY, and 
for the real and imaginary parts of their cross-correlation, ReXY and 
ImXY.  

In 2005, with the chosen orientation of the receiver, Q=XX--YY;  
For the receiver orientation used in 2004, U=XX--YY.  
It is useful to note that subsequent corrections may modify signs of Q 
or U, (and add a small fraction of V) but, essentially, only Q is 
dependent on the relative gains of the two receiver channels in 2005, 
whereas only U is dependent on the relative gains of the receivers in 
2004.    
Input parameters to  {\sc asap} for each observation include the 
receiver position angle  so as to distinguish these cases.  
Correction of any small phase error between cross-correlation inputs 
using the  {\sc asap} task `rotate\_xyphase' removes any small 
cross-contamination between V and either Q (in 2004) or U (in 2005).    
For the 2004 data, a phase rotation of -5$^\circ$ was required, whereas 
for the 2005 data, no correction was needed.  

From the  $Q$ and $U$ spectra, we also created spectra of total 
linearly polarized intensity, $L =\sqrt{Q^2 + U^2}$ (which we 
abbreviate to LINP in plot labels) and spectra of 
position angle of linear polarization, $\theta = 1/2\tan^{-1}{(U/Q)}$ 
(subsequently referred to as `polarization position angle', 
abbreviated to `ppa').    
\\
Concise graphical display of full polarization spectral data presents a 
difficult choice, and some degree of compromise.
\\
We choose to display the results as two panels of spectra for each 
transition, showing: 
\\
1.  	Spectra of  RHCP and LHCP, respectively  (I+V)/2 and (I--V)/2, 
overlaid with the linear polarization, $\sqrt{Q^2 + U^2}$. 
\\
2.  	Overlaid spectra of I with Q and U.
\\
We note that RHCP, LHCP, Q and U are a full represntation of the Stokes 
parameters; we have not chosen to display V since the individual RHCP 
and LHCP spectra are usually more informative for OH masers with large 
Zeeman splitting in commonly encountered magnetic fields of several mG.  
Furthermore, where the percentage of circular polarization is especially 
interesting (approaching 100 per cent), this is clear from the RHCP and 
LHCP spectra.   
\\
The added superposed plots of LINP and I are useful as an indication of 
fractional polarization.   However, the presence of weak linear 
polarization is most reliably seen on the Q 
and U spectra.  The value of linearly polarized intensity (derived from 
Q and U as noted above) has a positive noise bias;  we have chosen to 
limit the display to values exceeding   5 times the rms 
noise level, where the positive noise bias of real signals becomes 
insignificant, and very few spurious emission noise spikes remain.  
Features with high linear polarization are clearly evident from 
comparing this plot with the plot of total intensity. 
The polarization position angle is a noisy quantity which we have chosen 
not to show, but the plots of Q and U indicate it qualitatively, 
noting that ppa = $1/2\tan^{-1}{(U/Q)}$; in cases of special 
interest, the ppa  is discussed in the source notes.

\section[]{Results}

\begin {table*}

\caption{Polarization measurements of OH masers at 1665 and 1667 MHz. 
References to positions are: 
FC89 (Forster \& Caswell 1989); C98 (Caswell 1998);  A00 (Argon et al. 
2000) C03 (Caswell 2003); CG11 (Caswell \& Green 2011) and `text' 
refers to notes of Section 3.3.
Peak intensities refer to the stronger circular polarization; the 
intensity values shown in boldface correspond to the epoch for which the 
spectra are shown in Fig. 1. Lin(5,7) 
summarizes  linear polarization at 1665 and 1667 MHz, with  `P' denoting  
more than 50 per cent in at least one feature,  `p' denoting detectable 
but weaker polarization, and the absence of any  entry signifying 
no reliable detection of polarization.   Abbreviated (single letter) 
references to 
earlier polarization observations are N (NRT from Szymczak \& 
Gerard 2009);  V (VLBA from Fish et al.. 2006), v (vla from Argon et 
al. 2000) and c (Parkes data from Caswell \& Haynes 1983a,b).  
Column 13 heading `m/OH' refers to the 
intensity ratio of the peak of an 
associated 6668-MHz methanol maser to the highest OH peak. }

\begin{tabular}{llrlcccccclll}

\hline

\multicolumn{1}{c}{Source Name} & \multicolumn{2}{c}{Equatorial 
Coordinates} & \multicolumn{1}{l}{Refpos}  & \multicolumn{2}{c}{Vel. range}& 
\multicolumn{2}{c}{$\rm S_{peak}(2004)$} & \multicolumn{2}{c}{$\rm 
S_{peak}(2005)$} 
& \multicolumn{1}{l}{Lin(5,7)}  & \multicolumn{1}{l}{Refpol} 
& \multicolumn{1}{l}{m/OH}   \\

\ (~~~l,~~~~~~~b~~~)    &       RA(2000)        &       Dec(2000)       
&  &     $ \rm V_{L}$&$ \rm V_{H}$ &  $\rm S_{1665}$    &  $\rm 
S_{1667}$  &  $\rm S_{1665}$ & $\rm S_{1667}$     &       \\

\ (~~~$^\circ$~~~~~~~$^\circ$~~~) & (h~~m~~~s) & (~$^\circ$~~ '~~~~") & 
& \multicolumn{2}{c}{(\kms)} & (Jy) &  (Jy) & (Jy) & (Jy) \\

\hline


350.011$-$1.342	&	17 25 06.50	&	 $-$38 04 00.7	&	C98	&	$-$26.5	&	$-$17.5	&	\bf{4.3} & \bf{4.4}	&	11.0	&	4.5	&	5P;  7p	&	v	&	1/1.9	\\
350.015+0.433	&	17 17 45.44	&	 $-$37 03 12.9	&	C98	&	$-$35	&	$-$32	&	1.0	&	$<0.15$	&	\bf{1.1}& \bf{0.15}	&		&	c	&	1/6.5	\\
350.113+0.095	&	17 19 25.58	&	 $-$37 10 04.5	&	C98	&	$-$80.5	&	$-$65	&	\bf{33}	&  \bf{2.3}	&	34	&	2.6	&	5P;  7p	&	vc	&	$< 1/47$	\\
350.329+0.100	&	17 20 01.61	&	 $-$36 59 15.6	&	C98	&	$-$67	&	$-$63	&	\bf{0.3}&  \bf{0.15}	&	0.3	&	0.15	&	5P	&		&	$< 2.3$	\\
350.686$-$0.491	&	17 23 28.68	&	 $-$37 01 48.1	&	C98	&	$-$16.5	&	$-$13	&	0.8	&	0.2	&	\bf{0.9}&  \bf{0.2}	&		&		&	20	\\
351.160+0.697	&	17 19 57.35	&	 $-$35 57 52.4	&	C98	&	$-$15.5	&	$-$3.5	&	125	&	78	&	\bf{96}	&  \bf{80}	&	5P;  7P	&	vc	&	1/5.6	\\
351.417+0.645	&	17 20 53.39	&	 $-$35 47 01.8	&	C98	&	$-$13	&	$-$6	&	\bf{390}&  \bf{80}	&	400	&	79	&	5p;  7p	&	vc 	&	9.0	\\
351.581$-$0.353	&	17 25 25.25	&	 $-$36 12 45.1	&	C98	&	$-$102	&	$-$89	&	7.1	&	0.3	&	\bf{7.0}& \bf{0.2}	&	5P	&	vc	&	6.7	\\
351.775$-$0.536	&	17 26 42.56	&	 $-$36 09 17.6	&	C98	&	$-$36	&	8	&	190	&	22	&	\bf{85}	& \bf{22}	&	5P;  7P	&	Vvc	&	2.7	\\
352.161+0.200	&	17 24 46.28	&	 $-$35 25 20.2	&	C98	&	$-$43	&	$-$41	&	2.45	&	$<0.2$	&	\bf{2.4}& \bf{0.2}	&	5p	&	c	&	$< 1/8$	\\
352.517$-$0.155	&	17 27 11.34	&	 $-$35 19 32.2	&	C98	&	$-$56	&	$-$43	&	3.8	&	1.9	&	\bf{3.9}& \bf{2.2}	&	5p;  7p	&	c	&	2.5	\\
352.630$-$1.067	&	17 31 13.91	&	 $-$35 44 08.4	&	C98	&	$-$6.5	&	6	&	1.3	&	0.45	&	\bf{1.2}& \bf{0.55}	&	5p;  7P	&		&	15	\\
353.410$-$0.360	&	17 30 26.20	&	 $-$34 41 45.5	&	C98	&	$-$21	&	$-$18	&	24	&	0.2	&	\bf{26}	& \bf{0.3}	&	5P	&	vc	&	4.5	\\
353.464+0.562	&	17 26 51.56	&	 $-$34 08 24.8	&	C98	&	$-$46	&	$-$43	&	3.3	&	$<0.2$	&	\bf{4.4}& $\bf{<0.2}$	&	5P	&	c	&	2.7	\\
354.615+0.472	&	17 30 17.07	&	 $-$33 13 54.6	&	C98	&	$-$34	&	$-$13	&	4.5	&	2.7	&	\bf{5.5}& \bf{3.0}	&	5P;  7p	&	c	&	30	\\
354.724+0.300	&	17 31 15.52	&	 $-$33 14 05.3	&	C98	&	90	&	97	&	\bf{1.15}& \bf{0.3}	&	1.2	&	0.3	&		&		&	10	\\
355.344+0.147	&	17 33 29.05	&	 $-$32 47 58.2	&	C98	&	13	&	23	&	21.1	&	0.7	&	\bf{22}	& \bf{0.8}	&	5p	&	vc	&	1/2.2	\\
356.662$-$0.264	&	17 38 29.22	&	 $-$31 54 40.6	&	C98	&	$-$57.5	&	$-$40	&	1.5	&	0.3	&	\bf{1.15}& \bf{0.4}	&	5P;  7P	&	c	&	7.0	\\
357.968$-$0.163	&	17 41 20.36	&	 $-$30 45 05.5	&	C98	&	$-$10.5	&	$-$2	&	0.4	&	0.2	&	\bf{0.35}& \bf{0.2}	&	5p	&		&	120	\\
358.387$-$0.483	&	17 43 37.83	&	 $-$30 33 50.2	&	C98	&	$-$9	&	4.5	&	0.55	&	0.3	&	\bf{0.4}& \bf{0.4}	&		&		&	14	\\
359.137+0.032	&	17 43 25.62	&	 $-$29 39 17.3	&	C98	&	$-$7	&	0.5	&	7.4	&	3.5	&	\bf{7.5}& \bf{4.0}	&		&	vc	&	2.1	\\
359.436$-$0.103	&	17 44 40.54	&	 $-$29 28 15.1	&	C98	&	$-$53.5	&	$-$50.5	&	6.6	&	0.9	&	\bf{7.0}& \bf{0.7}	&	5p	&	vc	&	10	\\
359.615$-$0.243	&	17 45 39.07	&	 $-$29 23 29.1	&	C98	&	10	&	26	&	7.5	&	2.3	&	\bf{8.0}& \bf{1.9}	&	5p;  7p	&	c	&	4.9	\\
359.970$-$0.457	&	17 47 20.17	&	 $-$29 11 58.8	&	C98	&	14	&	23.5	&	15.0	&	0.7	&	\bf{19.8}& \bf{0.3}	&	5P;  7p	&	vc	&	1/8.3	\\
0.376+0.040	&	17 46 21.38	&	 $-$28 35 39.2	&	C98	&	28.5	&	40	&	\bf{6.3}& \bf{3.9}	&	6.0	&	3.4	&	5p;  7p	&	Nvc	&	1/3	\\
0.496+0.188	&	17 46 04.03	&	 $-$28 24 52.6	&	C98	&	$-$10	&	$-$5	&	\bf{0.4}& \bf{$<0.2$}	&	0.4	&	$<0.2$	&		&		&	62	\\
0.546$-$0.852	&	17 50 14.52	&	 $-$28 54 31.5	&	C98	&	4	&	20.5	&	\bf{4.2}& \bf{8.8}	&	3.8	&	8.6	&	7p	&	Nvc	&	14.8	\\
0.658$-$0.042	&	17 47 20.47	&	 $-$28 23 45.6	&	C98	&	65	&	77	&	210	&	27	&	\bf{220}& \bf{27}	&	5p	&	Nvc	&	$<1/220$	\\
0.666$-$0.035	&	17 47 20.12	&	 $-$28 23 06.2	&	C98	&	45	&	62	&	30	&	29	&	\bf{30}	& \bf{37}	&	5p;  7p	&	Nvc	&	$<1/30$	\\
2.143+0.009	&	17 50 36.13	&	 $-$27 05 47.2	&	C98	&	58	&	66.5	&	8.6	&	0.5	&	\bf{8.0}& \bf{0.9}	&	5p	&	Nvc	&	1/1.1	\\
3.910+0.001	&	17 54 38.77	&	 $-$25 34 45.2	&	C98	&	17	&	20	&	\bf{4.5}& \bf{0.3}	&	4.5	&	0.3	&	5p	&	c	&	1.1	\\
5.885$-$0.392	&	18 00 30.39	&	 $-$24 04 04.2	&	C98	&	$-$44	&	18	&	5.8	&	10.0	&	\bf{6.8}& \bf{10.0}	&	5P;  7p	&	Vv	&	1/10	\\
6.048$-$1.447	&	18 04 53.15	&	 $-$24 26 42.2	&	C98	&	10	&	12	&	6.6	&	0.65	&	\bf{7.5}& \bf{0.7}	&		&	Nv	&	$< 1/25$	\\
6.795$-$0.257	&	18 01 57.72	&	 $-$23 12 34.6	&	C98	&	13	&	24	&	4.6	&	1.3	&	\bf{4.6}& \bf{1.3}	&	5p	&		&	20	\\
8.669$-$0.356	&	18 06 19.01	&	 $-$21 37 32.8	&	C98	&	38	&	42	&	\bf{2.5}& \bf{n}	&	2.4	&	n	&	5p	&	c	&	4.0	\\
8.683$-$0.368	&	18 06 23.46	&	 $-$21 37 10.2	&	C98	&	35	&	45.5	&	\bf{3.4}& \bf{1.4}	&	3.9	&	1.4	&	5p;  7P	&	c	&	30	\\
9.619+0.193	&	18 06 14.92	&	 $-$20 31 44.0	&	C98	&	5	&	6	&	4.0	&	$<1.0$	&	\bf{3.2}& $\bf{<1.0}$	&	5P	&	NVvc	&	22	\\
9.620+0.194	&	18 06 14.87	&	 $-$20 31 36.7	&	C98	&	3	&	25	&	1.2	&	5.0	&	\bf{1.2}& \bf{5.0}	&		&	NVvc	&	$< 1/5$	\\
9.621+0.196	&	18 06 14.69	&	 $-$20 31 32.1	&	C98	&	$-$4	&	2	&	8.5	&	9.0	&	\bf{8.6}& \bf{9.0}	&	5p;  7P	&	NVvc	&	578	\\
10.444$-$0.018	&	18 08 44.88	&	 $-$19 54 37.9	&	C98	&	74	&	77	&	0.4	&	2.4	&	\bf{0.4}& \bf{2.4}	&	5P;  7P	&		&	10	\\
10.473+0.027	&	18 08 38.25	&	 $-$19 51 49.4	&	C98	&	43.5	&	71	&	1.2	&	1.4	&	\bf{0.9}& \bf{1.5}	&	5p;  7P	&		&	18.7	\\
10.480+0.033	&	18 08 37.87	&	 $-$19 51 16.1	&	C98	&	65	&	67	&	0.6	&	$<0.2$	&	\bf{0.5}& \bf{n}	&		&		&	44	\\
10.623$-$0.383	&	18 10 28.67	&	 $-$19 55 49.1	&	C98	&	$-$3.5	&	3.5	&	25	&	20	&	\bf{27}	& \bf{21}	&	5P;  7P	&	NVvc	&	$< 1/27$	\\
11.034+0.062	&	18 09 39.86	&	 $-$19 21 21.2	&	C98 (FC89)	&	19	&	28	&	6.0	&$<0.2$	&	\bf{6.0}& $\bf{<0.2}$	&	5p	&	Nc	&	1/10	\\
11.113+0.050	&	18 09 53.3	&	 $-$19 17 32~~  &	text	&	2.5	&	17.5	&	0.6	&	0.6	&	\bf{0.65}& \bf{0.7}	&	7p	&		&	$< 1.7$	\\
11.903$-$0.102	&	18 12 02.70	&	 $-$18 40 24.7	&	text	&	34	&	35	&	1.0	&	$<0.3$	&	\bf{1.0}& $\bf{<0.3}$	&	5P	&	 	&	11.3	\\
11.904$-$0.141	&	18 12 11.46	&	 $-$18 41 29.6	&	C98 (FC89)	&	39.5	&	45	&	10.0	&	1.6&	\bf{11.0}& \bf{1.7}	&	5p	&	Nc	&	5.9	\\
12.026$-$0.032	&	18 12 01.88	&	 $-$18 31 55.6	&	text	&	103	&	110	&	0.45	&	0.3	&	\bf{0.45}& \bf{0.3}	&		&	c	&	213	\\
12.200$-$0.033	&	18 12 23.44	&	 $-$18 22 49.3	&	text	&	46	&	48	&	 $-$	&	 $-$	&	\bf{0.8}& \bf{0.9}	&		&		&	15.5	\\
12.209$-$0.103	&	18 12 39.91	&	 $-$18 24 18.1	&	text	&	12	&	27	&	3.5	&	$<0.4$	&	\bf{5.5}& \bf{1.5}	&	5p 	&	Nvc	&	2.0	\\
12.216$-$0.119	&	18 12 44.45	&	 $-$18 24 24.6	&	C98 (FC89)	&	26	&	32	&	12.5	&	5.0&	\bf{14.0}& \bf{5.2}	&	5p;  7p	&	Nvc 	&	$< 1/14$ \\
12.680$-$0.183	&	18 13 54.79	&	 $-$18 01 47.9	&	C98	&	55	&	67	&	14.5	&	4.8	&	\bf{14.5}& \bf{5.6}	&	5p	&	Nvc	&	24	\\
12.889+0.489	&	18 11 51.49	&	 $-$17 31 30.8	&	C98	&	31	&	35.5	&	6.5	&	1.9	&	\bf{6.9}& \bf{1.9}	&	5P;  7P	&	Nv	&	10	\\
12.908$-$0.260	&	18 14 39.53	&	 $-$17 52 01.1	&	C98	&	28	&	42	&	49	&	55	&	\bf{47}	& \bf{75}	&	5P;  7P	&	Nvc	&	5.5	\\

\hline

\end{tabular}
\label{}
\end{table*}

\begin {table*}

\addtocounter{table}{-1}

\caption{\textit{- continued p2 of 2}}

\begin{tabular}{llrlcccccclll}

\hline

\multicolumn{1}{c}{Source Name} & \multicolumn{2}{c}{Equatorial 
Coordinates} & \multicolumn{1}{l}{Refpos}  & \multicolumn{2}{c}{Vel. range}& 
\multicolumn{2}{c}{$\rm S_{peak}(2004)$} & \multicolumn{2}{c}{$\rm 
S_{peak}(2005)$} 
& \multicolumn{1}{l}{Lin(5,7)}  & \multicolumn{1}{l}{Refpol} 
& \multicolumn{1}{l}{m/OH}   \\

\ (~~~l,~~~~~~~b~~~)    &       RA(2000)        &       Dec(2000)       
&  &     $ \rm V_{L}$&$ \rm V_{H}$ &  $\rm S_{1665}$    &  $\rm 
S_{1667}$  &  $\rm S_{1665}$ & $\rm S_{1667}$     &       \\

\ (~~~$^\circ$~~~~~~~$^\circ$~~~) & (h~~m~~~s) & (~$^\circ$~~ '~~~~") & 
& \multicolumn{2}{c}{(\kms)} & (Jy) &  (Jy) & (Jy) & (Jy) \\

\hline

13.656$-$0.599	&	18 17 24.27	&	 $-$17 22 13.4	&	text	&	40.5	&	63.5	&	\bf{31}	& \bf{2.3}	&	24	&	2.4	&	5P;  7p	&	N	&	1.1	\\
14.166$-$0.061	&	18 16 26.05	&	 $-$16 39 57.1	&	text	&	26.5	&	69	&	\bf{0.4}& \bf{3.0}	&	 $-$	&	 $-$	&	7p	&	c	&	$< 1/7.5$	\\
15.034$-$0.677	&	18 20 24.75	&	 $-$16 11 34.9	&	C98 (FC89)	&	19	&	23	&\bf{4.5}& $\bf{<1.0}$	&	5.0	&	$<1.0$	&		&		&	10.7	\\
16.585$-$0.051	&	18 21 09.21	&	 $-$14 31 48.4	&	FC89	&	56.5	&	65	&	0.8	&	0.35	&	\bf{0.65}& \bf{0.35}	&		&	Nc	&	55.4	\\
16.864$-$2.159	&	18 29 24.43	&	 $-$15 16 05.0	&	text	&	14.5	&	24	&	\bf{4.3}& \bf{0.65}	&	4.3	&	0.6	&	5p	&		&	6.5	\\
17.639+0.158	&	18 22 26.31	&	 $-$13 30 11.8	&	A00	&	19.5	&	42	&	3.2	&	0.8	&	\bf{3.5}& \bf{1.5}	&	5P;  7P	&	v	&	7.1	\\
18.461$-$0.004	&	18 24 36.36	&	 $-$12 51 07.5	&	text	&	46	&	55	&	4.5	&	1.1	&	\bf{4.6}& \bf{1.1}	&	5p;  7p	&	Nc	&	5.4	\\
18.551+0.035	&	18 24 38.17	&	 $-$12 45 15.4	&	text	&	34	&	37.5	&	0.45	&	1.6	&	\bf{0.4}& \bf{1.6}	&		&	c	&	$<1/2.3$	\\
19.471+0.170	&	18 25 54.53	&	 $-$11 52 39.5	&	text	&	9.5	&	16	&	\bf{3.0}& \bf{0.2}	&	4.0	&	0.3	&	5p	&	c	&	1.0	\\
19.473+0.170	&	18 25 54.91	&	 $-$11 52 31.5	&	text	&	17	&	22.5	&	\bf{24}	& \bf{5.8}	&	19.0	&	6.5	&	5P;  7p	&	Nc	&	1/1.4	\\
19.486+0.151	&	18 26 00.48	&	 $-$11 52 21.8	&	text	&	23	&	25	&	\bf{2.4}& \bf{1.3}	&	2.0	&	1.0	&	5p;  7p	&	Nc	&	2.5	\\
19.609$-$0.234	&	18 27 38.08	&	 $-$11 56 36.5	&	FC89	&	18	&	45	&	5.1	&	1.2	&	\bf{5.0}& \bf{1.1}	&	5p	&	c	&	1/7.7	\\
19.612$-$0.135	&	18 27 16.84	&	 $-$11 53 41.0	&	FC89	&	51.5	&	55.5	&	0.85	&	$<0.2$	&	\bf{1.1}& $\bf{<0.2}$	&	5P	&	Nc	&	$< 1/1.6$	\\
20.081$-$0.135	&	18 28 10.28	&	 $-$11 28 48~~	&	A00	&	42	&	51	&	\bf{11.0}& \bf{0.7}	&	11.5	&	0.8	&	5p	&	Nvc	&	1/3.7	\\
20.237+0.065	&	18 27 44.54	&	 $-$11 14 56.3	&	text	&	70.5	&	77	&	3.0	&	0.55	&	\bf{3.0}& \bf{0.6}	&	5p;  7P	&	Nc	&	27	\\
21.795$-$0.128	&	18 31 22.83 	&	 $-$09 57 26.8	&	text	&	36.5	&	40	&	 $-$	&	 $-$	&	\bf{0.7}& $\bf{<0.2}$	&	5p	&		&	$< 1.4$	\\
21.880+0.014	&	18 31 01.8	&	 $-$09 49 01~~	&	text	&	19.5	&	21.5	&	0.4	&	$<0.2$	&	\bf{0.35}& $\bf{<0.2}$	&		&		&	43	\\
22.435$-$0.169	&	18 32 43.83	&	 $-$09 24 32.8	&	text	&	24.5	&	35.5	&	9.7	&	0.2	&	\bf{10.0}& \bf{0.2}	&	5P	&	Nc	&	2.0	\\
23.010$-$0.411	&	18 34 40.26	&	 $-$09 00 37.5	&	FC89	&	64	&	84	&	2.6	&	1.2	&	\bf{2.4}& \bf{0.9}	&	5P	&	Nc	&	167	\\
23.437$-$0.184	&	18 34 39.26	&	 $-$08 31 39.6	&	FC89, text	&	105	&	107	&\bf{11.2}& \bf{0.5}	&	13.0	&	0.9	&		&	Nc	&	3.7	\\
23.440$-$0.182	&	18 34 39.19	&	 $-$08 31 25.7	&	FC89, text	&	100.5	&	105	&\bf{2.0}& \bf{1.1}	&	2.0	&	1.2	&	5p	&	Nc	&	8.0	\\
23.456$-$0.200	&	18 34 44.70	&	 $-$08 31 03.6	&	text	&	67	&	69	&	2.7	&	$<0.3$	&	\bf{2.6}& $\bf{<0.3}$	&		&		&	$< 1/3.7$	\\
24.329+0.145	&	18 35 08.09	&	 $-$07 35 03.6	&	CG11	&	63	&	112	&	\bf{0.2}& \bf{5.9}	&	0.2	&	4.6	&	7P	&	c	&	1.4	\\
24.494$-$0.039	&	18 36 05.84	&	 $-$07 31 19.5	&	CG11	&	108	&	117	&	n	&	n	&	\bf{1.1}& \bf{0.9}	&		&	N	&	11	\\
24.790+0.084	&	18 36 12.45	&	 $-$07 12 10.6	&	CG11; FC89	&	100	&	 114.5	&\bf{5.0}& \bf{0.8}	&	3.5	&	0.9	&	5P	&	Nc	&	16	\\
28.147$-$0.004	&	18 42 42.57	&	 $-$04 15 35.5	&	A00;  text	&	100	&	104	&\bf{4.5}& \bf{n}	&	4.2	&	0.3	&	5p	&	Nvc	&	14	\\
28.201$-$0.049	&	18 42 58.07	&	 $-$04 13 57.0	&	A00	&	92	&	100	&	\bf{12.0}& \bf{2.8}	&	12.5	&	2.8	&	5p;  7p	&	Nvc	&	1/3.4	\\
28.862+0.065	&	18 43 46.34	&	 $-$03 35 29.9	&	FC89	&	100	&	108	&	\bf{13.5}& \bf{3.4}	&	15.0	&	3.7	&	5p;  7p	&	c	&	1/12	\\
29.862$-$0.040	&	18 45 58.52	&	 $-$02 45 01.7	&	text	&	101	&	104.5	&	1.2	&	$<0.2$	&	\bf{1.3}& $\bf{<0.2}$	&	5P	&	N	&	52	\\
29.956$-$0.015	&	18 46 03.74	&	 $-$02 39 21.4	&	text	&	97.5	&	99	&	1.3	&	$<0.2$	&	\bf{1.4}& $\bf{<0.2}$	&		&	N	&	143	\\
30.589$-$0.043	&	18 47 18.81	&	 $-$02 06 16.9	&	A00	&	35.5	&	44.5	&	7.0	&	7.0	&	\bf{8.5}& \bf{7.5}	&	7p	&	Nvc	&	1/1.1	\\
30.703$-$0.069	&	18 47 36.76	&	 $-$02 00 54.5	&	A00	&	79	&	97.5	&	14.0	&	9.0	&	\bf{14.0}& \bf{10.0}	&	5P;  7P	&	Nvc	&	6.4	\\
30.820$-$0.060	&	18 47 48.00	&	 $-$01 54 24.4	&	text	&	100	&	108	&	3.0	&	1.5	&	\bf{3.0}& \bf{1.0}	&		&	Nc	&	6.0	\\
30.788+0.204	&	18 46 48.1	&	 $-$01 48 54~~	&	text	&	77	&	84	&	\bf{1.0}& $\bf{<0.2}$	&	1.0	&	$<0.2$	&		&	N	&	22	\\
30.819+0.273	&	18 46 36.73	&	 $-$01 45 21.0	&	text	&	99	&	105	&	\bf{3.5}& \bf{1.1}	&	4.0	&	1.1	&	5p	&	Nc	&	2.0	\\
30.897+0.161	&	18 47 09.14	&	 $-$01 44 17~~	&	text	&	107.5	&	108.5	&	\bf{2.0}&  \bf{text}	&	 $-$	&	 $-$	&		&		&	37	\\
31.243$-$0.111	&	18 48 45.10	&	 $-$01 33 13.3	&	FC89	&	18.5	&	26	&	\bf{2.0}& $\bf{<0.2}$	&	1.9	&	$<0.2$	&	5p	&	c	&	$< 1/38$	\\
31.281+0.061	&	18 48 12.48	&	 $-$01 26 30.0	&	FC89	&	103	&	108.5	&	\bf{2.2}& \bf{1.1}	&	2.2	&	1.8	&	5p	&	Nc	&	36	\\
31.394$-$0.258	&	18 49 33.05	&	 $-$01 29 09~~	&	text	&	81	&	89.5	&	1.4	&	0.4	&	\bf{1.4}& \bf{0.3}	&		&		&	$< 1/1.4$	\\
31.412+0.307	&	18 47 34.25	&	 $-$01 12 46.1	&	A00	&	86.5	&	113	&	\bf{2.3}& \bf{10.0}	&	2.8	&	10.5	&	5P;  7p	&	Nv	&	1.1	\\
32.744$-$0.076	&	18 51 21.89	&	 $-$00 12 05.5	&	FC89;  A00	&	25.5	&	41	&4.5	&	1.5	&	\bf{2.8}& \bf{1.4}	&	5p;  7P	&	Nvc	&	16	\\
33.133$-$0.092	&	18 52 08.01	&	 00 08 12.3	&	FC89	&	72	&	82	&	\bf{15.5}& \bf{0.95}	&	16.0	&	0.9	&	5p	&	Nc	&	1/1.3	\\
34.258+0.153	&	18 53 18.73	&	 01 15 00.3	&	FC89;  A00	&	55	&	63	&78	&	112	&	\bf{78}	& \bf{118}	&	5p;  7P	&	NVvc	&	1/4.2	\\
34.411+0.231	&	18 53 18.83	&	 01 25 18.2	&	text	&	53	&	57	&	2.5	&	1.8	&	\bf{4.5}& \bf{2.5}	&	5P	&		&	$<1/4$	\\
35.025+0.350	&	18 54 00.68	&	 02 01 19.3	&	FC89;  A00	&	40.5	&	51.5	&19.5	&	$<0.2$	&	\bf{19.5}& \bf{0.2}	&	5p	&	Nvc	&	2.9	\\
35.197$-$0.743	&	18 58 12.97	&	01 40 37.7	&	FC89	&	24.5	&	38	&	\bf{15.0}& \bf{0.9}	&	13.0	&	0.3	&	5P;  7P	&	Nvc	&	8.0	\\
35.201$-$1.736	&	19 01 45.60	&	  01 13 33.3	&	FC89	&	38.5	&	46	&	6.0	&	5.5	&	\bf{9.0}& \bf{6.0}	&	5p;  7p	&	Nvc	&	81	\\
35.578$-$0.030	&	18 56 22.54	&	 02 20 28.1	&	FC89;  A00	&	44.5	&	53	&	60	&	20	&\bf{60}& \bf{9.8}	&	5P;  7p	&	Vvc	&	$< 1/240$	\\
40.426+0.700	&	19 02 39.62	&	 06 59 12.0	&	text	&	7.5	&	17	&	5.0	&	0.2	&	\bf{7.8}& \bf{0.3}	&	5p	&	N	&	2.1	\\
40.623$-$0.138	&	19 06 01.64	&	 06 46 36.5	&	FC89;  A00	&	24.5	&	36.5	&	112	&	15.0	&\bf{119}& \bf{15.0}	&	5P;  7p	&	NVvc	&	1/9.2	\\

\hline

\end{tabular}
\label{}
\end{table*}

Source parameters and a summary of the new results are given in Table 
1.  Column 1 gives the Galactic coordinates, used also as a source name, 
and derived from the more precise equatorial coordinates given in 
columns 2 and 3.  Column 4 gives a reference to a position measurement 
for the OH emission, with `text' referring to text of Section 3.3, 
mostly related to previously unpublished measurements of 23 sources with 
the ATCA.  The velocity range of emission is given in columns 5 and 6, 
and in a few cases is larger than seen on the displayed spectra since 
it encompasses features at outlying velocities that have been prominent 
in the past but have subsequently weakened.   
The values of peak intensity of emission, for epochs 2004 and 2005, at 
both 1665 and 1667 MHz, are given in columns 7-10, listing the highest 
peak seen in the circular polarization spectra;  non-detections are 
given in the format $< 0.2 $ for no features above 0.2 Jy, etc.; a 
dash indicates no measurement available.  Boldface font identifies the 
epoch of the spectra selected for display in Fig. 1.  

A concise indication of linear polarization detectability from the 
present spectra is given in column 11, with 5P and 7P (upper 
case P) indicating the 
presence of a feature with more than 50 per cent at 1665 and 1667 MHz 
respectively and 5p and 7p (lower case p) indicating our clear detection 
of linear polarization, but not above 50 per cent in any feature.  
References to past published polarization spectra with 
comparable sensitivity are given in column 12, but noting that VLA and 
Parkes spectra (references v and c) are limited to circular 
polarization.  

Column 13 is an indication of the relative intensities of maser 
emission at the 6668-MHz transition of methanol and the stronger of the 
ground-state 1665 or 1667-MHz transition; this ratio, which we have 
evaluated from the highest peak spectral intensity from a methanol 
spectrum and the highest peak of OH emission (generally taken from 
the circularly polarized spectrum displayed here) is believed to be 
an indicator of the evolutionary stage of the high-mass star formation 
maser site (Caswell 1997, 1998) and is discussed further in Section 
4.1.


\subsection{Spectra}

Spectra of the 104 maser sites are presented in Figure 1, but only 92 
plots were required since, in a few instances, a single plot is 
sufficient for several adjacent sites that are in closely spaced 
clusters.

For the majority (81) of the plots, a velocity range of 30 \kms\ is 
sufficient to show all detected features, and display the fine detail 
present.  
Sources with very large velocity extents are displayed with larger 
ranges of up to 45 \kms\ (e.g. 351.775--0.536 and 14.166--0.061).
In two cases (5.885--0.392 and 24.329--0.039), a velocity 
range of at least 60 \kms\ was required, and has been split between two 
adjacent spectra  so as to allow recognition of fine detail over 
this large  range.  

Spectra have a channel separation of 0.488 kHz 
(equivalent to 0.088 \kms) and have not been smoothed, so the 
`resolution', full-width to half-maximum (FWHM), is 1.21 times the 
channel separation.  For comparison, we note 
that this is the same resolution  used for the VLBA data of W3(OH) by 
Wright et al. (2004a,b), who were able 
to derive simultaneous spectra for all four ground-state transitions (in 
the four IFs available), but limited to a velocity coverage of 11 
\kms\ (i.e. 128 channels over 62.5 kHz for each transition).  
In VLBA measurements by Fish et al. (2005), a wider velocity range was 
chosen, but at the expense of lower spectral resolution.  

For our data, and for a typical source observed with integration time of 
10 min, the rms noise level on a spectrum 
at full spectral resolution is 0.05 Jy.  Some targets have higher noise 
due to a high background sky noise, the most extreme example being  
15.034--0.677.  Other targets were observed with 
longer integration times, as long as 40 min for 14.166--0.061,  and have 
accordingly lower noise levels.  By chance, these two 
examples are close in Galactic longitude, and thus displayed side by 
side, so that the factor of 10 in noise level is very noticeable.  

In some spectra, most notably 15.034--0.677, there are absorption 
features which have velocity ranges of at least a few \kms\ and 
occasionally as much as 20 \kms.  Absorption features are spatially 
much broader (e.g. several arcmin) than the maser features (typically 
less than 0.01 arcsec) and thus, in single dish spectra, absorption 
features can be very prominent relative to the maser emission because 
they fill a much larger fraction of the beam.  Since they represent real 
structure, we have not attempted to remove these broad features in our 
reduction procedure.  
Narrow spikes of radio frequency interference are evident 
on 1667-MHz spectra of 12.026--0.032, 24.494--0.039, 29.862--0.040,  
29.956--0.015 and  30.820--0.060, as detailed in the notes for these 
sites.

\subsection{Other OH  data sets consulted for comparison}

Earlier OH observations have been consulted for each source, and reveal 
that many sources merit individual discussion highlighting unusual 
properties.  For compiling the resulting source notes, four 
datasets have been especially useful, all of them providing high 
spectral resolution and sensitivity comparable to ours.  The 
measurements from the NRT (Szymczak \& Gerard 2009) and from the VLBA 
(Fish et al. 2005) provide full Stokes polarization information, whereas 
the VLA (Argon et al. 2000) and earlier Parkes data (Caswell \& Haynes 
1983a,b) are limited to circular polarization.  
We now summarize some aspects  of these earlier datasets that are 
especially relevant to the comparisons presented in the source notes.

\subsubsection{NRT observations}

Our convention for the calculation and scaling of I, Q, U and V 
is the same as for the NRT results (Szymczak \& Gerard 2009);  our 
display is also similar to their  Fig. C1 but with a difference in 
the display of RHCP and LHCP, where our convention, I=R+L, contrasts 
with the NRT display which uses the less common convention, 
I=(R+L)/2, (with the consequence that individual RHCP or LHCP can 
be as much as twice as large as total intensity, I).  

The NRT coverage does not include Galactic longitudes 350$^\circ$ to 
360$^\circ$.  Between 0$^\circ$ and 40.6$^\circ$, we have 50 targets in 
common, amongst which we note in the NRT data 
an interchange of R and L labels for 0.546--0.852 at 1665 MHz (but not 
1667 MHz); and 24.51--0.05  for both 1665 and 1667 MHz.   However, all 
values for V are plotted with correct sign (with the same convention as 
ours, V=R--L).  

In Table B1, it appears that the listed polarization position angle for 
the strongly linearly polarized 1667-MHz feature of source 34.25+0.16 at 
velocity  58.62 \kms\ should be --42.31$^\circ$ 
rather than +42.31$^\circ$, presumably a printing error since no other 
discrepancies were evident.  

\textit{Source names and positions}.  

The NRT labelled source names are the 
approximate coordinates of the methanol targets, and thus sometimes 
significantly differ from ours, which used the best available OH maser  
positions.  
The equivalence of the NRT approximate 
position with the more precise position is usually obvious, but here we 
note a few instances where it may not otherwise be clear: 
10.96+0.02 is 11.034+0.062; 12.21-0.09 is 12.209-0.102; 13.72-0.52 is 
13.656-0.599;  19.49+0.14 is a blend of 19.486+0.151 (weak) and 
19.474+0.169 (stronger but off centre);   
22.34-0.16 - all features in the displayed spectrum are 
sidelobe responses to 22.45-0.17 (=22.435-0.169);  24.51-0.05 is 
24.494-0.039; 31.27+0.06 is 31.281+0.061. 

\textit{Blending of nearby sources}.

Szymczak \& Gerard (2009) remark on the blending of 0.658-0.043 with 
0.666-0.034, 12.21-0.09 with 12.216-0.117 and 43.165+0.015 with 
43.167+0.010.  In our notes of section 3.3, we draw attention to 
blending of several other sites, and incomplete velocity coverage (e.g. 
0.666-0.035).  The site 9.622+0.195 includes blended features from 
9.619+0.193 and 9.620+0.194, the latter with incomplete velocity 
coverage.  
For 29.86-0.05, features at velocity greater than +100 \kms\ are from 
this site (more precisely 29.862-0.040) and emission at lower velocities 
arises from 29.956-0.015, offset about 6 arcmin, chiefly in 
declination;  the spectrum of 30.82-0.05 (more precisely 30.820-0.060) 
shows a sidelobe response to 30.703-0.069 (offset by 7.2 arcmin).


\subsubsection{VLBA comparisons}

The 18 fields studied by Fish et al. (2005) provide observations of 9 
discrete sites in our observing list.
We note that their analysis, where some parameters are derived using  
fitting to individual spots, but others using peak channel brightnesses, 
can lead to some anomalies where there is a sweep of ppa across a 
feature;  the linearly polarized fraction can then be underestimated, 
and thus expectations that the VLBA high spatial 
resolution will lead to higher percentages of linear polarization are 
not necessarily realised.  The available total bandwidth was small and 
causes incomplete velocity coverage for several sources.  


\subsubsection{VLA comparisons}


The VLA dataset of Argon et al. (2000) provides an excellent set of 
reference spectra with good spectral resolution, although limited to 
the two circular polarizations.  They provide good comparisons with the 
present Parkes data at both 1665 and 1667 MHz for 23 sources, and for a 
further 18 sources at just 1665 MHz.  While making the comparisons, we 
noted two small errors which it is convenient to list here and minimize 
confusion when consulting this widely-used database.
 

For 351.582-0.352, there appears to be an error of about 1 \kms\ in the 
velocity labelling, both on the plot and in the listed features, e.g. 
the strongest feature listed at -93.86 is actually at -92.86 \kms.  

The source listed as 351.232+0.682 is spurious, and merely a sidelobe 
(about 2 per cent) of 351.161+0.697 (offset by several arcmin), 
clearly seen from their Fig. 34, which shows 1667-MHz spectra for each 
source (thus allowing a simple comparison), and from the identical 
velocities of listed features for each source in Tables 83 and 84.  


\subsubsection{Earlier Parkes data}

Many of the subsequent comparisons regarding variability relate to 
earlier Parkes data, with spectra from 1980 onwards (Caswell \& Haynes 
1983a,b) displaying good sensitivity and spectral 
resolution in the two circular polarizations.

\subsection{Source notes}

The source notes that follow draw attention to unusual features such as 
exceptionally high linear polarization, large velocity widths, unusual 
ratio of 1665 to 1667 MHz intensity, comparison with 
earlier data, and variability. 
In the case of linear polarization, we make several comparisons with  
measurements from the VLBA, which  cite polarized intensity and the 
polarization  position angle (ppa) in the range 0$^\circ$ to 
+180$^\circ$, 
rather than values of Q and U.  
For a quick qualitative comparison with our displayed values of Q and U, 
it is useful to recall that the values of Q and U for unit linearly  
polarized signal as a function 
of ppa are: ppa 0$^\circ$ (Q=+1); 45$^\circ$ (U=+1); 90$^\circ$ (Q=-1);  
135$^\circ$ (U=-1).  

Information on methanol maser emission at each site was summarized  
in Table 1 by listing the relative peak maser intensities of methanol 
and ground-state main-line OH.  The methanol values were 
taken from the Methanol Multibeam survey (Caswell et al. 2010b, Green et 
al. 2010), except for the 36 sites with Galactic longitude 
between 20 and 41$^\circ$ (a region where the MMB catalogue is in 
preliminary form and not yet published).  For these, available methanol 
data from the literature and our unpublished data are summarized in the 
source notes, and  values in Table 1 are based on these assessments.    
The comparison of methanol to OH intensity is now better than in the 
Caswell (1998) investigation owing to some improved positions of OH in 
the present paper, and many improved methanol positions 
from Caswell (2009) as well as from the Methanol Multibeam 
survey, allowing confirmation or rejection of some 
earlier apparent associations.  

Source notes are omitted for sites which are adequately described  
by the parameters of Table 1 and the spectra, with no known strongly 
unusual properties.

\subparagraph{350.011--1.342} 
  
Emission at 1665 and 1667 MHz shows negligible change between our 
observations of 2004 (shown here), our earlier unpublished 
Parkes observations of 1993, and those from the VLA, also taken in 1993
(Argon et al. 2000).  However there are marked differences from the 
discovery observation in 1985 (Cohen Baart \& Jonas 1988).  Our spectra 
from 2005 remained similar to those of 2004, except for the 1665-MHz 
feature at --18.1 \kms, for which RHCP flared from 2.5 Jy to 11 Jy, 
whereas LHCP and 
linearly polarized emission increased by less than a factor of 2.  All 
other features remained the same to within 10 per cent.  Our 2004 and 
2005 observations are the first to record linearly polarized emission 
for this source, with a high linearly polarized fractional emission of 
50 per cent in the 1665-MHz feature at --19.8 \kms.

\subparagraph{350.329+0.100} 

Weak features at 1665 and 1667 MHz are confined to the 
velocity range --67 to --63 \kms,  and high linear polarization at 
1665 MHz is seen in the  strongest feature, 0.3 Jy, at -63.5 \kms.  
The spectra have been aligned with those of 350.113+0.095 which is 
offset slightly more than 10 arcmin and causes a prominent 
sidelobe response in the velocity range --77 to --67 \kms.

\subparagraph{351.160+0.697} 

A long history of this source  shows that its many strong emission 
features at both 1665 and 1667 MHz from --16 to --3 \kms\ are still 
prominent, and of similar strength.  There is also deep absorption of 
the strong background emission.  Linear polarization is present in 
several features, and notably very high in the three strongest 1665-MHz 
features, remaining similar in 2004 and 2005.   



\subparagraph{351.417+0.645}  

This strong OH maser is well-known, and often referred to by its 
associated strong compact \HII\ region NGC6334f, with OH spectra showing 
deep absorption of the intense background emission.  
Since 1980, the maser emission has shown variations exceeding a factor 
of two for many features, but remains strong at 1665 and 1667 MHz, with 
peaks exceeding 250 and 75 Jy respectively.   
 
Linear polarization of several features is weak but persistent from 2004 
to 2005.

Detailed maps with the Australian LBA (Long Baseline Array) have been 
made of the accompanying strong excited-state OH 
emission at 6035 and 6030 MHz, showing a wealth of Zeeman pairs and 
a well-characterized magnetic field (Caswell Kramer \& Reynolds 2011a).


\subparagraph{351.581--0.353} 

The multiple 1665-MHz features from $-$100 to $-$88 \kms\ seen in 1980 
persist, but with considerable variability.  The 1667-MHz spectrum is 
dominated by absorption, with several very weak peaks of emission.  
High linear polarization is present at 1665 MHz in our 2005 spectrum (as 
displayed) at --96.5 \kms.  The two strongest maser features are at 
1665-MHz RHCP, at velocities --93.8 and --92.8 \kms, similar to our 
unpublished spectra of 1989; the feature at --92.8 \kms\ was the 
strongest in the 1980/1981 published spectrum of Caswell \& Haynes 
(1983).  

We note that the spectrum and table of features for the 1993 VLA 
observation published by Argon et al. (2000) indicate similar peaks but 
at velocities of --94.81 and --93.86 \kms.  The overall similarity with 
our spectrum, but shifted by approximately 1 \kms, suggests that the 
velocity scale of the VLA spectrum is incorrect.   

\subparagraph{351.775--0.536} 
  
This site is of special interest since the OH maser peak at some epochs 
has exceeded 1000 Jy, the largest known.  
Multiple features over the wide range -31 to +8 \kms\ occur at both 1665 
and 1667 MHz, and have remained prominent throughout the period 1980 to 
2004. The strongest feature at most epochs has been LHCP at 1665 MHz 
near -2 \kms, which in 1980 exceeded 1000 Jy, but is now only 10 Jy; its 
close 
neighbour, a RHCP feature at slightly more positive velocity, remains 
prominent but its peak has varied from 190 Jy in 2004 to 85 Jy in 
2005 whereas historically it has commonly been near 200 Jy. 

The large fractional linear polarization of 1667-MHz emission in our 
spectra from 2004 and 2005, at velocity -8.9 \kms, with U positive and Q 
negative  (average ppa approximately 67.5$^\circ$) is similar to that 
seen with the VLBA in 1996.  
The strongest 1667-MHz feature in our 2004 and 2005 spectra is at 
velocity -7 \kms, primarily LHCP but with significant (but lower) linear 
polarization (Q and U negative with Q stronger and thus ppa 
approximately  110$^\circ$).  The corresponding VLBA feature in 1996 was 
also chiefly LHCP polarized but with much lower linear polarization,   
most likely an indication of variability since there is no plausible  
reason why linear polarization in the higher spatial resolution data 
could be reduced instrumentally.  

At 1665 MHz, the feature at -9.2 \kms\ in our 2004 and 2005 data is 
stable with high linear polarization  (U and Q both negative, and thus 
ppa approximately 115$^\circ$) and it 
has similar high linear polarization in the 1996 VLBA  observations, 
with ppa 106$^\circ$.  
In contrast, emission near -2 \kms\ which at most epochs has been the 
strongest feature, was relatively weak in 2004, and even weaker in 2005,  
and primarily RHCP;  the VLBA 1996 data showed strong LHCP emission, 433 
Jy (at -1.98 \kms), weaker RHCP emission 131 Jy (at -1.87 \kms) and 
significant linear polarization of 147 Jy (at -1.92 \kms), clearly a 
complex feature with significant blending, and at least some parts variable.  

The velocity range of this source is very wide, much wider than 
the range covered by VLBA observations.  Outside of the VLBA range, a 
likely Zeeman pair near 
velocity -27 \kms, well-isolated from all other features, was first 
noted 
at both 1665 and 1667 MHz in the 1980 data of Caswell \& Haynes (1983); 
it  persisted in the 1991 VLA observation (Argon et al 2000) and remains 
prominent in our present 2004 and 2005 spectra.   
The most extreme velocity feature recorded was at 1665 MHz, 1993 
July 21 (unpublished Parkes data), at -35.8 \kms.

\subparagraph{352.630-1.067} 
  
Features are present at both 1665 and 1667 MHz but strong variations 
have occurred  since the ATCA measurements of 1996.  
The Q and U spectra show that pronounced linear polarization is present 
in the weak 1667-MHz features.

\subparagraph{353.410-0.360}  

Currently similar at 1665 MHz to 1980 observations but the main feature 
is three times stronger;  these first full polarization 
measurements show that significant linear polarization is present.  
The 1667-MHz spectrum is  
dominated by deep absorption, with a single weak feature of emission 
(LHCP) at both epochs 2004 and 2005.  

A complementary high resolution study of excited state OH maser emission 
at 6035 and 6030 MHz has been made with the Australian LBA (Caswell 
Kramer \& Reynolds 2011a),  showing many Zeeman pairs, all implying 
the same direction of the magnetic field.

\subparagraph{353.464+0.562}   

1665-MHz emission remained closely similar between 2004 and 2005, 
but several times stronger than in 1989.  The dominant feature is LHCP, 
and an almost equally strong adjacent feature has high 
linear polarization, with persistent ppa at our two epochs.   
There has been no confident detection at 1667 MHz.  

\subparagraph{354.615+0.472}

Emission is present over a wide velocity range spanning -34 to 
-14 \kms, but dominated by a RHCP triple-peaked  feature near -15.5 
\kms\ at both 1665 and 1667 MHz.  Significant linear polarization in 
several features is seen in these first full polarization observations.  

\subparagraph{354.724+0.300} 

Prominent absorption is present at both transitions, but maser 
emission is clearly recognised from its circular polarization;   
strongest 1665-MHz emission, wholly  RHCP, is accompanied at similar 
velocity by weaker emission at 1667 MHz, also wholly RHCP.  Weaker 
features at 1665 MHz are offset to more negative velocity, and wholly 
LHCP.  

\subparagraph{355.344+0.147} 

Prominent emission at 1665 MHz shows a clear Zeeman pattern of multiple 
features in a field of -4.3 mG, which has not changed since 1980 
(Caswell \& Haynes 1983a;  Caswell \& Vaile 1995).  Weak 
features at the 1667-MHz transition, seen for the first time in the 
present sensitive observations,  show a similar Zeeman pattern.  
Weak linear polarization at 1665 MHz is seen in some features.

\subparagraph{356.662-0.264} 
  
Earlier spectra from 1980 and 1989 showed a 1665-MHz feature (LHCP)  
and a slightly weaker 1667-MHz feature at slightly more negative 
velocity (-54 \kms), but RHCP.  The current spectrum shows both 
emission features weaker, but 1665-MHz emission is now highly linearly 
polarized with no significant circular polarization, 
and the 1667-MHz feature, while showing net RHCP, is also chiefly 
linearly polarized.  The sensitivity of the current 
spectrum allows weaker features to be seen extending over a wide 
velocity range.  

\subparagraph{357.968-0.163} 

Weak emission features with some circular polarization are clearly 
present at both 1665 and 1667 MHz, but somewhat confused by 
the broad absorption feature.

\subparagraph{358.387-0.483 }  



Positions for two 1665-MHz features detected with the ATCA were 
listed by Caswell (1998), showing a separation of 4 arcsec but believed 
to be a single large site; we cite it here as a single source at the 
mean position, $17^h43^m37.83^s,~-30^{\circ}33{\arcmin}50.2{\arcsec}$.  
Both of the earlier 
measured features prove to be LHCP;  a weaker feature near -4 \kms\ 
shows no net polarization and the emission near  velocity -1.2 \kms\ is 
suggestive of a RHCP and LHCP close pair.  A weak 1667-MHz feature is 
now seen, near velocity +4 \kms, of low intensity, and wholly LHCP.

\subparagraph{359.970-0.457} 

Emission features between 14 and 18 \kms\ at 1665 MHz, seen in 1980, 
persist, with a new RHC polarized feature extending the velocity range 
to +22.5 \kms, on the high velocity side of a deep absorption trough.   
The long-established 1665-MHz features at 14.2 and 15.6 \kms\ remain 
predominantly LHCP and RHCP, respectively, but in these first full 
polarization measurements, linear polarization is also seen, notably 50 
per cent in the latter.

Increased sensitivity allows us now to see weak 1667-MHz emission in the 
main velocity range, and also near 23 \kms\ in 2004,  with peak of 0.7 
Jy and 40 per cent  linearly polarized.  

\subparagraph{0.376+0.040} 
  
An increase in the source intensity, and higher instrumental 
sensitivity, now reveal 
multiple features at both 1665 and 1667 MHz over the velocity range 
28.5  to 40 \kms, all lying within a deep absorption trough.  There were   
no apparent changes between  2004 and 2005, and the 2003 epoch 
measurements with the NRT confirm the significant linear polarization.

\subparagraph{0.658-0.042, 0.666-0.035 and nearby locations}  
Our 12-armin beam combines the emission from the cluster of sites in the 
Sgr B2 region.  Multiple strong features are present over the wide 
velocity range 45 to 77 \kms, at both 1665 and 1667 MHz.  There is 
prominent wide absorption of the strong background continuum emission.  

The higher spatial resolution of VLA measurements  (Argon et al. 2000), 
although limited to circular polarization, clearly demonstrate that our 
spectra are dominated by 0.658-0.042 in the velocity range 66 to 77 
\kms, and by 0.666-0.035 in the range 45.5 to 64 \kms.  Weaker sites, 
0.678-0.027, only at 1665 MHz and with peak of 6 Jy near 70 \kms, and 
0.672-0.031, with peaks of only 4 and 2.5 Jy at 1665 and 1667 MHz, in 
the velocity range 45.5 to 56 \kms, contribute only weakly to the 
overall emission of the two dominant sites.

The 1667-MHz feature of 0.666-0.034  at 52 \kms\ shows the most 
significant linear polarization, of 20 per cent in a predominantly RHCP 
feature;  it lies outside the velocity range covered by the NRT (the 
only other high sensitivity linear polarization measurements towards the 
source).

\subparagraph{5.885-0.392} 

The remarkably wide velocity range of emission seen at 1665 and 1667 
MHz has been known for many years, but only limited  portions have 
been studied in depth, and the full wealth of detail remains to be 
interpreted.  

We show the spectrum from 2005, noting that the spectrum from 2004 is 
similar.  The observations are the first with full polarization coverage 
of the whole  velocity range, and the  display is spread over two 
adjacent panels to capture this wide range with adequate detail;  
it emphasises that the 1665 and 1667-MHz emission is comparably strong 
at positive velocities, whereas at negative velocities, 1667-MHz 
emission is even stronger, and 1665-MHz emssion is markedly weaker.  
Prominent linearly polarized emission is present at 1665 MHz, with an 
especially strong feature at +10.9 \kms, a feature also displaying 
strong LHCP; for this feature, a comparison with  VLBA measurements 
taken in 2001 (Fish et al. 2005) is in good agreement with our results.  
We also find significant polarization  evident in features at -18.2 
and -19.2 \kms, velocities not covered by the VLBA, nor 
by any previous linear polarization measurements.  
1667-MHz emission shows only weak linear polarization at both high and 
low velocities.

We note that an astrometric parallax distance (with uncertainty 7 per 
cent, based on water masers at this site) has recently been reported 
(Motogi et al. 2011a), yielding 1.28 kpc, much smaller than 
previous estimates between 1.8 and 3.8 kpc.  This 
resolves a previous puzzle whereby the angular extent of the maser spot 
distribution and 
associated continuum \HII\ emission is large and the corresponding 
overestimated linear extent was  improbably large;  the linear extent is 
now estimated to be 45 mpc, no longer an extreme value, and comparable  
to several other similar objects (Caswell et al. 2010a).

\subparagraph{6.795-0.257}  

Predominantly LHCP emission features are present, stronger at 1665-MHz  
and  weaker at 1667-MHz;   some linear polarization 
is present at 1665 MHz.

\subparagraph{8.669-0.356 and 8.683-0.368} 

Significant changes have occurred over the past 20 years.  ATCA data 
of 1995 suggest that the first source is weaker and accounts merely for 
some 1665-MHz emission features near 40 \kms\ and at lower velocity.  
The strong narrow RHCP 1665-MHz feature at 42.8 \kms, and perhaps all of 
the emission at 1667 MHz, most likely arises only from the the second 
source (offset by 1 arcmin).  Linear polarization exceeding 50 
percent is present at 1667 MHz. 

At 8.669-0.356, Forster \& Caswell (2000) show compact \HII\ emission 
and there is no clearly associated mid-IR source in GLIMPSE (Gallaway et 
al 2013) so we interpret the site as evolved, with methanol maser 
emission in decline.  In contrast, 
8.683-0.368 has no compact continuum, but has a mid-IR point source 
counterpart,  indicative of a younger object with higher methanol to OH 
ratio, and no obvious radio \HII\ region.  A detailed interpretation 
based on thermal lines of other molecular species is in agreement with 
this (Ren et al. 2012).

\subparagraph{9.619+0.193, 9.620+0.194 and 9.621+0.196}  

The three distinct sources, spread over 15 arcsec, have 1665-MHz 
emission peaks near velocities 5.5, 22.5 and 1.4 \kms\ 
respectively, according to ATCA measurements of 1996 (Caswell 1998).  
Emission from the first site is probably confined to this single 
1665-MHz feature (now seen to be chiefly RHCP at 6 \kms).  

Some features in our spectra may be confused, with emission contributed 
from more than one site.  
However, the second source, 9.620+0.194, has a wide velocity range, 
accounting not only for all 1665 and 1667-MHz emission near +22 \kms, 
but also for 1665 and 1667-MHz emission near 4 \kms, and for 1667-MHz 
emission near 7 \kms.  
The emission near +20 \kms\ is not within the velocity range of 
published data from either the NRT, VLA or VLBA.  
At 9.620+0.194, there is no detectable methanol maser nor any uc\HII\  
region.  

The third site, 9.621+0.196,  accounts not only for emission at 1.4 \kms 
(probably extending to 2 \kms), but for all features at negative 
velocity;  it coincides with the most intense known maser on the 
6668-MHz methanol transition, peaking, like the OH,  at velocity 1.4 
\kms.  

Strong linear polarization is evident, especially at 1667 MHz, where it 
exceeds 50 percent, confirming VLBA and NRT measurements.

\subparagraph{10.444-0.018, 10.473+0.027 and 10.480+0.033}   

There are no previous detailed OH spectra published for these three 
sites.  The sites are spread over 3 arcmin and show  1665-MHz  
emission peaks at velocities 75.5, 51.5 and 66 \kms\ respectively,  
according to ATCA measurements of 1996 (Caswell 1998).  Emission from 
the first site is 6 times stronger at 1667 than at 1665 MHz, and is 
confined to a narrow velocity range from 74 to 77 \kms.  
The third site, 10.480+0.033, is detected only at 1665 MHz, confined to 
a narrow single feature.  Emission from the second site is stronger at 1667 
than at 1665 MHz, and extends over the range 50 to 68 \kms, at least.

With regard to the present observations, the 1667-MHz spectrum is thus 
wholly   accounted for.  However, it is not clear where 1665-MHz 
features at velocities 70.5 \kms\ (LHCP) and 44.5 \kms\ (RHCP) arise 
from, although the 
second site (with a known wider velocity range) seems most likely, and 
the site then has one of the largest velocity spans of our sample.  

The first site shows strong linear polarization at both 1665 and 1667 
MHz.  At the second site, features of 1667-MHz emission near 65 \kms\ 
show 50 per cent linear polarization, in contrast to the feature at 51.8 
\kms\ which is wholly LHCP with no detectable linear polarization.

\subparagraph{10.623-0.383}  

Strong emission, at both 1665 and 1667 MHz, has remained very stable 
since 1982.  The prominent linear and circular polarization at both 
transitions in 2004 and 2005 shows excellent agreement with the NRT and 
VLBA measurements taken in 2003 and 2001 respectively.

\subparagraph{11.034+0.062}  

Emission is limited to 1665 MHz: a broad RHCP feature which has changed 
only slightly since 1982, a weak LHCP feature at 24 \kms, and even 
weaker features near 26 \kms\ (RHCP) and 27 \kms\ (LHCP).  
An apparent weak (0.4-Jy) LHCP feature at velocity 17.1 \kms\ 
is not at this position, but arises from a previously unreported source 
11.113+0.050, offset 5 arcmin (see next note).

\subparagraph{11.113+0.050} 

This source was first noted in Parkes observations in 2004 while 
targeting 11.034+0.062.  From subsequent ATCA observations  (2005 May), 
a position measurement of the strongest emission, at 1667 MHz, at 5 
\kms, yielded 
11.113+0.050 ($18^h09^m53.3^s,~-19^{\circ}17{\arcmin}32{\arcsec}$  with 
rms errors of 8" (=0.6s) and 4.2"), offset 5 arcmin from 11.034+0.062; 
thus each site lies near the half-power point of the Parkes beam when 
targeting the other.  There is no reported methanol maser 
at this site (Green et al. 2010)

\subparagraph{11.904-0.141 and 11.903-0.102} 

11.904-0.141 is a well-known source with emission in the range 
39.5 to 45 \kms.  Our 2005 spectrum shows it to be similar, though 
stronger, than seen in 1982.   

The second source is a weak solitary 1665-MHz feature at 34.3 \kms.  We 
discovered it on our 2004 and 2005 spectra and noticed its absence from 
the NRT spectrum of 11.90$-$0.16 (Szymczak \& Gerard 2009) despite the 
NRT 
observation epoch (2005 August) lying between our measurements.  We 
suggest that the absence of emission from the Nancay spectrum is not due 
to variability, but because it arises from a different site, offset in 
RA from the centre of the narrow Nancay beam.
A likely location is the methanol maser site 11.903-0.102 (Green et al.
2010) with methanol emission in the range 32 to 36.7 \kms, at an
offset in RA from the Nancay target by about 3 arcmin.  Any Nancay
response to OH at this location would thus be reduced in amplitude by 90
per cent, whereas the expected reduction for the Parkes, larger, 12 
arcmin, beam, at an offset from the pointing target of 2 arcmin, is less 
than 10 per cent.  We therefore list 11.903-0.102 as an additional OH 
maser site, citing the precise methanol maser position, pending precise
measurement of the OH position.

\subparagraph{12.026-0.032}

This source had a peak of 4 Jy in 1982 (Caswell \& Haynes 1983b) but, 
because of variability, no precise position measurement was possible 
until our present measurements with the ATCA (OH rms position 
uncertainty 0.4 arcsec) confirming its coincidence with a methanol 
maser.  
It remains weak, but clearly detectable at both 1665 and 1667 MHz.  
Note that two narrow spikes of interference are evident on the 1667-MHz 
spectrum at velocities 105.9 and 107.9 \kms.

\subparagraph{12.200-0.033} 

This is a weak new source at 1665 and 1667 MHz discovered while 
targeting 12.216-0.119 (about 6 arcmin away, but at quite different 
velocity) and with ATCA rms position uncertainty 0.4 arcsec.  A methanol 
maser coincides (Caswell 2009;  Green et al. 2010).

\subparagraph{12.209-0.102 and 12.216-0.119}

The two sources are shown in a single spectrum since they are 
spatially close, but are clearly distinct in velocity, with only slight 
overlap.  Their spatial separation is 70 arcsec and 6.5 arcsec in RA and 
declination respectively, as found from our ATCA measurements (including 
new ones) and those of Argon et al. (2000).  NRT spectra 2003 and 2005 
are similar to our 2004 and 2005 spectra, and corroborate the linear 
polarization  detection.  

The Parkes spectrum 12.22-0.12 (Caswell \& Haynes 1983b) is a 1982 epoch 
measurement of both sites and shows the absence of the currently 
dominant features of 12.209-0.103 near 15 and 16 \kms.  
The Argon et al. (2000) spectrum taken 1993 does not cover these 
velocities.  
The principal features of 12.216-0.119 have persisted from 1982 to 2005, 
with intensity variations typically less than a factor of 2.

\subparagraph{12.889+0.489}  

A detailed discussion of this source is given by Green et al. (2012a) 
where flux density variations measured with the ATCA in 2010 and 2011 
were shown to match the 29.5-day periodicity of the associated methanol 
maser.  We note that the high 
linear polarization seen in many features of our Parkes spectra closely 
match those measured more than 5 years later with the ATCA.

\subparagraph{12.908-0.260} 

Despite great variability since 1982, at each subsequent observing 
epoch, there have always been very strong features at both 1665 and 
1667 MHz, within the velocity range 28 to 42 \kms.  

In the velocity range of strongest persistent emission, 35 to 41 \kms, 
where methanol maser emission is also strong indicative of this being 
the systemic velocity, there is widespread but modest linear 
polarization, corroborating the NRT results.    

Of greater interest is the remarkable flaring 1667-MHz emission, 
velocity +29 to +32 \kms, and thus blue-shifted relative to systemic.    
The strongest emission, seen in 2005 (displayed spectrum), was an 
order of magnitude weaker in 2004 (6 Jy RHCP, 6 Jy LHCP), but 100 per 
cent linearly polarized at both epochs.  The flaring emission was 
present in the NRT spectrum of 2003 June, and highly 
polarized, but less than 1 Jy total intensity at that epoch.

\subparagraph{13.656-0.599}  

The present position, in Table 1, is a new ATCA measurement (rms 
uncertainty 0.4 arcsec), following from the discovery report by MacLeod 
et al. (1998).  

Our 2004 and 2005 observations provide the first good OH main line 
spectra from Parkes.  Features between 55 and 63.5 \kms, at both 1665 
and 1667 MHz, with peaks of up to several Jy remain similar to the 
spectrum by  MacLeod et al taken before 1998.  The total range, best 
seen at 1667 MHz, extends from 40.5 \kms\ to 63.5 \kms, and is closely  
centred on 
the methanol maser emission from the site, near 50 \kms, and likely to 
be a good approximation to the systemic velocity.   

Remarkably, 1665-MHz emission near velocity +48.5 \kms\ which peaked 
pre-1998 at about 1 Jy in each circular polarization (MacLeod et al. 
1998) now has a peak of approximately 30 Jy in each circular 
polarization (no significant net circular polarization) and is nearly 
100 percent linearly polarized. NRT measurements (2005oct) confirm 
the linear polarization, but the 
intensity was underestimated since it was not observed at the best 
position.  

Elsewhere, at velocities further from the centre,  
both 1665 and 1667-MHz features, in both 2004 and 2005, show only 
circular polarization, systematically with RHCP components at slightly  
higher velocity than nearby LHCP components, and thus (interpreted as 
Zeeman pairs) indicative of the same magnetic field direction.

\subparagraph{14.166-0.061}  

A longer than usual integration time of 40 min was used for our 2004 
observations of this weak maser so as to achieve a very low noise 
level;  it was not observed in 2005.  
The only previous detailed spectrum was from a 1982 Parkes observation 
(Caswell \& Haynes 1983b), when it had flared since an earlier 1975 
measurement. Variability continued, with non-detection by Forster \& 
Caswell (1989) and eventual detection with the ATCA in 2005, when a 
precise position was obtained (rms position uncertainty 0.4 arcsec), as 
given in Table 1.  
1667-MHz emission is stronger than 1665-MHz by an order of magnitude, 
and the velocity range, from 26.5 to 68 \kms, is one of the largest 
known.  It has no 1612-MHz OH counterpart (Sevenster et al. 2001) and 
is thus unlikely to be an AGB star.  It does have a water maser 
counterpart (Forster \& Caswell 1989), but has no detected methanol 
counterpart (Caswell et al 1995).  The absence of methanol and the wide 
velocity range suggests it is approaching the end of its maser emitting 
phase.

\subparagraph{15.034-0.677}

As remarked in Section 3.1, the sky background temperature in this 
direction is the highest for any source in our list,  and causes high 
rms noise on the spectra, despite a 20 min integration.  Current spectra 
(2004 displayed, and 2005) show stable spectra with no net circular 
or linear polarization.  The relatively smooth absorption spectra seen 
at 1667 MHz have been stable over all high spectral resolution  
measurements since first recorded in 1976 (Haynes, Caswell \& Goss 
1976), and are assumed as a baseline to derive the peak intensity of 
1665-MHz emission given in Table 1.  1665-MHz spectra are barely 
distinguishable from those in 1989 (unpublished Parkes data), but 
significant changes have occurred since 1976 when the feature at 21.5 
\kms\ was stronger and showed net RHCP, and was used for the VLA 
position measurement by Forster \& Caswell (1989).

\subparagraph{16.864-2.159} 

The position given in Table 1 is from new ATCA observations, with rms 
uncertainty 0.4 arcsec.  
The strong solitary 1665-MHz feature is primarily RHCP, but with 10 per 
cent linear polarization; its intensity  has doubled since 1993.    
1667-MHz emission is also present, again mainly RHCP, but a weak 
LHCP feature in the displayed 2004 spectra was confirmed in the 2005 
spectra.  Weak 1665-MHz LHCP emission (0.2 Jy) seen in 1989 and 1993 
(unpublished Parkes spectra) is currently below our detection limit.

\subparagraph{17.639+0.158} 

1665-MHz emission remains similar to 1993, except there is now no 
emission between 40 and 42 \kms.    
1667-MHz emission matches 1665-MHz in main features, and weak new 
emission near 29.2 \kms\ in 2005 was undetectable in 2004, less than 
half the 2005 value.  
Both 1665 and 1667-MHz emission displays strong  linear polarization of 
the 20 \kms\ feature, and a Zeeman pair of circularly polarized 
features implying a magnetic  field of -1 mG at velocity 20.7 \kms. 
Nearby 1720-MHz emission is discussed in detail by Caswell 
(2004a) revealing dominant 1720-MHz emission at velocity 28 \kms\  
(extending with  weaker emission to 35 \kms), offset by 4 arcsec from 
the position of 1665-MHz emission (Argon et al. 2000);  there is also 
weak 1720-MHz emission corresponding to a Zeeman pair near 20.5 \kms, 
magnetic field -2 mG, that agrees better in position with the 1665-MHz 
position.  1667-MHz absorption of -0.3 Jy at 22.5 \kms\ matches 
1720-MHz at the same velocity, with similar depth.

\subparagraph{18.461-0.004} 

The position in Table 1 is a new ATCA measurement with rms uncertainty 
0.4 arcsec.  
The full velocity range is 42 to 55 \kms\ at 1665 MHz, with linear 
polarization of up to 30 per cent in some features, as also seen from 
NRT observations.  There is weaker  
accompanying  1667-MHz emission.  Emission remains generally similar to 
1982.  
Note that a 1612-MHz  single feature is present at this position, with 
velocity 49.9 \kms\ (Sevenster et al. 2001).

\subparagraph{18.551+0.035}  

This is a new source offset by 6 arcmin from the previous source 
18.461-0.004, with ATCA position measured from the same dataset.  
A feature of 1.6 Jy at 1667 MHz is accompanied 
by weaker 1665-MHz emission at velocity 36.5 \kms.

\subparagraph{19.486+0.151, 19.473+0.170 and 19.471+0.170} 

Our new measurements reveal that OH maser emission in this direction 
is distributed over an 80 arcsec extent, and arises from three distinct 
OH sites that coincide respectively with the more precisely positioned 
methanol maser sites 19.486+0.151, 19.472+0.170n and 
19.472+0.170 (Green et al. 2010);  slight discrepancies in the Galactic 
source names are caused by larger uncertainties in the OH positions, 
which were measured by the ATCA in a short baseline hybrid 
configuration,  

The current overall spectrum is generally comparable to that in 1982, 
with major 1665 and 1667-MHz features still recognisable;  however, the  
1665-MHz feature at velocity 20.2 \kms\ flared from 3 Jy to 17 Jy 
in 1990 and as high as  24 Jy in 2004 (displayed) followed by a slight 
decrease in 2005.  The total velocity range is now seen to be 10 to 26 
\kms.  

In detail, we find that 19.486+0.151 accounts only for weak 
1665 and 1667-MHz features between 23 and 25 \kms, but is also the site 
of 6030 and 6035-MHz excited-state OH emission (Caswell 2003) and 
6668-MHz methanol emission.  The position in Table 1 is from ATCA 
measurements of 1665 and 1667-MHz features, with rms uncertainties of 3 
arcsec (0.2 s) and 2 arcsec.  

The strongest emission, the 1665-MHz flare of 24 Jy at 20.2 \kms\ 
and 1667-MHz 7 Jy at 19.1 \kms, together with emission adjacent in 
velocity, arises from 
$18^h25^m54.91^s,~-11^{\circ}52{\arcmin}31.5{\arcsec}$, with rms 
uncertainty 2.2 arcsec (0.15 s), 1.4 arcsec, 
a site with approximate Galactic coordinates 19.474+0.169,  
and probably coinciding with the more precisely determined position of 
methanol maser 19.472+0.170n (Green et al. 2010). 
Taking all position information into account, we adopt the preferred
name of 19.473+0.170.    

The  third source, with 1665-MHz emission of 2 Jy at 14.2 \kms, is at 
$18^h25^m54.53^s,~-11^{\circ}52{\arcmin}39.5{\arcsec}$ with rms errors 
4.3 arcsec (0.29 s), 2.6 arcsec, and corresponding to 19.471+0.170, 
matching methanol maser 19.472+0.170 (Green et al. 2010).  
The  weak 1665-MHz feature of 0.3 Jy at 10.1 \kms\ is also most likely 
at this position, although  not yet well determined.    

\subparagraph{19.609-0.234}  
  
The major 1665-MHz feature of 4 Jy at 41 \kms\ is unchanged since 1982.  
Our present increased sensitivity shows that it is accompanied by weaker 
1667-MHz emission, and at both transitions is predominantly RHCP, and    
extends to 18 \kms.  

There is associated methanol emission with mid-velocity 39 \kms\ which,  
regarded as the systemic velocity for this site, implies that the OH 
emission near 19 \kms\ is a highly blue-shifted outflow.

\subparagraph{19.612-0.135} 

19.612-0.135 is a pair of 1665-MHz features at 53 and 55 \kms,  with no 
evident 1667-MHz emission.  
It is now slightly weaker than in 1982 or 1989.  The prominent 
double-peaked  feature near +55 \kms\ has strong linear polarization 
that has been  stable over the 2004, 2005 period (including NRT 
observations 2005).    

Although the methanol maser 19.612-0.134 is nearby (Green et al. 2010), 
its precise position suggests a separation of more than 5 arcsec from 
the precise OH position measured by Forster \& Caswell (1989), and hence 
interpreted  here as indicating a likely non-detection of methanol at 
the OH site.

\subparagraph{20.081-0.135}  

Remarkably, the current spectrum is almost indistinguishable from 1982 
and 1993 for all 1665-MHz features in the range 42 to 51 \kms, and 
shows high levels of circular polarization.  Current sensitivity reveals 
accompanying  1667-MHz emission.  
Coincident with the OH is a methanol maser (Walsh et al. 1998).

\subparagraph{20.237+0.065} 

There is clear evidence of long term variability, with  the 1665-MHz 
main peak 3 times weaker in 1982 or 1989 than values at epochs 2004, 
2005 and 2003 (NRT).  
Circular polarization dominates, but several features show linear 
polarization, mostly weak, but exceeding 50 per cent for one 1667-MHz 
feature, with corroboration 2004, 2005 (displayed) and 2003  (NRT).   

The ATCA OH position (cited in Table 1) was first reported by Caswell 
(2003) and, coincident, is a methanol maser (Caswell 2009), the stronger 
of a close pair.

\subparagraph{21.795-0.128}

This source was newly discovered in 2004 during observations of the 
following source (21.880+0.014), offset 9.8 arcmin.  The position cited 
in Table 1 from ATCA observations has rms uncertainties of 11 and 6 
arcsec in RA and Declination.  There is no known methanol maser at the 
site.  
Interestingly, Sevenster et al. (2001) report a 1612-MHz OH single 
feature maser, 21.797-0.127, at velocity 40.9 \kms, peak flux density 
5.9 Jy; its position (from the VLA) is  RA 
$18^h31^m22.95^s,~-09^{\circ}57{\arcmin}21.1{\arcsec}$, with realistic  
rms uncertainty of 2 arcsec.  It lies within the uncertainty ellipse of 
our 1665-MHz OH position and seems likely to be coincident.

\subparagraph{21.880+0.014}  
 
The OH discovery resulted from a Parkes follow-up in 1993 of a 
strong methanol maser (Caswell et al. 1995).  The weak 0.35-Jy 1665-MHz 
emission  seen in our 2005 spectrum (displayed) remains similar to  
1993 and our  2004 observations.  It has remained too weak to position 
with the ATCA, and in Table 1 we cite the position of the assumed 
associated methanol maser.  
Note that the apparent emission near velocity 38 \kms\ is a weak 
response to 21.795-0.128, offset 9.8 arcmin and thus near the edge of 
the Parkes beam for this observation.  

\subparagraph{22.435-0.169}

1665-MHz emission in the velocity range 23.5 to 35.5 \kms\ has    
remained similar since  1982 and 1989, except that the 
currently strong 9-Jy LHCP feature at 25.6 \kms\ was then only 3 Jy.  
More recently it has been stable over the period 2005, 2004 and 2003 
(NRT), and has shown persistent linear polarization.   
1667-MHz emission is very weak.  
A new 1665-MHz position from ATCA observations (rms position 
uncertainties 3.3 arcsec, 1.5 arcsec) confirms the coincidence 
with 6035-MHz emission and with methanol (Caswell 2009), but the 
slightly more precise methanol position is treated as the most probable 
position, as cited in Table 1.

\subparagraph{23.010-0.411}  

1665-MHz emission is weaker than in 1982 or 1989, and with many changes 
but still covering the velocity range 66 to at least 75 \kms; 1667-MHz 
emission is weak but spanning a wider velocity range, from 64 to 
84 \kms\ across an absorption feature.  Despite variability 
there has been persistent 1665-MHz linear polarization in 2005, 
2004, and 2003 (NRT), and especially high at 66.8 \kms\ in 2005. 
1667-MHz emission has been more variable,  and the linear polarization 
seen by the NRT was below our sensitivity threshold in 2005, and 
marginal in 2004.   
There is an associated methanol maser (Caswell 2009).


\subparagraph{23.437-0.184 and 23.440-0.182} 

The position of the strongest 1665-MHz feature at velocity +106.0 \kms\ 
(Forster \& Caswell 1989, 1999) corresponds to 23.437-0.184.  As noted 
by Forster \& Caswell (1999), there are weaker RHCP features at 
velocities below +105 \kms, extending to +100 \kms, and overlapping 
an absorption feature, which  are offset 14 arcsec to the north 
(see also Forster \& Caswell 2000), corresponding to 23.440-0.182, and 
we interpret all emission at velocity below +105 \kms\ to be at this 
site, although this is tentative in view of spectral variability, and 
there is probably some overlap in velocity.  

The weak 1665-MHz  feature at velocity  +101 \kms\ shows weak 
but persistent  linear polarization (especially negative Q) at our 
observing epochs 2004, 2005,  and at the 2003 epoch of NRT  
observations.  
1667-MHz emission near  106 and 103.5 \kms\ extends to 98.5 \kms, as 
confirmed by the NRT spectra and is likely to 
include features at both sites.  Both sites are accompanied by 
methanol maser emission (Caswell 2009).    

\subparagraph{23.456-0.200} 

This source was newly discovered in our 2004 observations while 
targeting the previous sources (23.437-0.184 and 23.440-0.182).  Its 
spatial offset from them is approximately 1.5 arcmin, and its velocity 
range is quite distinct.
The position cited in Table 1 is from our ATCA measurements, with rms 
uncertainties of 5.9 and 3.4 arcsec in RA and Declination respectively.  
Emission is detected only as RHCP at 1665 MHz, with no change 
recorded between our 2004 and 2005 epoch observations.   
There has been no report of methanol maser emission here.

\subparagraph{24.329+0.145}  

We refer to Caswell \& Green (2011) for an extensive discussion of this 
source, which highlights its coincidence with a methanol maser and 
interprets the most prominent OH emission  as a 
blue-shifted outflow.  The 1667-MHz emission displayed here is in good 
agreement with the spectra shown by Caswell \& Green (2011) in every 
respect, including linear polarization.  Our spectra are shown 
in two panels so as to cover both the outflow velocity and the 
systemic velocity near +113 \kms\ (as interpreted from associated 
methanol maser emission).  The 1665-MHz spectra from 2004 reveal LHCP 
emission at +111.8 \kms\ which persisted in our 2005 spectra, and  is 
close to velocities where transient emission had been reported (+115 \kms)
at two epochs in 1993.  The feature at +106.9 \kms\ was below 
our sensitivity threshold in 2005.  The implied variability for features 
near the systemic velocity is similar to earlier reports summarised by 
Caswell \& Green (2011).  
Absorption between +105 and +120 \kms\ is indicative of more extended OH 
clouds in the vicinity of the maser and its inferred young high-mass 
star host.

\subparagraph{24.494-0.039}  

This site is a neighbour of the previous one (24.329+0.145, offset 
spatially by 15 arcmin) and close in velocity;  both sites are   
discussed extensvely  by Caswell \& Green (2011).  The three  
features of the 1667-MHz spectrum of 2010 shown by Caswell \& Green 
(2011) are evident in our 2005 spectra shown here, and we note that 
the very narrow spikes at 106.9 and 108.7 \kms\ in our present spectrum 
are terrestrial interference, not astronomical. The 1665-MHz emission 
is also unchanged;  the present spectrum is not affected by sidelobes 
from the site 24.790+0.083 (the following source) that were evident in 
the spectrum shown by Caswell \& Green (2011).   
Coincidence with a methanol maser was confirmed  by Caswell \& Green 
(2011).

\subparagraph{24.790+0.084 }  

Together with the two previous sources (offset 20 arcmin), this  site is 
also discussed in detail by Caswell \& Green (2011).  Compared 
with their spectra in 2010, our spectra from 2004, although showing 
emission over the same velocity range,  have much lower peak 
1665-MHz intensity (demonstrating the continuing variability at 1665 
MHz), and show the weak accompanying emission 
at 1667 MHz.  The record of strong variability now extends over the long 
period  from 1982 through 1990, 2003, 2004 and 2005 to 2010.  The 
velocity range of prominent 1665-MHz emission in our displayed spectrum 
(epoch 2004) is 105 to 113 \kms, and at 1667 MHz  extends to 
114.5 \kms.  In 1982, features in the velocity range 100 to 110 \kms\ 
were prominent.  
NRT spectra from 2003 agree well with ours, and 
corroborate in detail the linear polarization of the 1665-MHz emission, 
reaching 80 per cent for the strongest feature at +111 \kms.  
Coincidence of methanol is discussed by Caswell \& Green (2011).

\subparagraph{28.147-0.004 and 28.201-0.049}

28.147-0.004 is a site with weak OH emission detectable only at 1665 
MHz, and confined to velocities between 100 and 103.5 \kms, as evident 
from Argon et al. (2000) and NRT observations.  We do not show a 
spectrum precisely at this target since it is offset by only 4.3 arcmin 
(mostly in RA) from the displayed source 28.201-0.049, and can be seen 
on that spectrum (but note the reduced amplitude of 28.147-0.004 
resulting from the offset, so that its apparent emission should be 
increased by a factor of 1.4; the correction has been applied to peak 
values in Table 1).
Linear polarization of nearly 50 per cent is present as confirmed by 
the NRT.  
The OH is coincident with a methanol maser (see Caswell 2009)

The remaining emission on the spectrum, at velocities below 100 \kms, is 
from 28.201-0.049. The strongest 1665-MHz feature of 1982, RHCP 15 Jy at 
velocity 95 \kms,  increased through 1989 to 26 Jy and decayed through 
1993 to now only 3 Jy, and thus weaker than the other  features at both 
1665 and 1667 MHz that remain in the velocity range 92 to 100 \kms.    
There is coincident methanol (Caswell 2009).

\subparagraph{28.862+0.065} 

The current spectra at both 1665 and 1667 MHz have changed little since 
1982 and 1990.  
Remarkably, all features are highly LHCP with the exception of a weak 
RHCP 1665-MHz feature at 107.5 \kms.     
There is weak 1665-MHz linear polarization at 102.8 \kms, persistent 
from 2004 to 2005.  
The OH position has subarcsecond uncertainty, and the nearby methanol, 
with less precise position.  
most likely coincides since it agrees to within 1.5 arcsec in 
declination and 7 arcsec in RA, well within the current uncertainty of 
about 10 arcsec (Caswell et al. 1995).

\subparagraph{29.862-0.040 and 29.956-0.016}  

Note that for both these sites, the 1667-MHz spectra show interference 
spikes at velocities 107.9 and 109.9 \kms, and no 1667-MHz maser 
emission is present.  

1665-MHz emission at both sites is of similar intensity.  The sites are 
separated by approximately 6 arcmin (half the half-power beamwidth for 
the present measurements). The aligned 
spectra displayed here clearly show that features between 100 
and 108 \kms\ are accounted for by 29.862-0.040, while features between  
97.5 and 99 \kms\ arise from 29.956-0.015.  
New ATCA position measurements for the OH were taken with a 
short baseline hybrid array.  The resulting position for 29.862-0.040 
given in Table 1 has position rms uncertainties of 
5.4 and 2.9 arcsec in RA and declination respectively.  
It seems likely to match a methanol maser site,  
29.864-0.043, with corresponding position estimate of 
$18^h45^m59.58^s,~-02^{\circ}44{\arcmin}59.9{\arcsec}$ 
(derived from the Walsh et al (1998) position for a  feature  at 
+101 \kms, after correction for a suspected declination error of 2.4 
arcsec at this declination).

In the case of 29.956-0.016, the uncertainty in the  OH position was 
larger, 15 arcsec, owing to unresolved absorption which led to a poorly 
determined 
baseline.  Near this OH source, the methanol maser 29.956-0.016 has 
been measured to much higher precision (Minier Conway \& Booth  2001) 
and is coincident with the OH to within 15 arcsec;  since 
this seems likely to be a valid association, we have cited the 
methanol position in Table 1.  

At 29.862-0.040, the NRT OH spectrum (epoch 2005) shows excellent 
agreement with our 2005 spectrum, and both datasets indicate the 
remarkable 100 per cent linear polarization for the two features. 
Archival Parkes data (with no information on linear polarization) from 
1989 show similar total intensity but, in 1993, a peak less than half 
the present value.   

The OH at 29.956-0.016 has shown very little change since Parkes 
archival spectra in 1993, which showed a similar single, wholly RHCP, 
feature at 98.2 \kms, with peak of 0.9 Jy.  Although there is no 
targeted NRT spectrum at  29.956-0.016, the NRT observation towards  
29.86-0.05 also provides a good spectrum of 29.956-0.016 since the  
offset is less than 1 arcmin (4s) in RA (and 6 arcmin in declination 
where the beamwidth is larger).  It  agrees well with the Parkes 2005 
and 2004 present measurements.

\subparagraph{30.589-0.043}

Strong emission is present at 1665 and 1667 MHz.  Our 2004 and 2005 
spectra show no major changes relative to spectra from 1982, 1990 
(Parkes archival data) or 1993 (Argon et al. 2000).  Fractional linear 
polarization is low, The OH position agrees well with a prominent  
uc\HII\ region (Argon et al. 2000).  A small discrepancy with the 
methanol maser position (Walsh et al. 1998) is wholly in declination and 
likely to reflect a large uncertainty of the methanol estimate, similar 
to other sources nearby that needed a shift south by 7 arcsec to 
correspond with more precise position measurements.  

\subparagraph{30.703-0.068}  

This source is dominated by a feature at 91.2 \kms, mainly RHCP at both 
1665 and 1667 MHz, with peaks of 14 Jy and 10 Jy respectively.  At 1665 
MHz it spans at least 86 to 97.5 \kms, with additional 1667-MHz features 
(not seen in 1982) now peaking at 82 \kms\ (extending down to 79 \kms\ 
in 2004 but not seen in the displayed spectrum of 2005).  
Observations in 1993 January (Argon et al. 2000) surprisingly fail to 
show features at velocity lower than 85 \kms, and the highest velocity 
emission is poorly covered, owing to the plot limit at +97 \kms.  

Comparison of our spectra in 2004 and 2005,  and the NRT in 
2003 May, show all features mildly variable.  Linear polarization is 
persistent and exceeded 50 per cent on several features at both 
transitions in the displayed spectrum of 2005.

The methanol position (Walsh 1998) has an apparent declination offset 
from the OH position similar to that of the previous and following 
source, suggesting the association is valid and there is a small 
systematic  error of several arcsec in the methanol position.  

\subparagraph{30.820-0.060}  

Our display range overlaps that of the previous site 
30.703-0.068 whose strong emission is also seen, but with 
amplitude reduced by the 
offset of about 6 arcmin.  The major features of 30.820-0.060 are near 
106 \kms, chiefly LHCP 1665 MHz, and weaker at  1667 MHz. Weak 1667-MHz 
interference is seen at velocity 108.1 and 110.1 \kms.

The Nancay spectrum (labelled 30.82-0.05) is similar to ours, but with 
the 106.2 \kms\ feature slightly stronger, and corroborating negligible  
linear polarization.  

Our OH position from new ATCA measurements in a short baseline hybrid 
configuration is cited in Table 1 and has rms uncertainties of 10 and 5 
arcsec in RA and Declination.  To within this precision, it coincides 
with a methanol maser at  
$18^h47^m47.0^s,~-01^{\circ}54{\arcmin}26.3{\arcsec}$ (with 
uncertainty less than 1 arcsec from MMB unpublished data) corresponding 
to 30.818-0.057, which is most likely a better estimate of the  OH 
position.


\subparagraph{30.788+0.204, 30.819+0.273, and 30.897+0.161}

The displayed plot shows the large velocity range +75 to +110 \kms, so 
as to show not only 30.788+0.204 (with 1665-MHz emission between 77 and 
84 \kms) but also 30.819+0.273 (with 1665 and 1667-MHz emission between 
100 and 105 \kms) and 30.897+0.161 with 1665-MHz emission at 107.8 
\kms;  
the second and third sites are offset from the target position 
by 4.6 arcmin and 7 arcmin respectively, and thus the apparent 
intensities on the displayed plot need intensity correction factors of 
1.54 and 2.5 to compensate for the reduced gain at the offset position.  

30.788+0.204 was first observed at Parkes in 1993 and the two major 
features, RHCP at 78 \kms, and LHCP at 81 \kms\ remain prominent but 
with RHCP now slightly stronger and LHCP slightly weaker.  No detectable 
changes are seen between our 2004 data (as displayed), our 2005 October 
data, or the NRT data from 2005 September.   

The position cited in Table 1 is a precise 6668-MHz methanol position;  
the OH position from our ATCA data has rms uncertainty of 21 and 12 
arcsec in RA and Declination respectively, yielding the same 
declination, and RA larger by 0.7s, and thus coincident with the 
cited methanol position to within the OH position uncertainty. 

For the site 30.819+0.273,  both 1665 and 1667-MHz spectra show features 
between 100 and  105 \kms, with peak at 1665 MHz of 3.5 Jy (after 
correction for the offset position).  Additional unpublished spectra 
centred at 30.819+0.273 confirm this, and show essentially no change 
between 2004 and 2005 or the NRT spectra in 2003.  The 1665-MHz emission 
when first observed at Parkes in 1993 (archival data) showed similar 
spectra with emission dominated by RHCP features peaking at 3 Jy.  
The OH position (in Table 1) determined with the ATCA has rms 
uncertainties of 2.5 and 1.3 arcsec in RA and Declination respectively 
and coincides with a methanol maser.

The 1665-MHz emission at velocity 107.9 \kms, predominantly 
LHCP, most likely arises from a third site, 30.897+0.161 where a 
methanol maser was first reported by Schutte et al (1993) with peak 
flux density between 50 and 100 Jy near velocity 102 \kms; the best 
estimate of its position is  
$18^h47^m09.14^s,~-01^{\circ}44{\arcmin}17{\arcsec}$, where we use the 
methanol position from Walsh et al. (1998), after applying a declination 
correction of 7 arcsec which we found necessary for other sources at 
this declination observed in that project.  The OH emission was first 
noted at Parkes in 1993, with LHCP peak of 1 Jy, somewhat weaker than 
its 2004 value, after allowing for the offset in position by 7 arcmin of 
our displayed 2004 spectra. 
We reiterate  that the position cited in Table 1 is from a methanol 
measurement, and the coincidence of OH is currently uncertain to 
several arcmin.

\subparagraph{31.243-0.111} 

Archival spectra of 1665-MHz emission obtained in 1982 showed a peak of 
5.3 Jy which has subsequently been  decaying through 3 Jy (1993)  
to the present peak near 2 Jy (2004 and 2005).  Linear polarization is 
weak but significant.  There is no known methanol here, with upper limit 
0.5 Jy (Caswell et al. 1995).

\subparagraph{31.281+0.061} 

Several features at 1665 and 1667 MHz were present in 1982, spanning 
velocities 103 to 109 \kms, and with absorption extending to 111 
\kms;  features in the same velocity range and of similar intensity 
persist through 1989 to the present but differ in detail.  
At 1665 MHz, a 1-Jy flare at 107.6 \kms\ occurred in 2005 October, seen 
slightly weaker in 2005 September on the NRT spectra.  
Weak linear polarization is evident in all observations.  
Methanol and OH positions in RA agree well, and the discrepancy in 
declination of 7 arcsec is similar to that of nearby sites and seems 
likely to be chiefly an error in the methanol position (Walsh et al. 
1998;  note that the position from Minier et al. 2001 is merely taken 
from Walsh et al. 1998)

\subparagraph{31.394-0.258}

First reported by Cohen et al. (1988), emission is mainly 1665-MHz LHCP, 
with 1.5 Jy peak in 1993 and still similar;  weak 1667-MHz emission is 
also present.  The new ATCA position (Table 1) has rms uncertainties of 
6 and 4 arcsec in RA and declination.  There are no reports of methanol 
emission here.

\subparagraph{31.412+0.307} 

The present spectra at 1665 and 1667 MHz extend over a range 86 
to at least 114 \kms, as also seen 
in 1993 at 1665 MHz, and extending beyond the 108 \kms\ limit of the 
Argon et al. (2000) spectra.  
Despite marked changes since 1993, our spectra from 2004 and 2005, and 
2004 September spectra from the NRT have remained stable. Prominent 
linear polarization is seen at 95 \kms, 1665 MHz.  At the OH position, 
there is also methanol maser emission (Walsh et al. 1998)

\subparagraph{32.744-0.076} 

Features from 25 to 40 \kms\ in 1982, with peaks of 2 Jy and 1 Jy at 
1665 and 1667 MHz respectively, have remained similar through 1989 to 
2004 and 2005.  Flaring features in 2004,  1665-MHz LHCP at velocity 
33.4 \kms, and 1667-MHz RHCP at 30.9 \kms, subsided in 
2005 (our displayed spectrum and the NRT spectra of 2005 April).  
At the OH position there is a methanol maser (Caswell et al. 1995), 
coincident to within its 10 arcsec position uncertainty.  

\subparagraph{33.133-0.092}  

Features from 72 \kms\ to 82 \kms\ in 1982 showed peaks of  8 Jy at  
1665 MHz and 3 Jy at 1667 MHz; spectra of  2004, 2005 and 
2004 July (NRT) mutually agree, and are stronger at 1665 MHz but 
weaker at 1667 MHz.  
The association of a methanol maser (Caswell et al. 1995) at 
$18^h52^m07.3^s,~+00^{\circ}08{\arcmin}05{\arcsec}$   
holds to within 10 arcsec, the methanol positional accuracy.

\subparagraph{34.258+0.153}  

OH maser spots are spread over a large extent of 4 arcsec, with complex 
continuum emission from an \HII\ region covering a similar extent (Argon 
et al. 2000; Fish et al. 2005).    
A methanol maser also coincides (Caswell 2001). 

The extensive previous OH observations include those from the VLBA, VLA, 
NRT and Parkes.  Peaks now at 1665 MHz (60 Jy) and at 1667 MHz (100 Jy) are
slightly stronger than in 1982 and 1990, but generally similar.  
Linear polarization at 1667 MHz is seen to be especially high (2004 and 
2005), in agreement with the VLBA measurements (Fish et al. 2005) in 
2001 January and NRT 2003 March, with a feature at 58.7 \kms\ with ppa 
140$^{\circ}$;  and a 
secondary feature at 57.5 \kms\ with ppa 60$^{\circ}$.   

\subparagraph{34.411+0.231}

The displayed spectra are aligned with those of the very strong previous 
source, 34.258+0.153 which is separated by only 10 arcmin and is 
detectable at this target position with amplitude reduced by an order of 
magnitude.  It is then evident that 34.411+0.231 accounts only for the 
emission from 53 to 57 \kms\ at 1667 MHz and from 53 to 55.5 \kms\ at 
1665 MHz.  An early Parkes observation in 1993 showed peaks of 2 Jy and 
1.5 Jy at 1665 and 1667 MHz respectively.  The 2004 observation was 
similar but an increase is clear in the 2005 spectra as displayed.  
Prominent linear polarization is present at 1665 MHz.  The position 
obtained with the ATCA (Table 1) has rms uncertainties of 4 and 2 arcsec 
in RA and Declination respectively.  
A nearby water maser has a precise parallax distance of 1.56 kpc, 
despite an uncertain absolute position (Kuruyama et al 2011); it seems 
likely that the water, 34.394+0.221,  
coincides with a 6668-MHz  methanol maser 
(Pestalozzi, Minier \& Booth 2005), but is offset from the OH by at 
least 1 arcmin, indicating an association 
merely in the same cluster rather than coincidence with a common 
exciting star.

\subparagraph{35.025+0.350}  

No significant changes occurred between epochs 2005 (displayed), 2004, 
and the NRT spectra of 2003 March, with all showing persistent linear 
polarization; similar spectral features were  
recognisable in 1993 and 1982, although with intensity changes by 
factors of 2.

Precise methanol maser position measurements (Cyganowski et al. 2009). 
yield a position 
$18^h54^m00.66^s,~+02^{\circ}01{\arcmin}19.3{\arcsec}$, in good 
agreement with the OH.  

\subparagraph{35.197-0.743}  

Features at 1665 MHz span velocities 24.5 to 38 \kms\  at various 
epochs.  Great variability has occurred between 1982, 1990, 1993 January 
(VLA), 1993 December (MERLIN) and 2005, with spectra essentially so 
different as to be unrecognisable except for position and velocity 
range. 

Several weak features are seen at 1667, fading from 2003 (NRT) and 
2004 to 2005.  Linear polarization is evident at 1667 MHz, and 
at 1665 MHz, in both cases agreeing well with the NRT data. 
However the details have changed greatly from those observed with 
MERLIN in 1993 (Hutawarakorn \& Cohen 1999).  

Precise methanol maser position measurements establish an astrometric 
parallax distance for the site, 2.19 kpc with 10 per cent uncertainty 
(Zhang et al. 2009). and the OH emission, which is spread over a 2 
arcsec region (Forster \& Caswell 1999; Argon et al. 2000) encompasses 
the methanol masing region, consistent with a common source of 
excitation.

\subparagraph{35.201-1.736} 

There have been considerable changes in spectra at both 1665 and 1667 
MHz between 1982 and 2005.  More recently there has been closer 
agreement between 2003 (NRT) and 2005, although the significant linear 
polarization (approaching 50 per cent) of the prominent flaring feature 
seen in the displayed (2005) 1665-MHz spectrum at 43.2 \kms\ was much 
lower in the pre-flare emission in 2004 and not seen in 2003.

Precise methanol maser position measurements establish an astrometric 
parallax distance for the site, 3.27 kpc with 15 per cent uncertainty 
(Zhang et al. 2009). The associated OH emission region  overlaps the 
methanol masing region.

\subparagraph{35.578-0.030}  

Strong features at both 1665 and 1667 MHz have slowly increased from 
1982 through 1990 and 1993 (VLA) to current peaks of 60 Jy (but RHCP  
now stronger than LHCP) at 1665 MHz, velocity  49.0 \kms.  The 1667-MHz 
peak was 20 Jy (LHCP) at 50.3 \kms\ in 2004 but 
halved in 2005.  

The VLBA measurements at 1667 MHz show the major feature mainly 
LHCP, with weaker linear polarization, at ppa 60$^{\circ}$ in agreement 
with our spectra showing positive U  and smaller negative Q.  
The 1665-MHz emission peak is a complex blend of three features, and 
close comparison suggests that there is again good agreement between 
the VLBA and the present data.  

The OH maser is accompanied by continuum emission that is compact on 
a scale of several arcsec, but with complex structure (Argon et al. 
2000), and is presumably an associated HII region.  There has been no 
detection of any 6668-MHz methanol maser nearby  (Caswell et al. 1995 
and subsequent unpublished searches).  

\subparagraph{40.426+0.700}  

Current 1665-MHz emission remains similar to the earliest Parkes 
unpublished archival spectrum of 1993.  

A position measurement for 1665-MHz emission with the ATCA gave a 
position 
$19^h02^m40.02^s,~+06^{\circ}59{\arcmin}11.6{\arcsec}$ with rms 
uncertainties 2.5 and 1.5 arcsec.  Within these errors, it coincides 
with the better determined position of 6035-MHz emission, which we have 
preferred to cite in Table 1, with the intention of providing the most 
likely precise position.  
It agrees well with the 6668-MHz methanol position (Caswell 2009).

\subparagraph{40.623-0.138}  

Since 1982, the 1665-MHz peak has increased from 80 Jy to 110 Jy, and 
at 1667 MHz remains below 20 Jy, but secondary features have increased.  

Full polarization comparisons are possible between the VLBA and the NRT 
(as noted by Szymczak \& Gerard 2009), and with our data.   A specific 
example, the 1665-MHz strongest feature, at 32.6 \kms, has high linear 
polarization of 74 per cent at ppa -78$^{\circ}$ i.e +102$^{\circ}$ in 
the NRT data.  Our 2005 data show a similarly high percentage linear, 
almost wholly negative Q, and thus ppa near 90$^{\circ}$, in good 
agreement.  The VLBA lists 46 per cent linear polarization with ppa 
117$^{\circ}$, in somewhat poorer agreement, in part likely to arise 
from the epoch difference of several years.  

The OH position closely agrees with the precise position of a 6668-MHz 
methanol maser (Caswell 2009).   


\section[]{Statistics from the present data}

We first recall the selection parameters of the present data set. It has 
been limited to maser sites believed to be associated with regions of 
young massive star formation in a portion of the Galactic disc 
accessible to both northern and southern observatories, thus allowing 
extensive comparisons with other data.  Many of these sources had been 
discovered by unbiased surveys, but others are from targeted 
observations, especially the more  northerly sources.  The sample is 
also limited to sites with well-determined positions (usually with 
arcsecond accuracy), but is not complete, with a bias amongst northern 
sites towards the stronger sources known for several decades.  However,  
none of the statistics that follow are expected to suffer any major bias 
from these limitations.  
A well-known statistic, that the ratio of 1665 to 1667-MHz intensity 
(peak or integrated) usually favours 1665-MHz emission (but with some 
clear counter-examples) is evident from our data, as expected.  
We will re-visit this statistic, after the analysis of 150 more 
southerly sources, in Paper II.

\subsection[]{Comparisons with masers of water and methanol}

High spatial resolution studies of the association of water masers with 
OH  (Forster \& Caswell 1989) showed that, in addition to the 
frequent occurrence of coincident water and OH maser sites, there were 
often additional sites of water maser emission slightly separated from a 
target OH maser.  A search for water towards a larger sample of OH 
masers (Breen et al. 2010b) showed that 79 per 
cent of the OH masers had a closely associated water maser.  We note 
that most of the OH masers studied here are contained in these earlier 
water studies, and thus similar conclusions apply to the present OH 
masers.  

Turning to methanol, previous statistics with respect 
to associations of masers of OH (of the SFR or massive YSO variety) with 
methanol suggest that 80 per cent of OH  masers have an accompanying 
methanol maser (Caswell 1998).  Following the procedures of that study, 
in Table 1, a simple comparison of methanol to OH is made, listing the 
ratio:  peak methanol intensity to peak OH intensity.  As remarked in 
Section 3.3, the comparison has been improved since the 
Caswell (1998) investigation owing to some improved OH positions in the 
present paper, and recently improved methanol positions.   

We conclude that 87 of the 104 OH masers presented here have closely 
associated methanol masers, and 17 do not.  It has been noted that OH 
masers without methanol (or with very low ratio of methanol to OH 
intensity) are more commonly associated with detectable ultra-compact 
HII regions (Caswell 1996, 1997).  This is interpreted as an 
evolutionary trend (Caswell 1996, 1997, 1998; Breen
et al. 2010a), such that maser sites where OH outshines the methanol  
are in the later stages of evolution.  

We draw attention to the need for a northern hemisphere, sensitive, 
unbiased survey for OH masers, at least matching, and preferably 
surpassing, the southern hemisphere counterpart (Caswell 1998).  
Although follow-up to the unbiased methanol surveys recently completed 
will lead to the detection of many new OH masers (Green et al. 
2012b), this 
will not wholly achieve the aims of an unbiased OH survey since  
the OH class that have little or no accompanying methanol will be 
greatly under-represented.  
In the southern hemisphere, an improved unbiased survey is planned with 
the GASKAP survey (Dickey et al. 2012).

\subsection{Maser site velocity ranges}

The velocity range of emission for masers in star formation regions can 
be a useful diagnostic in several ways.  

In the case of masers at the methanol 6668-MHz transition, large samples 
(of several hundred) show a distribution of velocity range which has a 
small median value of about 6 \kms\ (Caswell 2009).  A source  with 
range of 26.5 \kms\ (Caswell et al. 2011b) is the only one exceeding 25 
\kms.  There is a weak  trend for velocity range to be an increasing 
function of intensity;  although a sensitivity effect limits the 
measurable range for weak sources (Caswell et al. 1995), it none the 
less  seems likely that a major contribution to the 
overall trend is one of evolution, where more evolved sources are 
stronger and have wider velocity ranges (Breen et al. 2010a).  However, 
there has been no clear evidence for significant outflows, and the 
mid-range 
velocity appears to be a good estimate of the systemic velocity for 
all sources (see also Szymczak Bartkiewicz \& Richards 2007; Pandian 
Menten \& Goldsmith 2009;  Green \& McClure-Griffiths 2011).  

In the case of water masers, a large sample of several hundred masers 
shows a median velocity range of 15 \kms\ (Breen et al. 2010b);  Breen 
et al. draw attention to those masers with features offset 
from the systemic velocity by more than 30 \kms, interpreted as high 
velocity outflows. 
These high velocity outflows most likely occur in collimated
jets.  In particular, there is an unusual class of such high
velocity outflows favouring blue-shifts (Caswell \& Phillips 
2008; Caswell \& Breen 2010; Caswell Breen \& Ellingsen 2010c;  Motogi 
et al. 2011b, 2012);  these primarily occur at sites where there is 
neither OH emission nor readily detected uc\HII\ continuum emission, 
consistent with an interpretation that they are confined to an early 
evolutionary phase of massive star formation.  
This association of wide velocity ranges of water masers with an early 
evolutionary phase is in marked contrast to methanol masers, where 
the rare, wider, velocity ranges are believed to be the signature of a 
late evolutionary phase.

The velocity ranges for OH masers have been less studied, but a median 
value of 9 \kms\ has been reported for OH 1665 or 1667-MHz masers 
(Caswell 1998).  The present sample of sensitive spectra with good 
velocity coverage allows us to revisit the statistics  for this maser 
variety.  
The median for our sample of 101 is 8.3 \kms, essentially the same as 
found by Caswell (1998) for a larger (partially overlapping) sample.  We 
first single out sites with a velocity spread exceeding 25 \kms, of 
which we find seven.

Considering them individually, we note that 351.775-0.536 has a likely 
systemic velocity near -3 \kms, estimated from the median of the narrow 
range of methanol emission velocities;  thus the large OH range, from 
-36 to +8 \kms\ arises largely from weak blue-shifted features.  The 
site has no reliably detected continuum emission (with rms noise level 
of 0.27 mJy at 8.4 GHz in maps from Argon et al. (2000)) and is likely 
to be at an early evolutionary phase.

In contrast, 5.885-0.392, with relatively weak methanol emission and a 
strong \HII\ region, is a well-studied evolved site, probably 
approaching the end of its lifetime as a maser emitter. 
The velocity range of OH emission, from -44 to +18 \kms\ is the largest 
in our sample and its interpretation  as a general expansion driven by
the  \HII\ region (Stark et al. 2007) is plausible.  A 
noticeable asymmetry, relative to the mean recombination
line velocity centred near +9 \kms,  indicates a favouring of blue over
red-shifted outflow.  An absence of strong red-shifted emission might
be caused by the extremely strong \HII\ region obscuring any possible 
red-shifted emission flowing out from the far side, and providing 
increased seed radiation for masers on the near side which are 
blue-shifted towards us. Alternative variations include biconical 
outflows (Zijlstra et al. 1990), and a combination where the \HII\ 
region is interacting with a pre-existing outflow (Hunter et al. 2008).

10.473+0.027 has an OH velocity range from +43.5 to +71 \kms, and the 
associated strong methanol maser has a range from 57.5 to 77.6 \kms, 
but the site has not been studied in depth and cannot be reliably 
classified.  14.166-0.061 
with a range from +26.5 to +68 \kms, is an OH maser that is weaker at 
1665 than 1667 MHz, and with no known methanol counterpart, but with a 
water maser counterpart, and little other information.  19.609-0.234 
with OH range +18 to +45 \kms\ has a matching methanol counterpart with 
velocity range +36.0 to +42.0, and thus the likely systemic velocity of 
the site is near +39.0 \kms.  On this interpretation, the OH emission 
near +40 \kms\ is at the systemic velocity, stronger at 1665 than 1667 
MHz, while emission near +20 \kms\ represents a blue-shifted outflow, 
equally strong at 1667 and 1665 MHz.  24.329+0.145 has been studied in 
detail (Caswell \& Green 2011) and is dominated by a blue-shifted 
outflow at 1667 MHz, with weak emission also seen at 1665 MHz but near 
the systemic velocity only.  Towards 31.412+0.307, the OH emission 
velocity range, from +86.5 to +113 \kms, has accompanying methanol 
emission over a slightly smaller range, +90 to 108, with almost the same 
central velocity, and thus no indication of any preferred outflow.

Summarising the above details: 

5.885-0.392, 31.412+0.307, 10.473+0.027 and 14.166-0.061 all seem 
likely to be evolved sites where there is a general outflow, although 
with some bias to blue-shifts perhaps accounted for by effects 
from a central \HII\ region.  

Most remarkable amongst our OH masers are 351.775-0.536, 19.609-0.234, 
and 24.329+0.145, with asymmetric strong outflows, all interpreted as 
blue-shifted relative to  the systemic velocity, and with no readily 
detected uc\HII\ region.  We interpret these three to be at an early 
evolutionary stage.  

For 24.329+0.145, a detailed analysis (Caswell \& Green 2011) links the 
OH blue-shifted outflow to an accompanying water maser blue-shifted 
outflow, and thus membership of the distinct class  
of water masers with blue-shifted outflow characteristics  
recognised by Caswell \& Phillips (2008). 24.329+0.145 appears to be an 
especially rare object where the dominant blue-shifted outflow is 
displayed by both the  water and OH maser emission.

\subsection{Variability of OH masers}

Details of variability are given in the source notes of Section 3.3.  
Very few sources show features with high variability between our 2004 
and 2005 observations, with only five showing more than a factor of 2 
variation in their strongest feature at either 1665 or 1667 MHz.  They 
are:  350.011-1.342, 351.775-0.536, 12.209-0.103, 35.197-0.743 and 
35.578-0.030.  

On a longer timescale, high stability is evident at 10 of the 82 sites 
where comparisons are possible with earlier data over several decades. 
Over these long periods, dramatic changes, mostly strong `flares', have 
been seen for 6 sources:  351.775-0.536, 12.908-0.260, 13.656-0.599, 
19.473+0.170, 22.435-0.169 and 28.201-0.049.  

These statistics, as well as the reference spectra shown here, will be 
useful in planning future variability studies.  We note that the weakly 
varying source 12.889+0.489 with likely periodicity of 29.5 days is 
within our sample, and there are, doubtless, other examples to be 
discovered which will require dedicated monitoring programs (Green 
et al. 2012a).

\subsection{Circular polarization of the OH emission}

Many earlier studies of OH maser sites note the occurrence of circular 
polarization at 1665 and 1667 MHz, caused by Zeeman splitting in magnetic 
fields of several mG.
From large, uniformly studied, samples of sources (e.g. Szymczak \& 
Gerard 2009), the majority of features, at most sites, exhibit 
significant polarization, and some features are found to be essentially 
100 per cent polarized.  The same is true of the present 
sample, with the noteworthy exception of 15.034-0.677 which shows 
neither circular nor linear polarization above 
our sensitivity level detection limit.  The fact that 15.034-0.677 is 
associated with extremely strong continuum \HII\  emission (M 17), the 
strongest background emission towards any source in our survey, may be 
linked to this.  More detailed discussion of circular polarization  
statistics will not be addressed here since it can be better dealt 
with when combined with the additional 150 
sources that will shortly be published for the most southern portion of 
our survey, Part II.

\subsection{Linear polarization of the OH emission}

Our measurements reveal some remarkable examples of linear polarization 
greater than 90 per cent, and in three cases these occur in features 
stronger than 5 Jy.  13.656-0.599 is an outstanding example with a 
1665-MHz feature at 48.4 \kms\ showing over 90 per cent linear 
polarization of a feature with total intensity varying between 60 and 50 
Jy at our two epochs.  The NRT observations corroborate our polarization 
measurements (but we note that their offset target position causes a 
severe underestimate of total intensity).  
12.908-0.260 exhibited several adjacent 1667-MHz features near 30 \kms, 
that flared in 2005 with the strongest peak reaching 150 Jy, relative to 
peaks of 6 Jy in our 2004 measurement, and only 1 Jy with the NRT in 
2003.  Fractional linear polarization has been high at all epochs, and 
indistinguishable from 100 per cent in our measurements.  
12.889+0.489 also shows prominent linear polarization which appears to 
exceed 90 per cent in a 1665-MHz feature near 32.5 \kms, but is blended 
with strong adjacent features.  
Additional to these three outstanding objects in our data, we draw 
attention to one remarkable 1665-MHz feature in the NRT data set,  in 
133.95+1.06, W3(OH), at -47.41 \kms, with large linearly polarized flux 
density of 135.9 Jy, corresponding to 98.9 per cent.  Surprisingly, the 
VLBA data of Wright et al. (2004a) in 1996 show no similarly large 
fractional polarization for any feature.

Turning to the more general statistics on the occurrence of linear 
polarization,  amongst the  
sample of nearly 100 sources studied by Szymczak \& 
Gerard (2009), some linear polarization was detectable 
in 80 per cent of 1665-MHz spectra  and 62 per cent of 1667-MHz 
spectra;  the smaller fraction at 1667 MHz is most likely due to the 
relatively weaker emission, and thus the observational sensitivity 
threshold has greater impact at 1667 MHz.  In our sample, we find linear 
polarization in more than 74 of 89 1665-MHz 
spectra (excluding those weak sources with total intensity peaks less 
than 0.5 Jy), i.e. 83 
per cent; and in 39 of 70 1667-MHz spectra (again excluding spectra 
with only weak emission), i.e. 56 per cent.  Approximately half the 
sources are common to the NRT sample and, for these, our percentages are 
87 per cent at 1665 and 50 per cent at 1667 MHz, where the slight 
difference from the NRT statistics can readily be accounted for by  
variability and slightly different noise levels.  We find similar 
statistics for sources not in the NRT sample, percentages of 79 per 
cent and 62 per cent at 1665 and 1667 MHz respectively.  

Narrowing our focus to sources where very strong polarization occurs  
in at least one feature: with criterion similar to Szymczak 
\& Gerard (2009), we find features that are at least 50 per cent 
linearly polarized in 35 per cent of the 1665-MHz spectra and 23 per 
cent of the 1667-MHz spectra.  
These sources are identifiable in column 11 of Table 1, and we now 
compare them with the similar results of Szymczak \&
Gerard (2009), in Table  B1 of their Appendix.

We first note that 16 sources are listed with high linear polarization  
by the NRT and by us.  Four other sites are listed with strong linear 
polarization by the NRT but not by us 
and are considered individually:

12.209-0.103 has many features in the spectra at both 1665 and 1667 MHz 
that show significant linear polarization, with good agreement between 
the NRT data and ours.  The listed 1665-MHz feature in NRT Table B1 
is a minor feature that lies just above the 0.5 Jy and 50 per cent 
threshold, whereas it lies just below the threshold in our spectra.

23.44-0.18 has only low linear polarization for the strong spectral 
features.  The listed 1665-MHz feature in NRT Table B1
is a minor feature that lies just above the 0.5 Jy and 50 per cent
threshold, whereas it lies just below the threshold in our 
spectra.  

31.27+0.06 has a listed 1665-MHz feature in NRT Table B1 which 
is a weak shoulder feature that lies just above the 0.5 Jy and 50 per 
cent threshold, whereas it lies just below the threshold in our spectra.

35.200-1.736 has NRT features of high linear polarization listed at 
1665 and 1667 MHz.  We also detect similarly strongly polarized features 
but they lie just below the 50 per cent threshold and are not listed in 
our Table.  

Two sites are listed with high linear polarization by us but not NRT.  
They are: 

23.010-0.411 at 1665 MHz for features weak in 2003 (NRT) but stronger 
and polarized in 2004 and 2005.

32.744-0.076 at both 1665 and 1667 MHz.  Similar polarization visible  
in the NRT spectra is not listed in their Table B1 because it is 
slightly weaker.  

Finally, many of the spectra with no perceptible linear polariation have 
only weak features, for which only high percentage linear polarization 
would be above our detection threshold.  With this in mind, our results 
agree in detail with those of Szymczak \& Gerard (2009) for sources 
in common and, statistically, for the sources observed only in 
our observations.  Overall, the mutual corroboration lends high 
confidence in the reliability of both data sets.

\section[]{Polarization patterns}

The  displayed spectra from our present observations show that 
most of the individual spectral features are elliptically polarized, 
with the circular polarization dominating, and often with unpolarized 
emission also present.  
Linear polarization is occasionally very pronounced, often for one or a 
few features in a multi-feature spectrum.  

In most spectra there are examples of apparent Zeeman patterns.  However, 
compared with single dish (low spatial resolution) spectra of 1720, 6035 
and 6030-MHz masers (Caswell 2003, 2004a), they are less easily 
recognised because:  (i) there are generally more features, leading to 
confusion; (ii) confusion is exacerbated by the quite large splitting; 
and (iii) the amplitudes of Zeeman pair components are commonly very 
unequal.  
None the less, the full polarization spectra of the OH 1665 and 
1667-MHz transitions can be very informative, provided that the spectral 
resolution is high, as demonstrated for the similar statistics from 
the NRT sample of spectra by Szymczak \& Gerard (2009).  

Szymczak \& Gerard (2009) review the puzzle that, despite the common 
presence of two $\sigma$ components (predominantly  circularly  
polarized in the opposite sense, and separated in frequency) which  
indicate a Zeeman pattern, the expected linearly polarized $\pi$ 
component midway between them is hardly ever seen.  
As noted by Szymczak \& Gerard (2009), a variety of explanations 
can account for these puzzles.  
However, it is  strange that, within some spectra, there are 
indeed highly linearly polarized features, but they seem not to be part 
of any Zeeman pattern.  A few such examples  occur in the sample 
of 150 masers that result from combining the present survey with the NRT 
survey.  We note that the sample size will soon be doubled when analyses 
of our observations extending over the southern sky are complete,  
and we therefore defer further discussion of these puzzles.  
High spatial resolution from VLBI measurements will then be needed, and 
we reiterate the remarks of Szymczak \& Gerard (2009), that high 
spectral resolution is especially vital for the linear polarization 
studies since NRT measurements demonstrate the common 
occurrence of large changes of ppa across the linearly polarized 
features.  One contributor to such swings of position angle is Faraday 
rotation within the masing source.  However, this may not be the 
dominant cause, and future investigation of OH masers at the higher 
frequency 6035-MHz transition may cast some light on this.

\section[]{Conclusion}

Full polarization spectra of OH masers are a vital step in their  
subsequent exploration with high spatial resolution.  The large set of 
spectra in the present Parkes study, and the similar study with 
the Nancay Radio Telescope (Szymczak \& Gerard 2009) show excellent 
agreement for the 50 sources in common, greatly expand the 
total of spectra now available, and provide reliable statistics of large 
samples.
The linear polarization statistics, in particular, have 
resulted in the capture of several rare features 
with extremely high linear polarization, of which we noted three 
exceptional examples, 
12.908-0.260, 13.656-0.599 and 12.889+0.489;  further study of such 
sources may advance our poor understanding of the maser polarization 
properties, and their deviation from expectations (e.g. the absence of 
full Zeeman triplets, despite the common presence of Zeeman pairs).  
Our broader statistics, investigating the fraction of sources where at 
least one feature displays linear polarization exceeding 50 per cent, 
found this fraction to be 35 per cent for the 1665-MHz emission and 23 
per cent for the 1667 MHz emission.  

The maser site velocity spreads for OH have been compared with those of 
methanol and of water masers.  The largest spans in velocity, exceeding 
25 \kms, are present in only a few OH masers, and appear to be of two 
distinct varieties:  the first variety arise in the later stages of 
the maser evolution, with a general expansion over a wide 
angle possibly driven by the enclosed \HII\ region;  the second variety 
appear similar to the collimated 
outflows from water masers, with indications of a preponderance of blue 
shifts, such as the remarkable source 24.329+0.145.  These 
investigations will be pursued for a larger southern 
sample and require high spatial resolution studies to confirm the 
effects and interpret the cause.  We note that future VLBI measurements 
at high spatial resolution require accompanying high spectral resolution 
to prevent depolarization caused by the common presence of a sweep of 
polarization position angle across the frequency width of the feature.

Finally, our data set has allowed preliminary characterisation of 
variability at each maser site, including recognition of variability 
in the past, and is an excellent yardstick for recognising future 
variability. With new  monitoring programs able to select promising 
candidates from the present study, there are prospects for identifying 
periodic variables, with corresponding significance for the history and 
physical properties at such sites.

\section*{Acknowledgments}


We thank Warwick Wilson for implementing the correlator 
enhancements, John Reynolds and Parkes Observatory staff for 
enabling the non-standard observing procedures, and Malte 
Marquarding for implementing, within the ATNF spectral analysis package 
{\sc asap}, the spectropolarimetric reduction features developed for the 
present data.

\bsp

\begin{figure*}
 \centering
\includegraphics[width=15.5cm]{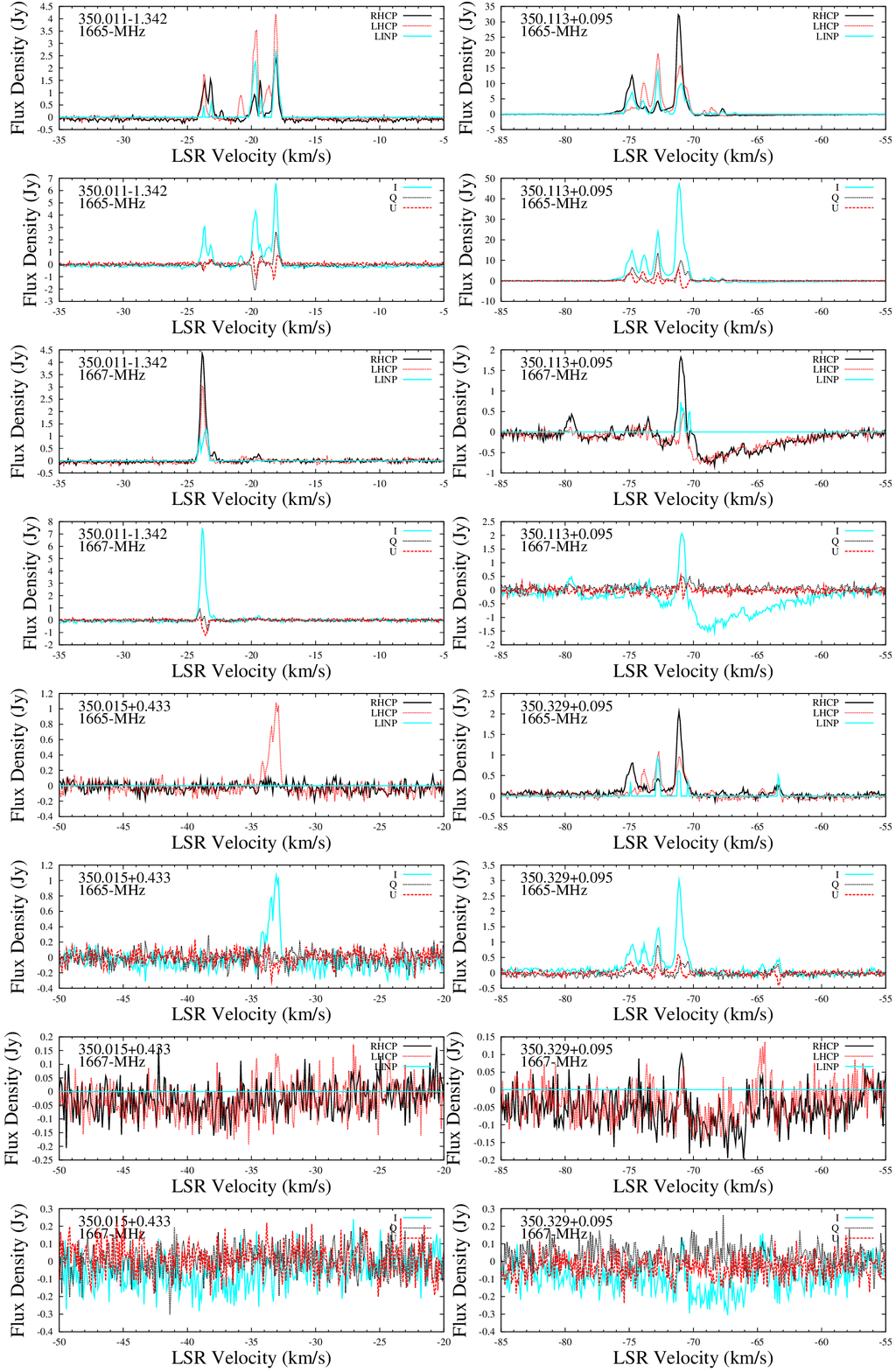}
\caption{Spectra of OH masers at 1665 and 1667 MHz. Within each 
panel, the source name and transition are given, and the  
polarization parameters are plotted as: overlaid spectra of  
RHCP and LHCP with linear polarization;  
and overlaid spectra of Q and U with I.}   

\label{fig1 part 1}

\end{figure*}

\begin{figure*}
 \centering

\addtocounter{figure}{-1}

\includegraphics[width=15.5cm]{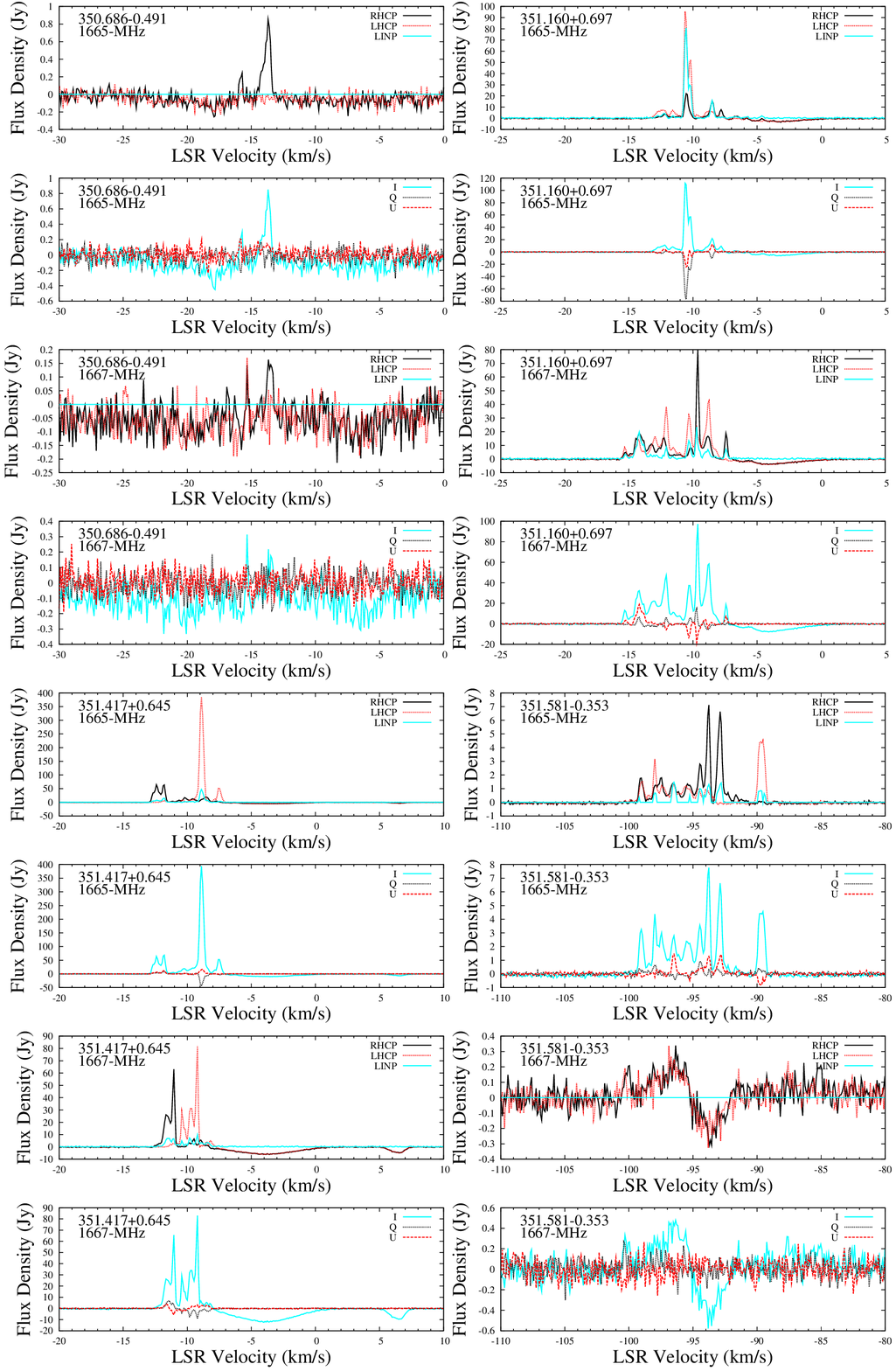}

\caption{\textit{- continued p2 of 23}}

\label{fig1p2} 

\end{figure*}

\begin{figure*}
 \centering

\addtocounter{figure}{-1}

\includegraphics[width=15.5cm]{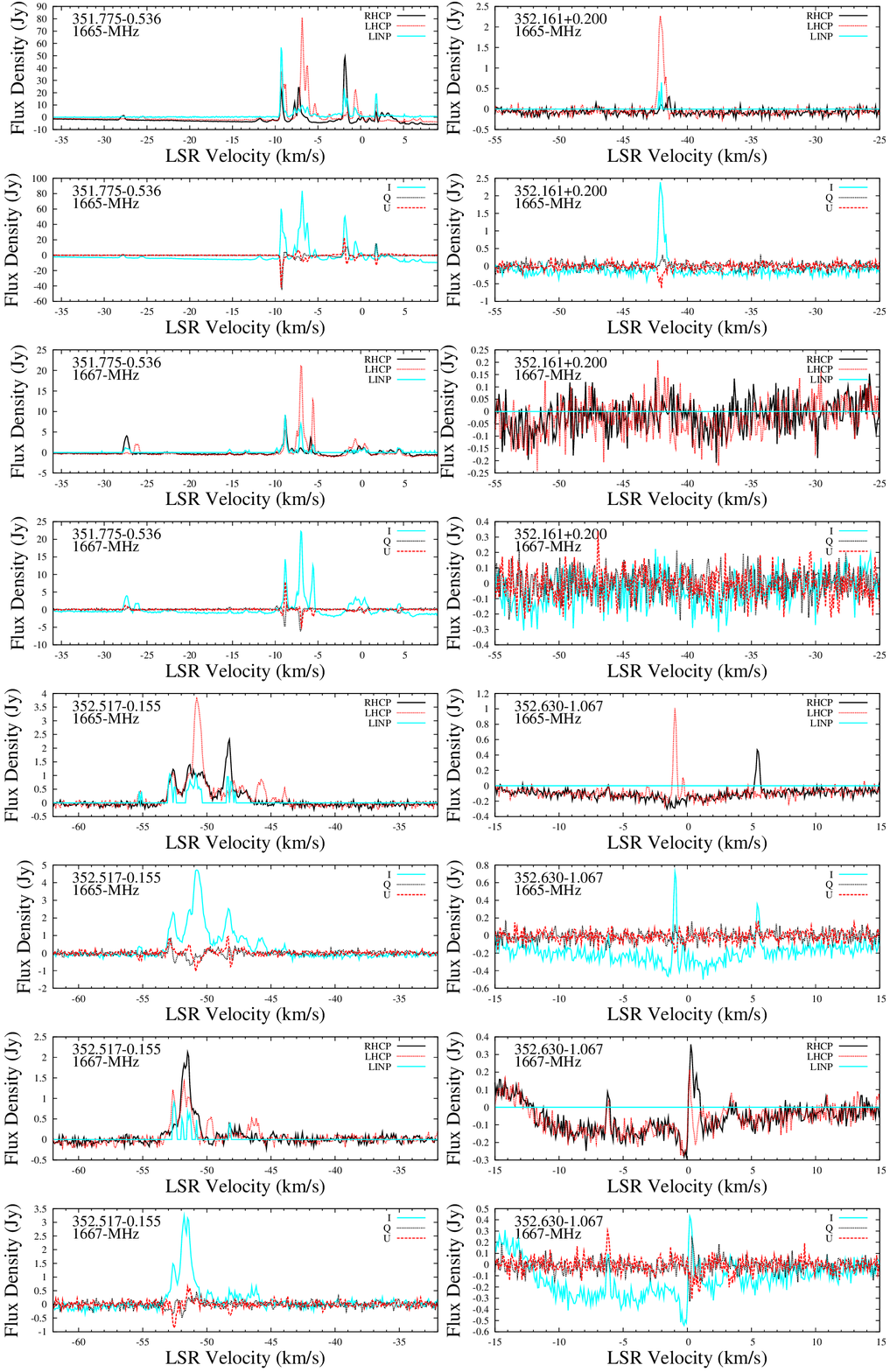}

\caption{\textit{- continued p3 of 23}}

\label{fig1p3} 

\end{figure*}

\begin{figure*}
 \centering

\addtocounter{figure}{-1}

\includegraphics[width=15.5cm]{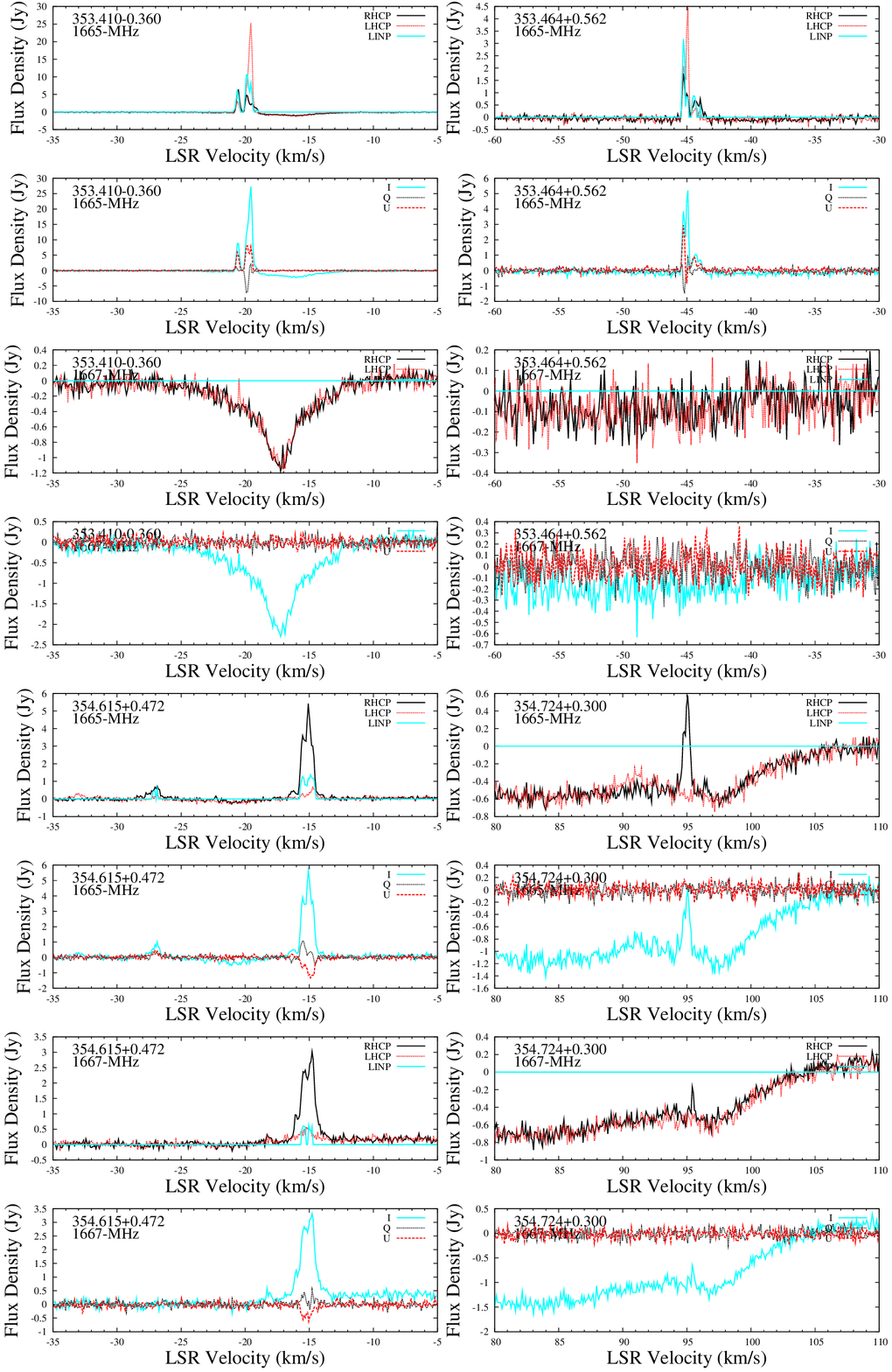}

\caption{\textit{- continued p4 of 23}}

\label{fig1p4} 

\end{figure*}

\begin{figure*}
 \centering

\addtocounter{figure}{-1}

\includegraphics[width=15.5cm]{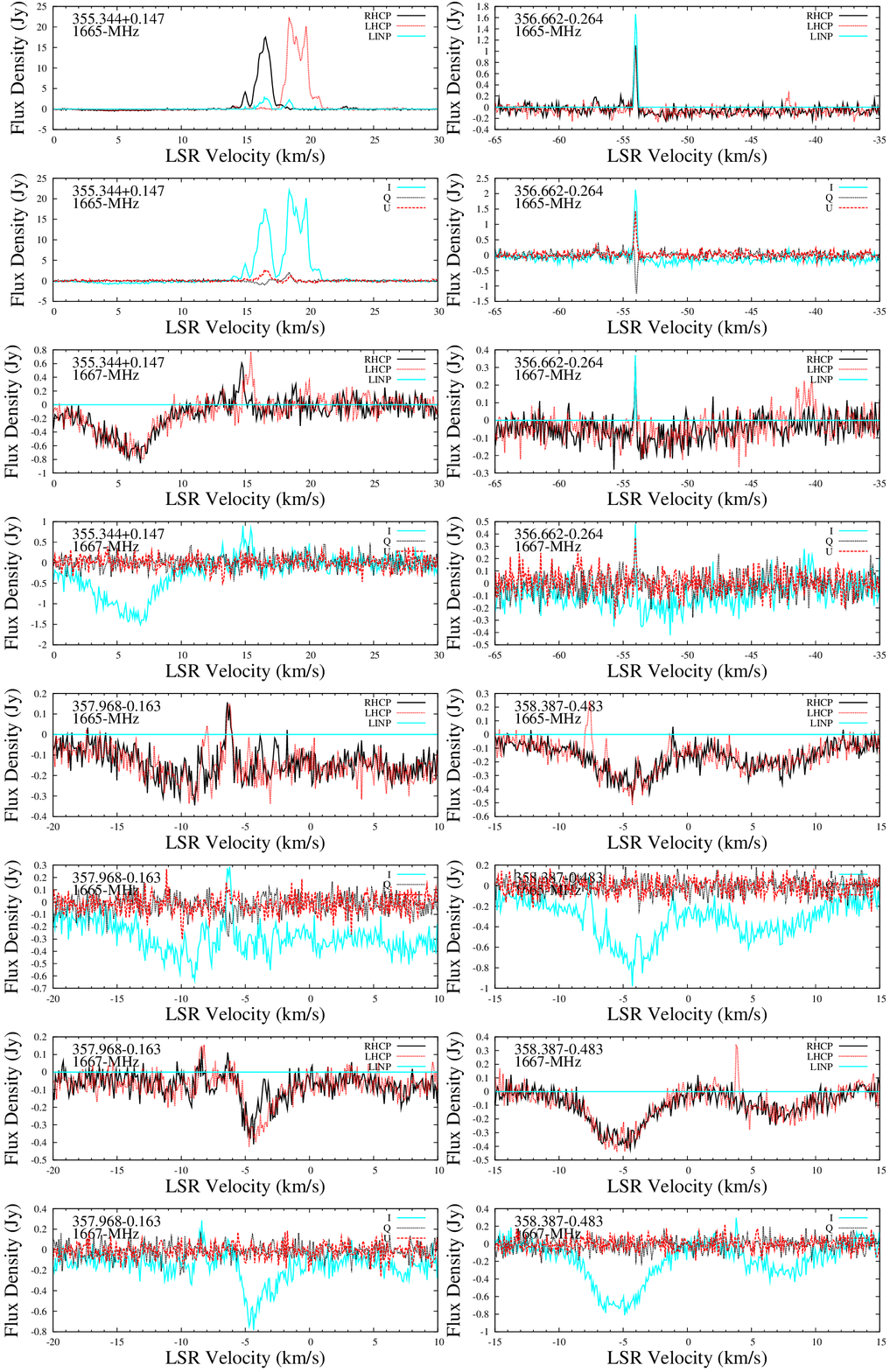}

\caption{\textit{- continued p5 of 23}}

\label{fig1p5} 

\end{figure*}

\begin{figure*}
 \centering

\addtocounter{figure}{-1}

\includegraphics[width=15.5cm]{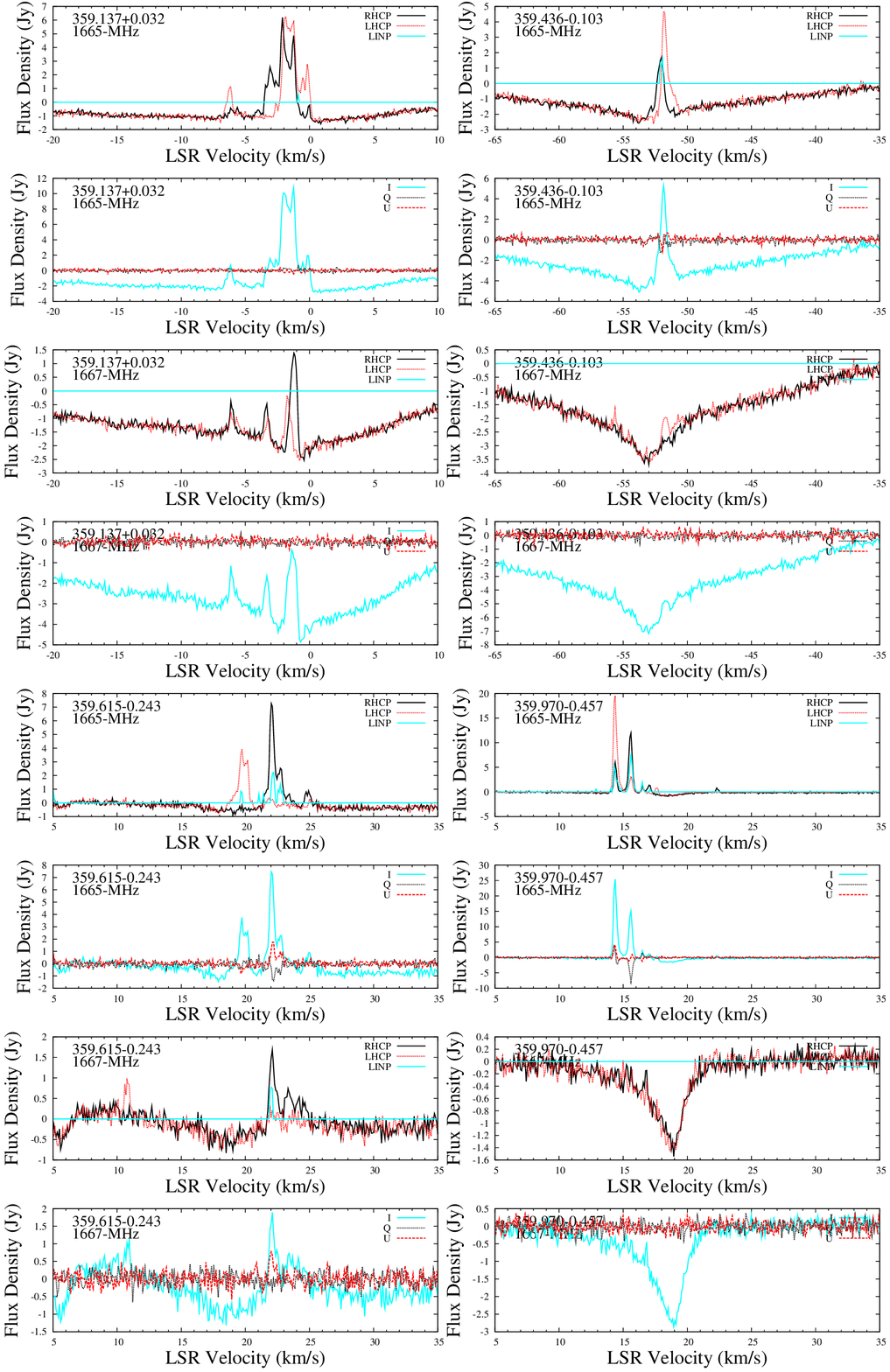}

\caption{\textit{- continued p6 of 23}}

\label{fig1p6} 

\end{figure*}

\begin{figure*}
 \centering

\addtocounter{figure}{-1}

\includegraphics[width=15.5cm]{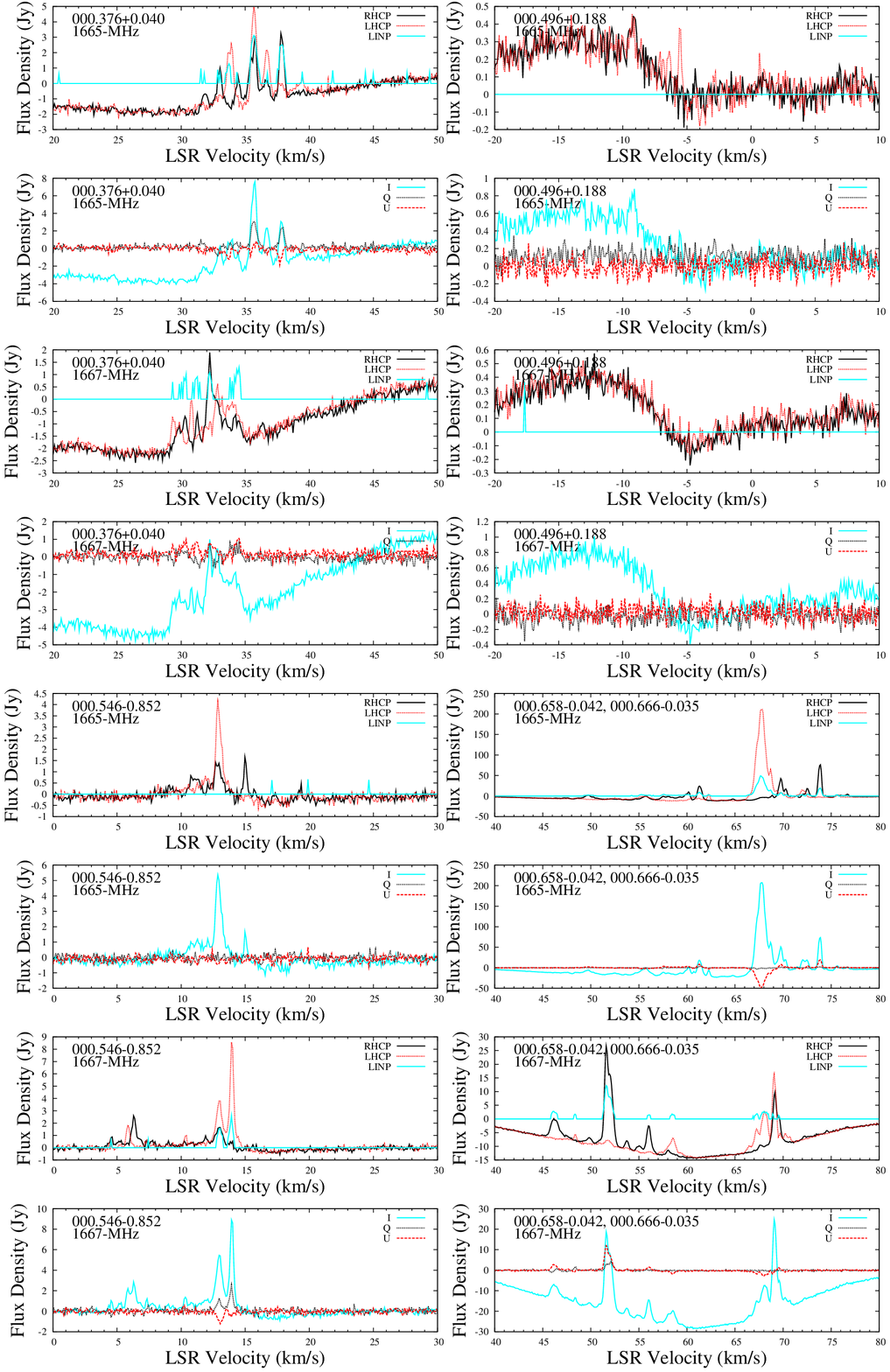}

\caption{\textit{- continued p7 of 23}}

\label{fig1p7} 

\end{figure*}

\begin{figure*}
 \centering

\addtocounter{figure}{-1}

\includegraphics[width=15.5cm]{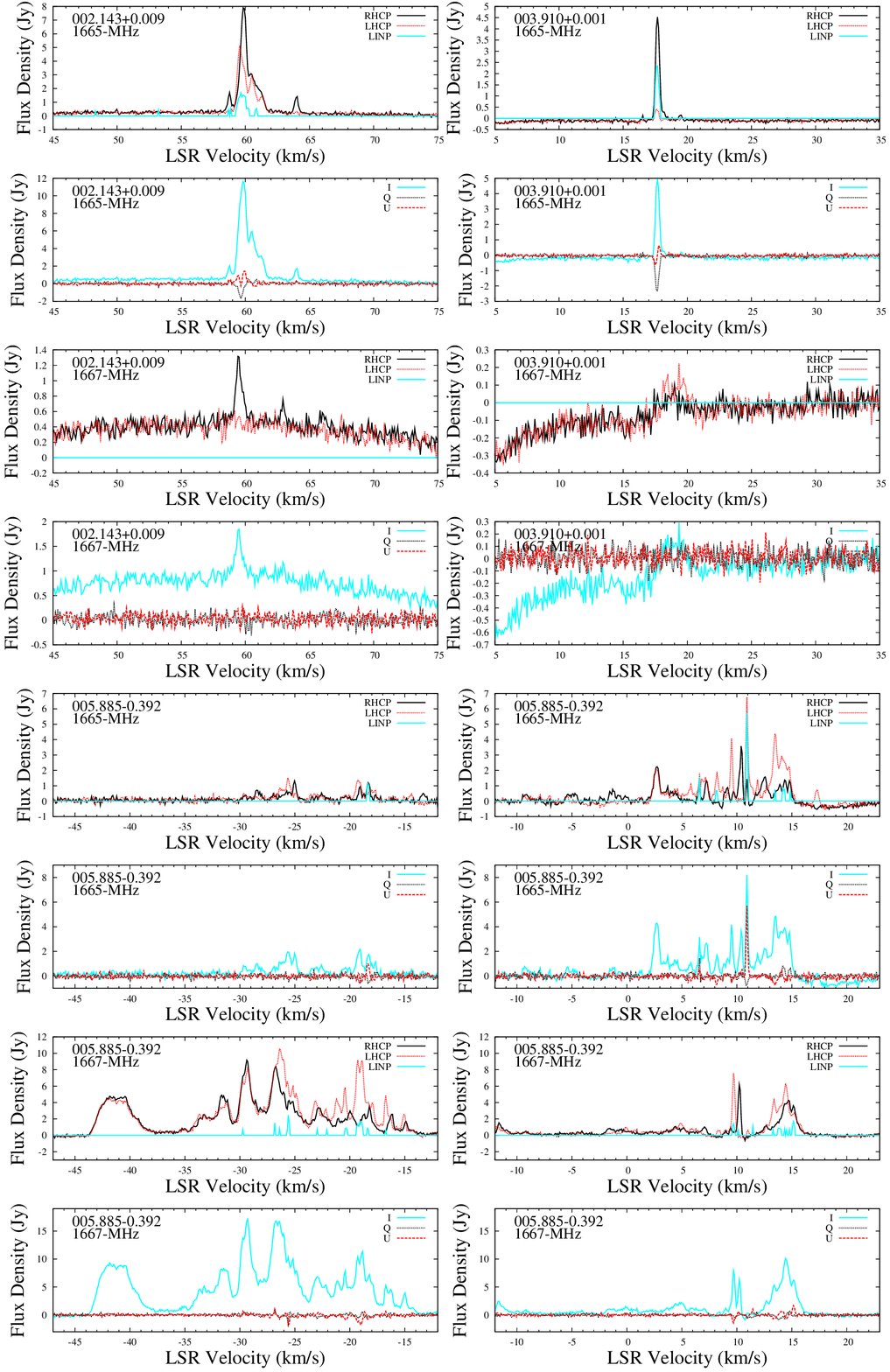}

\caption{\textit{- continued p8 of 23}}

\label{fig1p8} 

\end{figure*}

\begin{figure*}
 \centering

\addtocounter{figure}{-1}

\includegraphics[width=15.5cm]{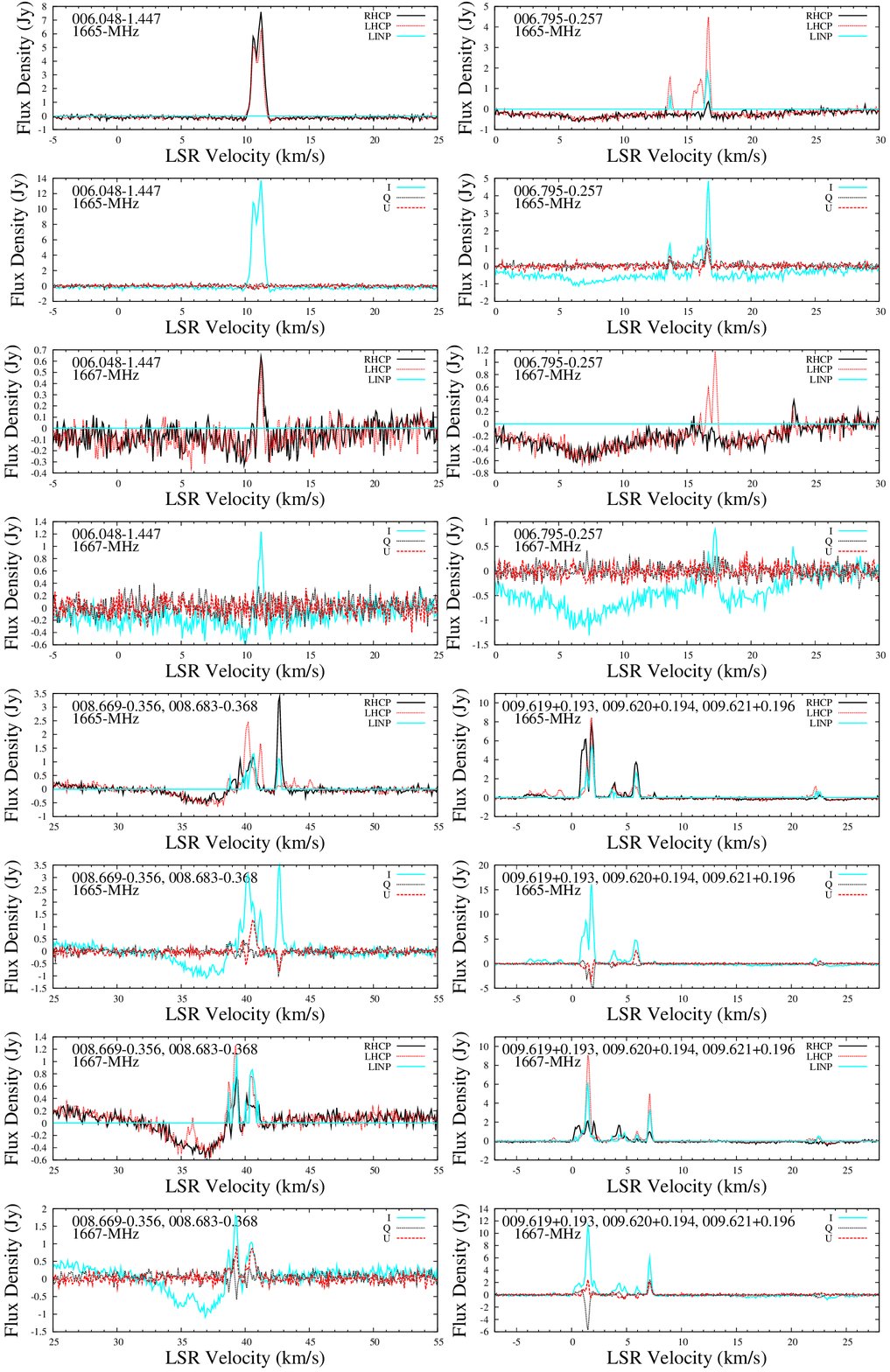}

\caption{\textit{- continued p9 of 23}}

\label{fig1p9} 

\end{figure*}

\begin{figure*}
 \centering

\addtocounter{figure}{-1}

\includegraphics[width=15.5cm]{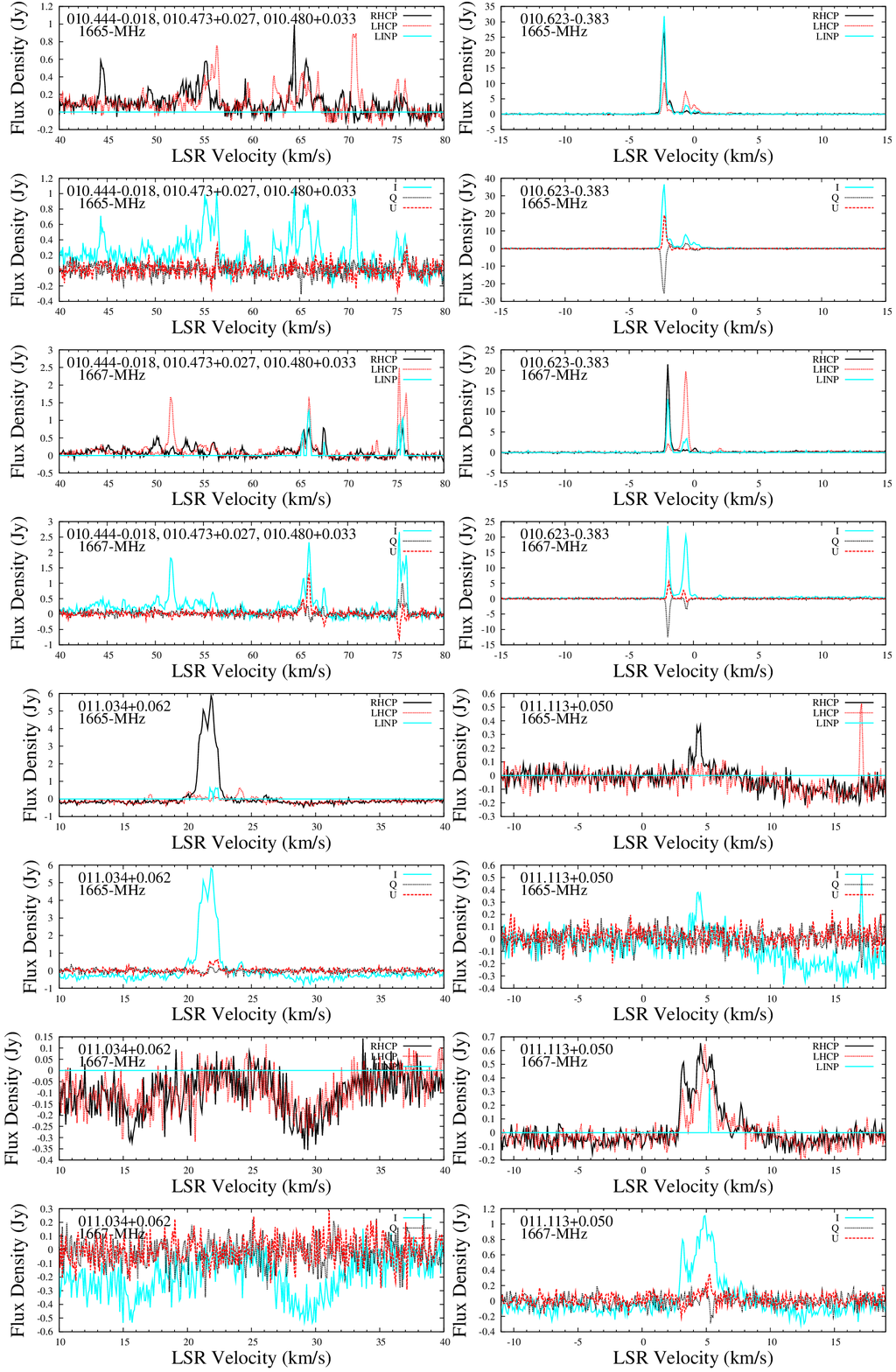}

\caption{\textit{- continued p10 of 23}}

\label{fig1p10} 

\end{figure*}


\clearpage

\begin{figure*}
 \centering

\addtocounter{figure}{-1}

\includegraphics[width=15.5cm]{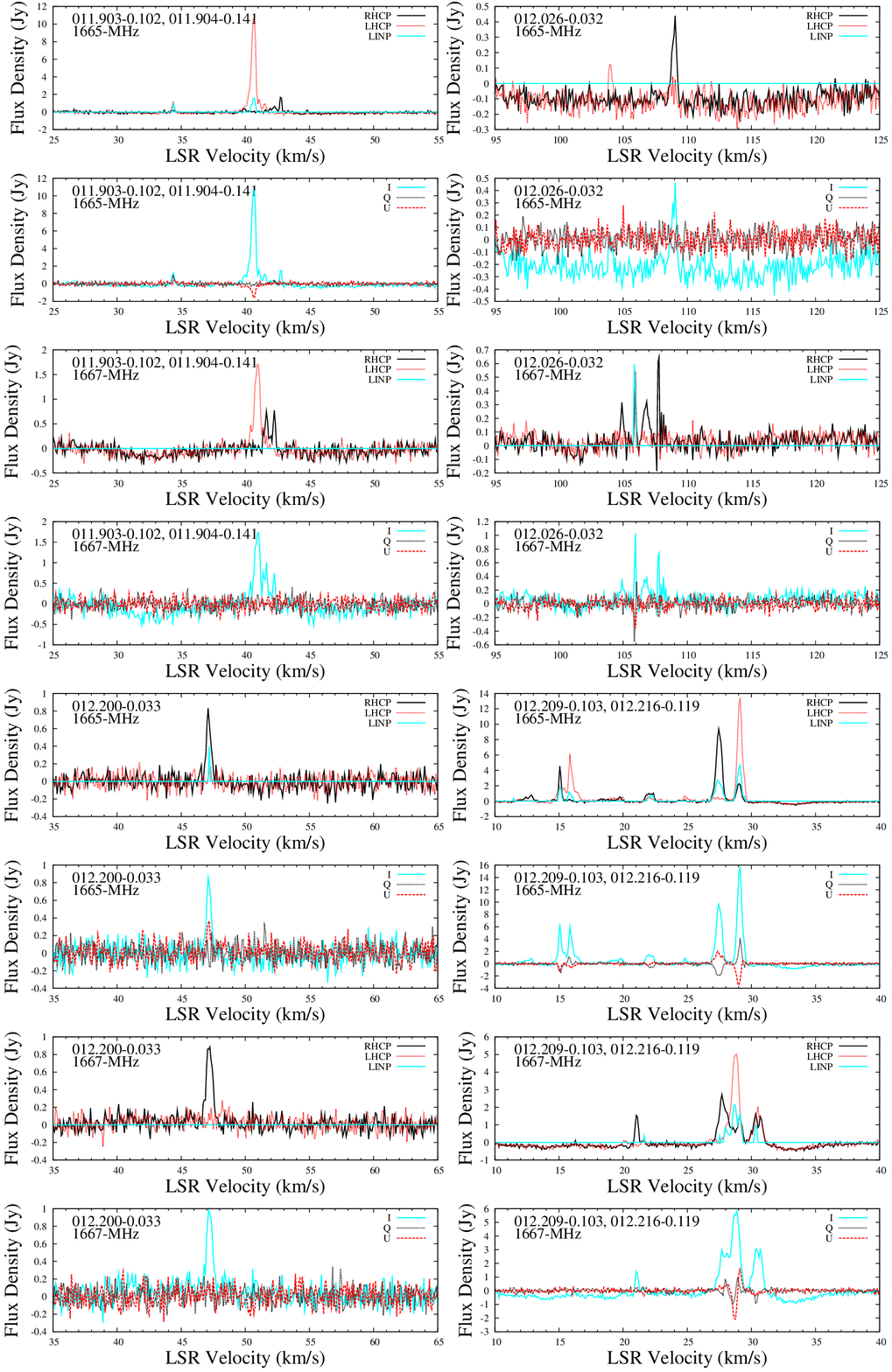}

\caption{\textit{- continued p11 of 23}}

\label{fig1p11} 

\end{figure*}

\begin{figure*}
 \centering

\addtocounter{figure}{-1}

\includegraphics[width=15.5cm]{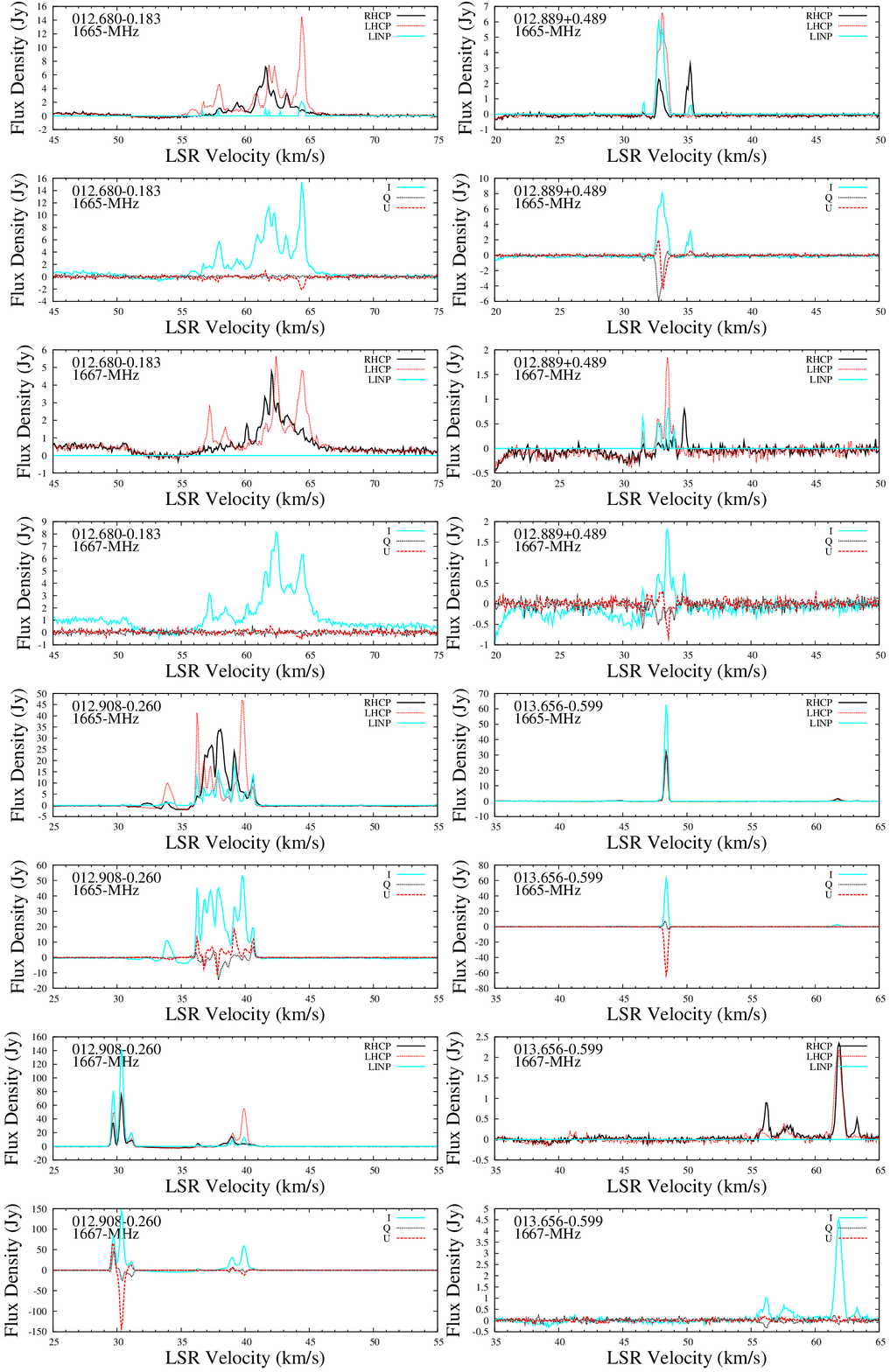}

\caption{\textit{- continued p12 of 23}}

\label{fig1p12} 

\end{figure*}

\begin{figure*}
 \centering

\addtocounter{figure}{-1}

\includegraphics[width=15.5cm]{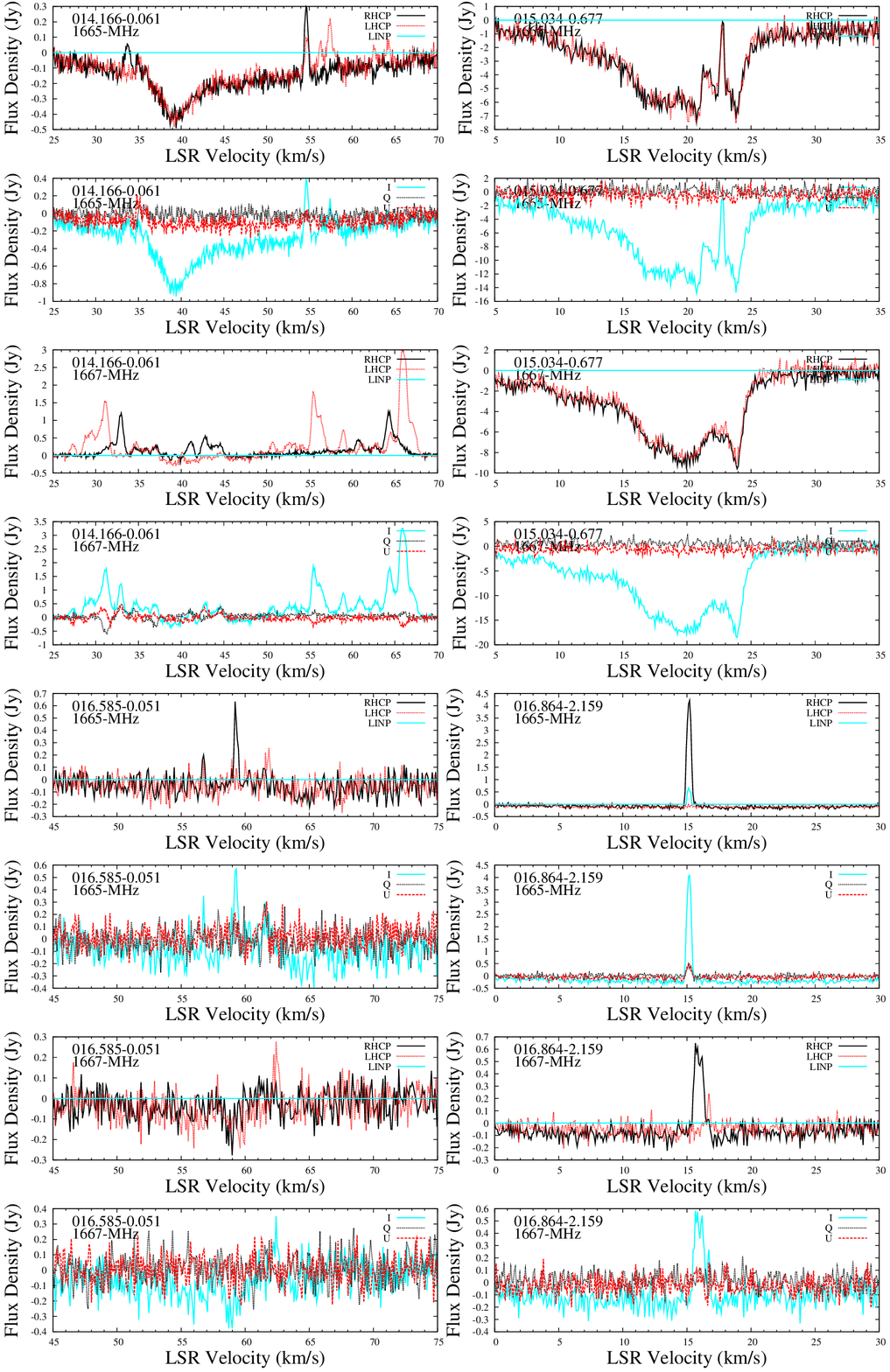}

\caption{\textit{- continued p13 of 23}}

\label{fig1p13} 

\end{figure*}

\begin{figure*}
 \centering

\addtocounter{figure}{-1}

\includegraphics[width=15.5cm]{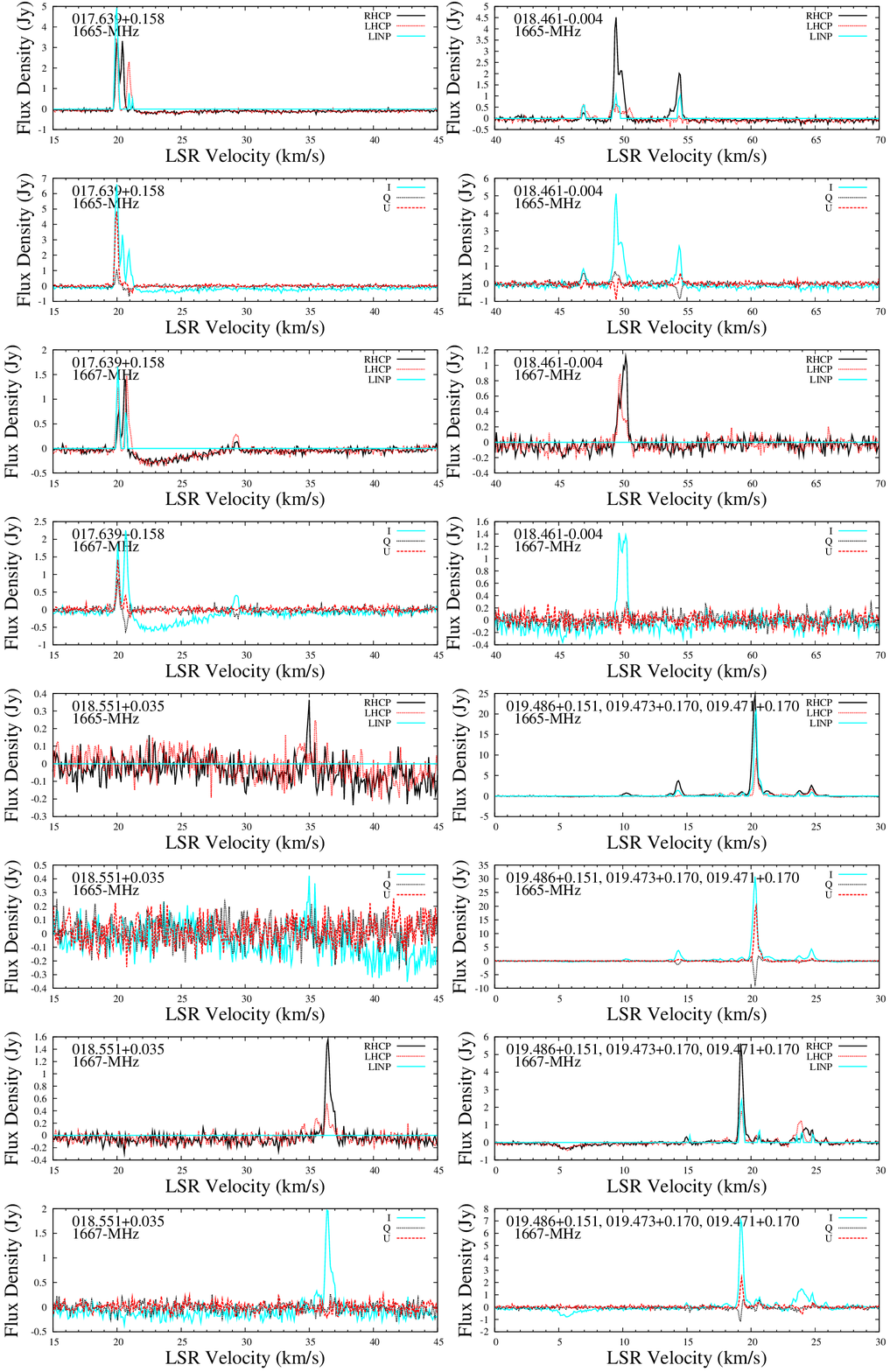}

\caption{\textit{- continued p14 of 23}}

\label{fig1p14} 

\end{figure*}

\begin{figure*}
 \centering

\addtocounter{figure}{-1}

\includegraphics[width=15.5cm]{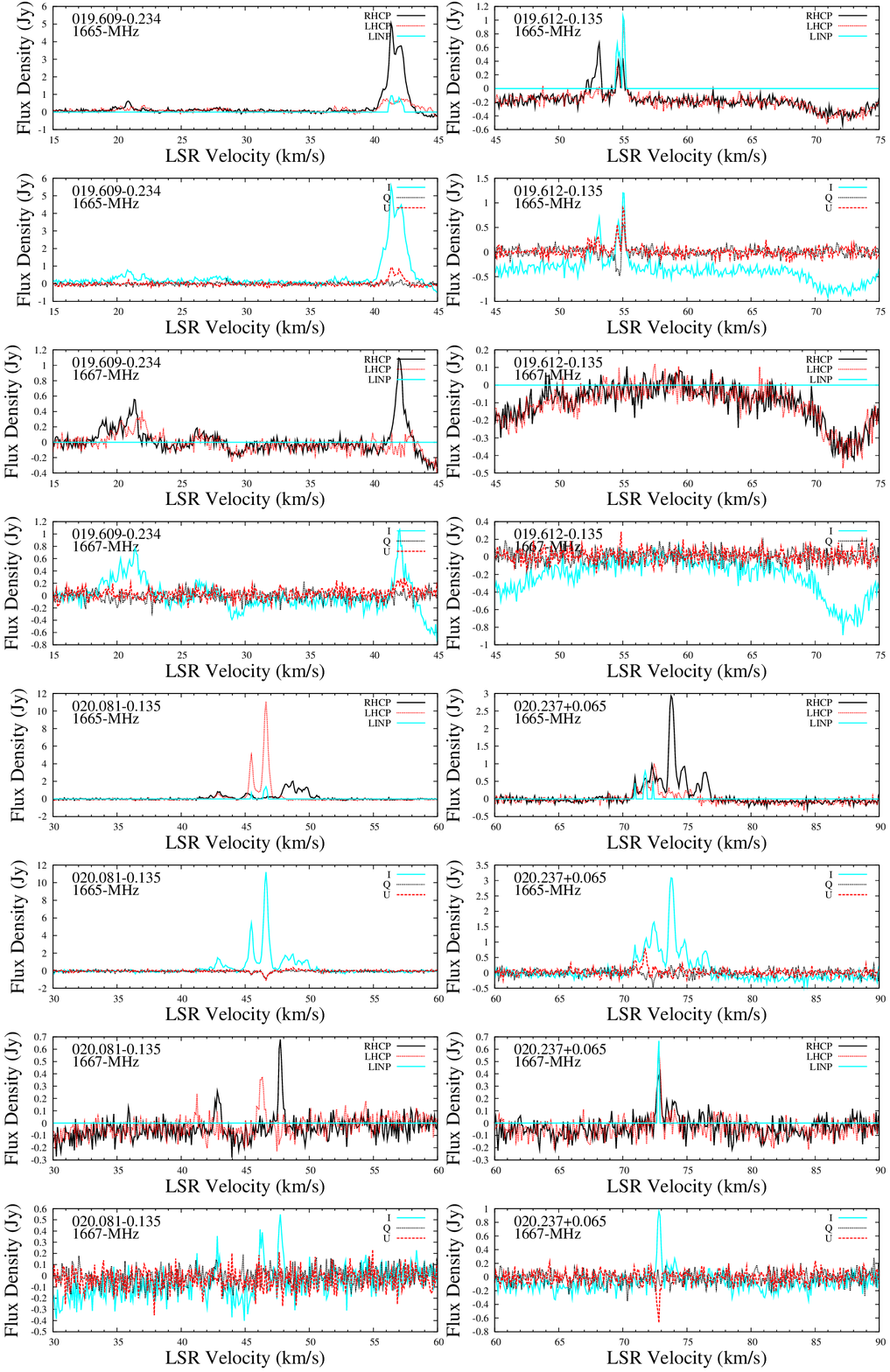}

\caption{\textit{- continued p15 of 23}}

\label{fig1p15} 

\end{figure*}

\begin{figure*}
 \centering

\addtocounter{figure}{-1}

\includegraphics[width=15.5cm]{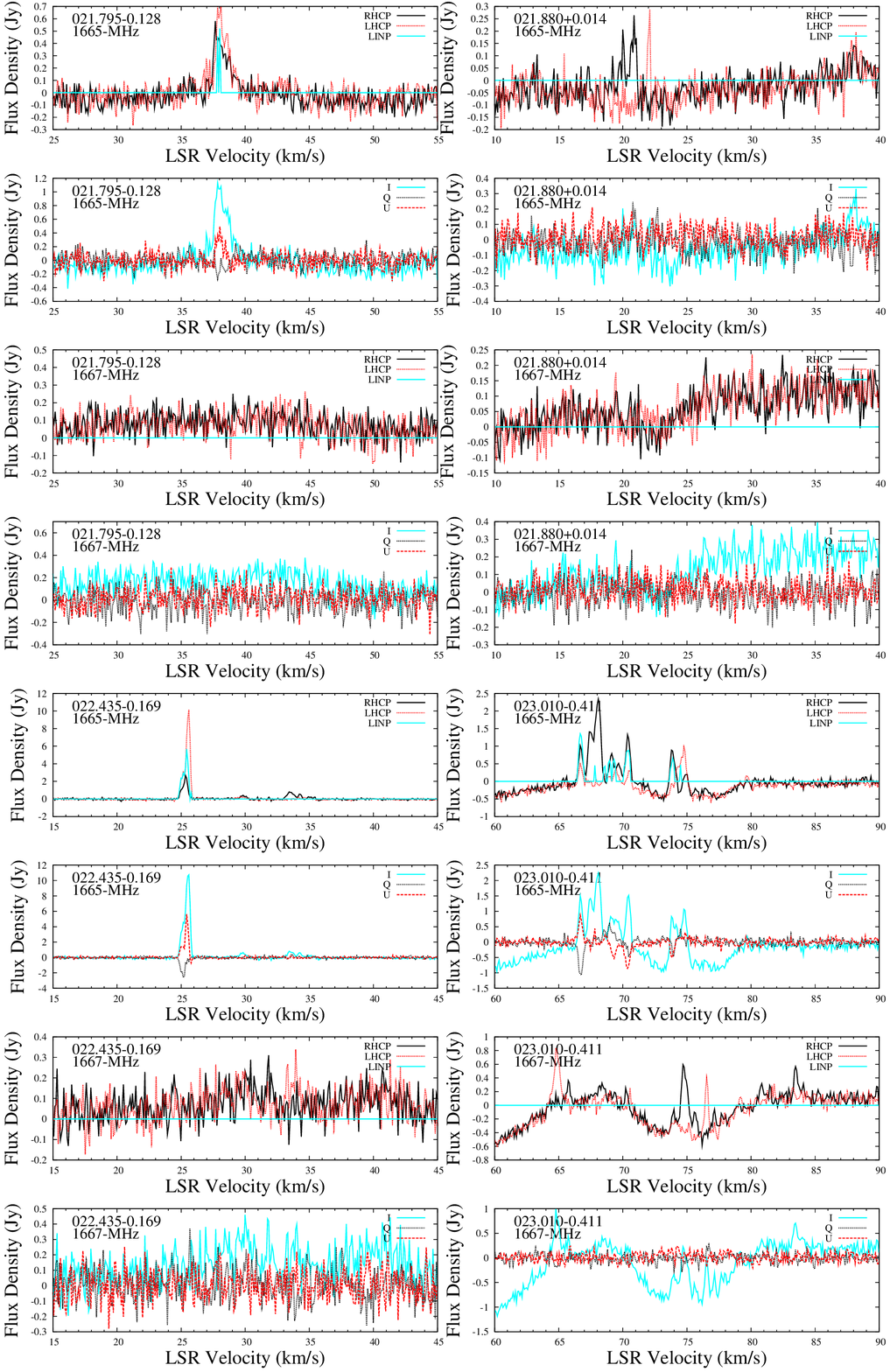}

\caption{\textit{- continued p16 of 23}}

\label{fig1p16} 

\end{figure*}

\begin{figure*}
 \centering

\addtocounter{figure}{-1}

\includegraphics[width=15.5cm]{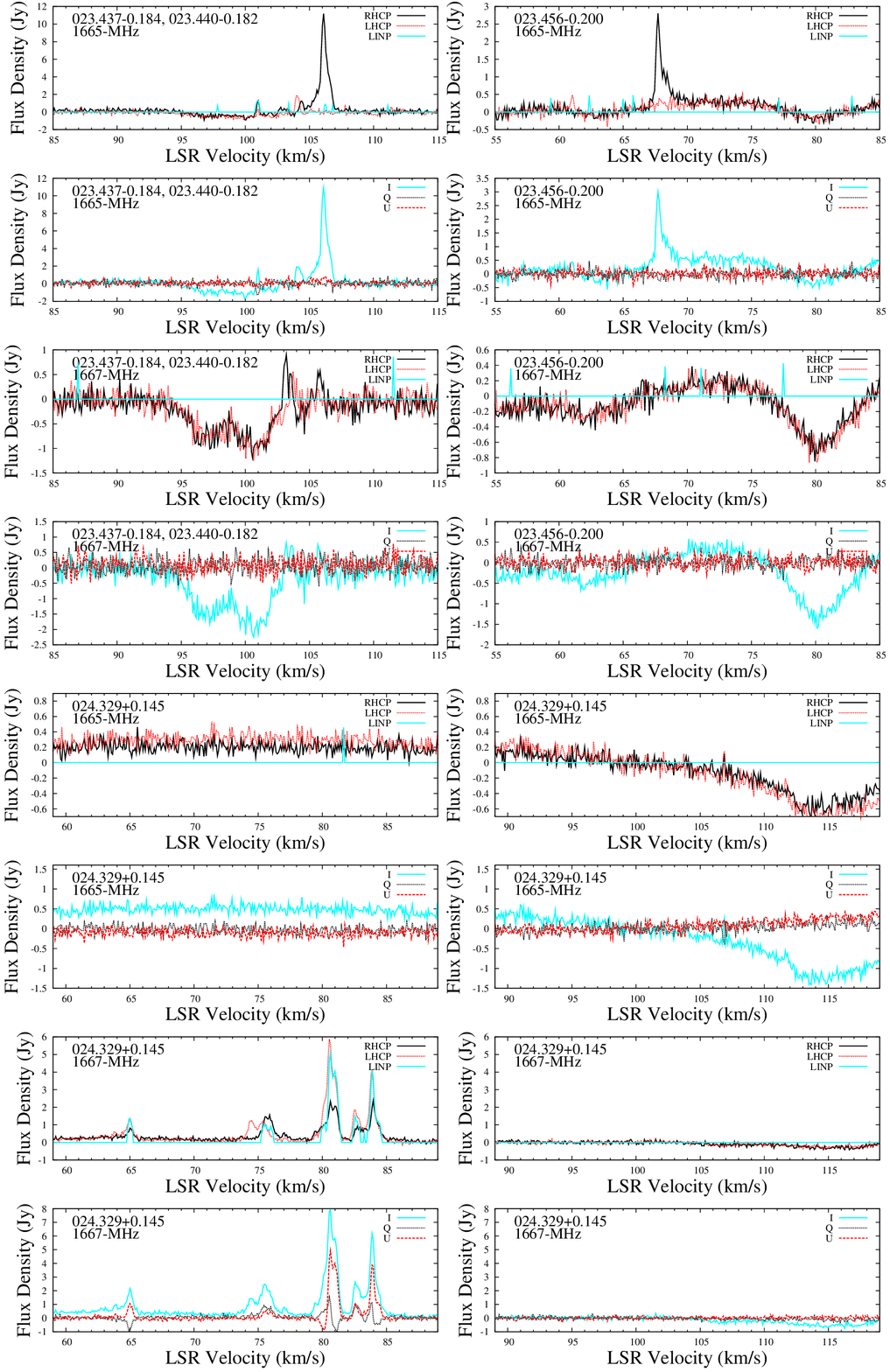}

\caption{\textit{- continued p17 of 23}}

\label{fig1p17} 

\end{figure*}

\begin{figure*}
 \centering

\addtocounter{figure}{-1}

\includegraphics[width=15.5cm]{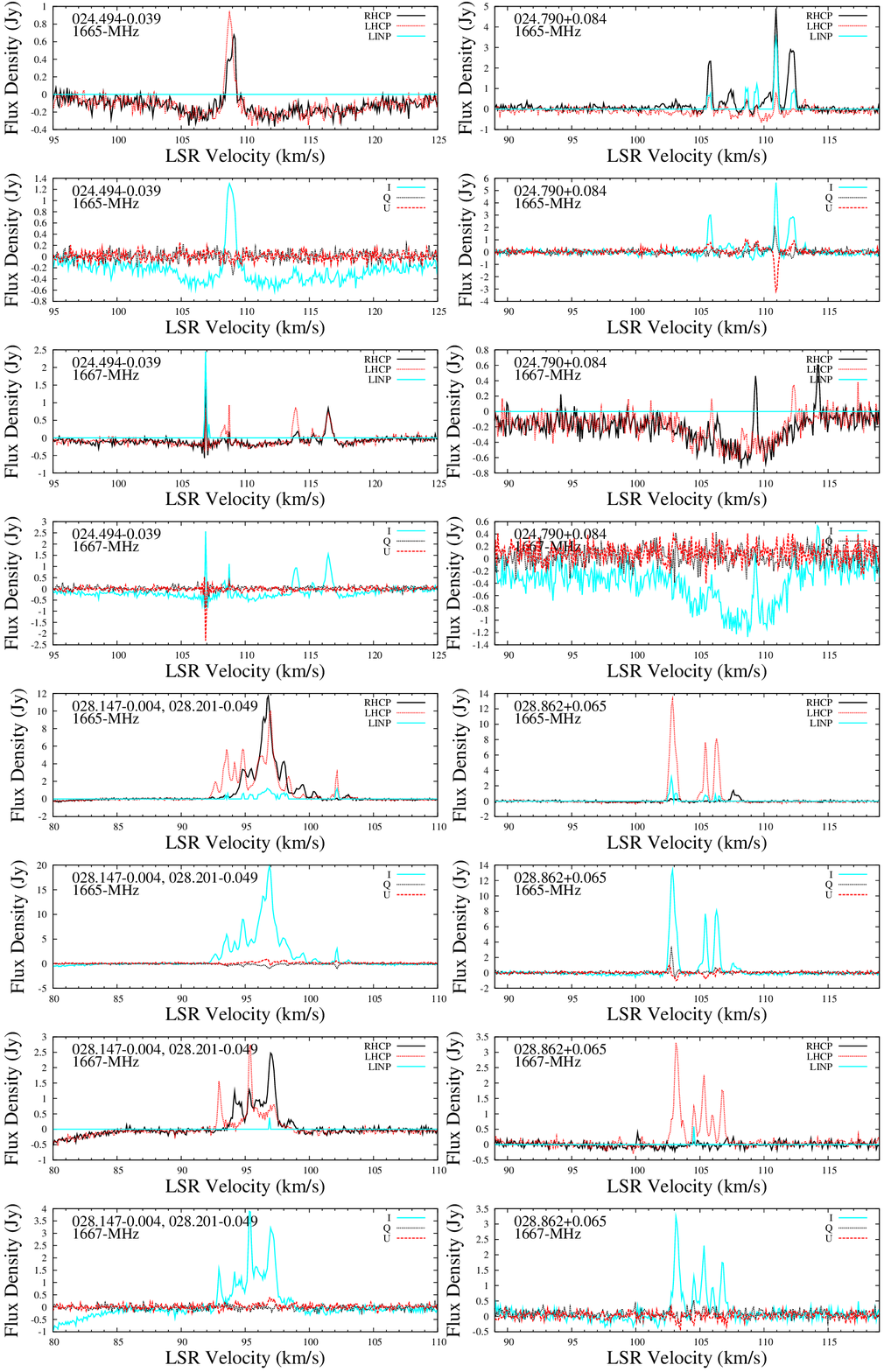}

\caption{\textit{- continued p18 of 23}}

\label{fig1p18} 

\end{figure*}

\begin{figure*}
 \centering

\addtocounter{figure}{-1}

\includegraphics[width=15.5cm]{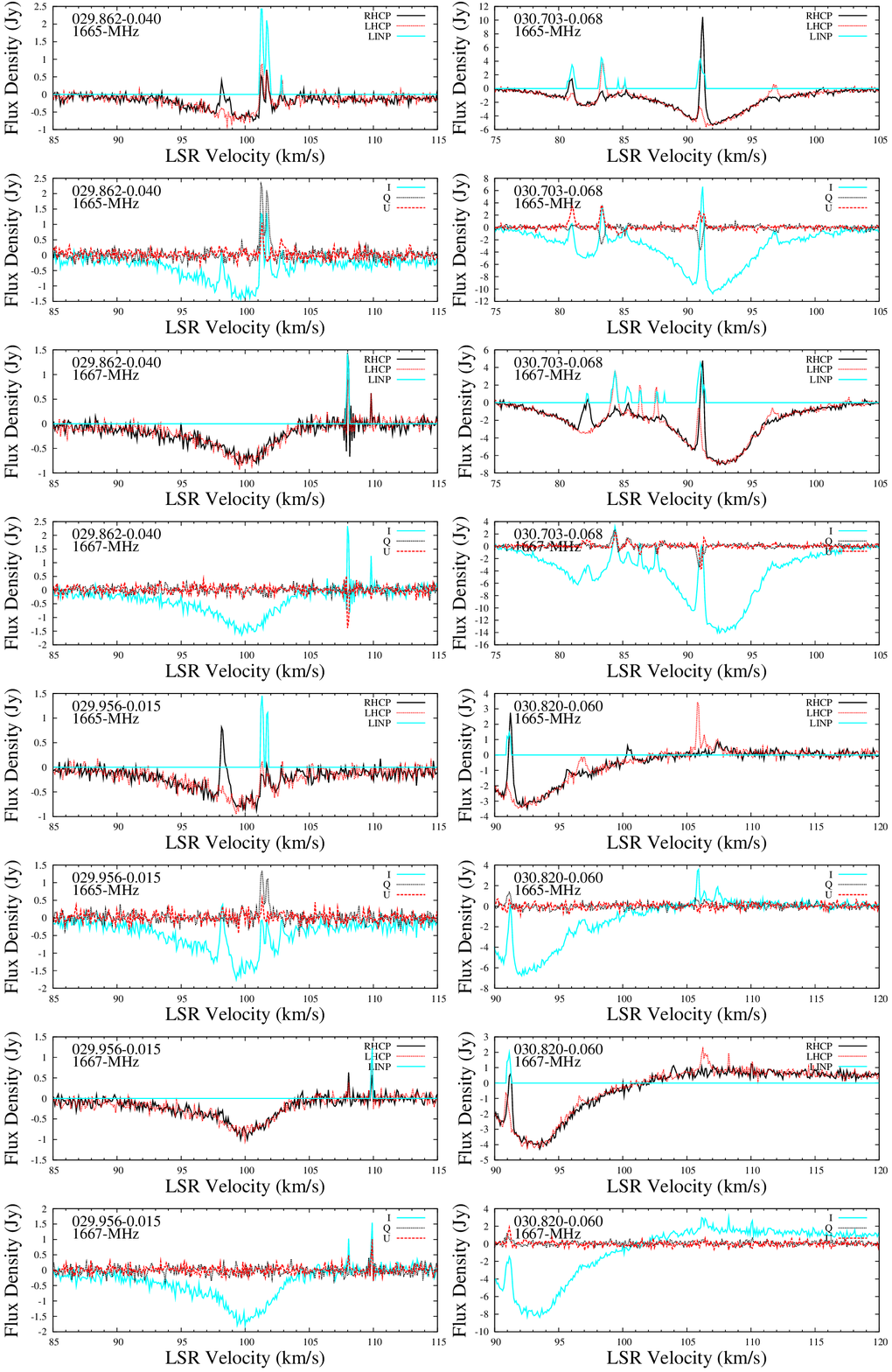}

\caption{\textit{- continued p19 of 23}}

\label{fig1p19} 

\end{figure*}

\begin{figure*}
 \centering

\addtocounter{figure}{-1}

\includegraphics[width=15.5cm]{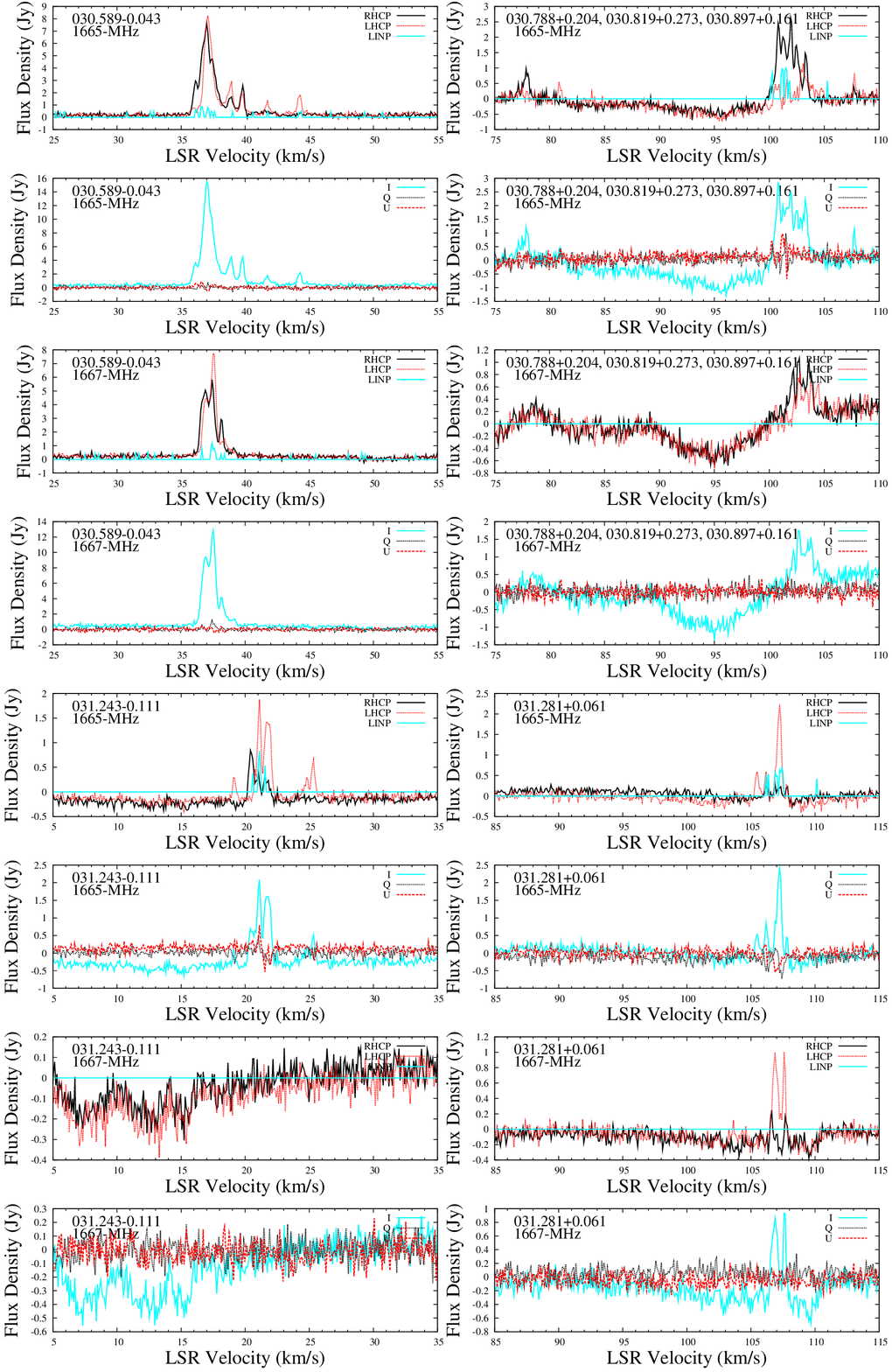}

\caption{\textit{- continued p20 of 23}}

\label{fig1p20} 

\end{figure*}

\begin{figure*}
 \centering

\addtocounter{figure}{-1}

\includegraphics[width=15.5cm]{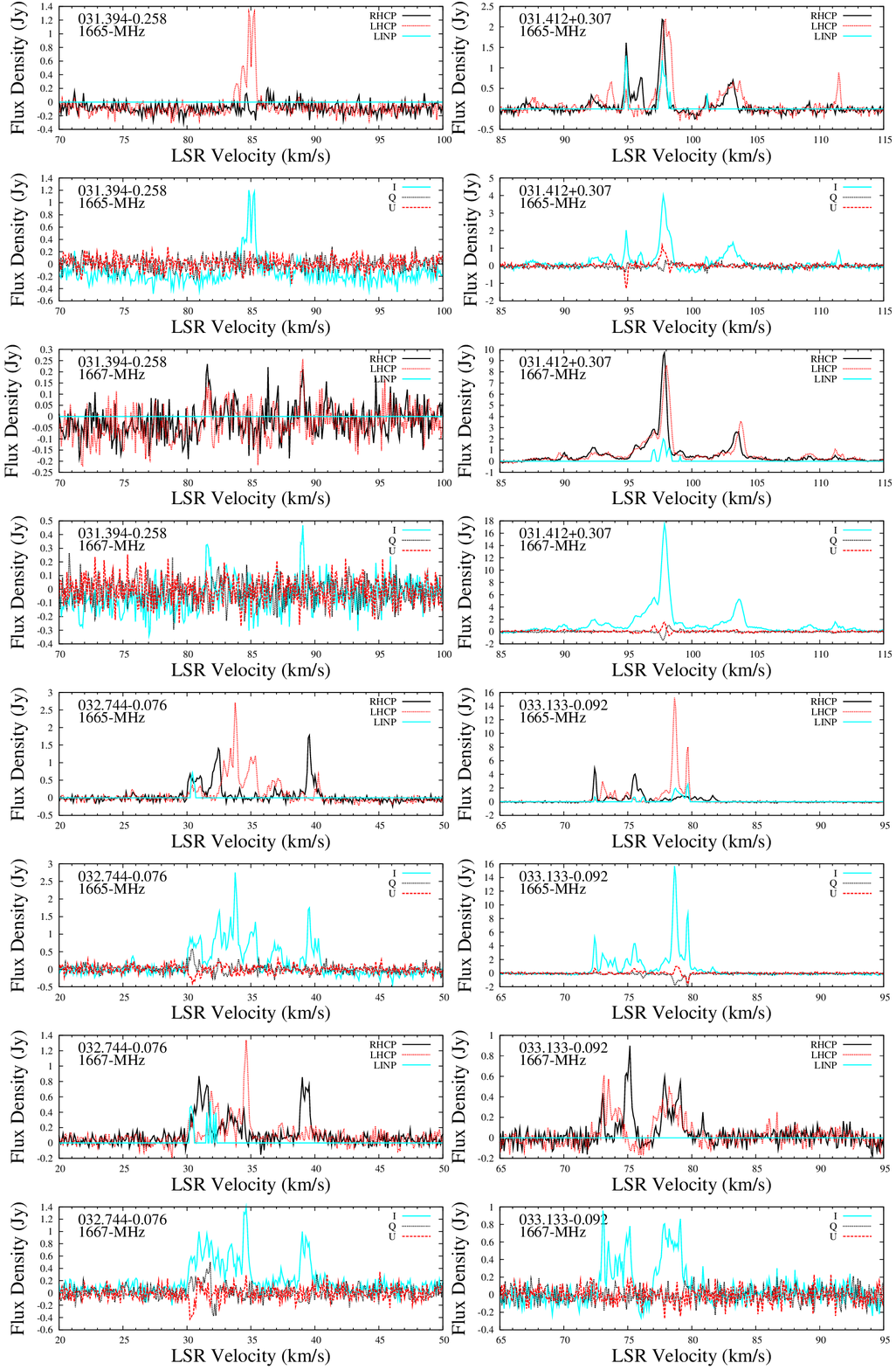}

\caption{\textit{- continued p21 of 23}}

\label{fig1p21} 

\end{figure*}

\begin{figure*}
 \centering

\addtocounter{figure}{-1}

\includegraphics[width=15.5cm]{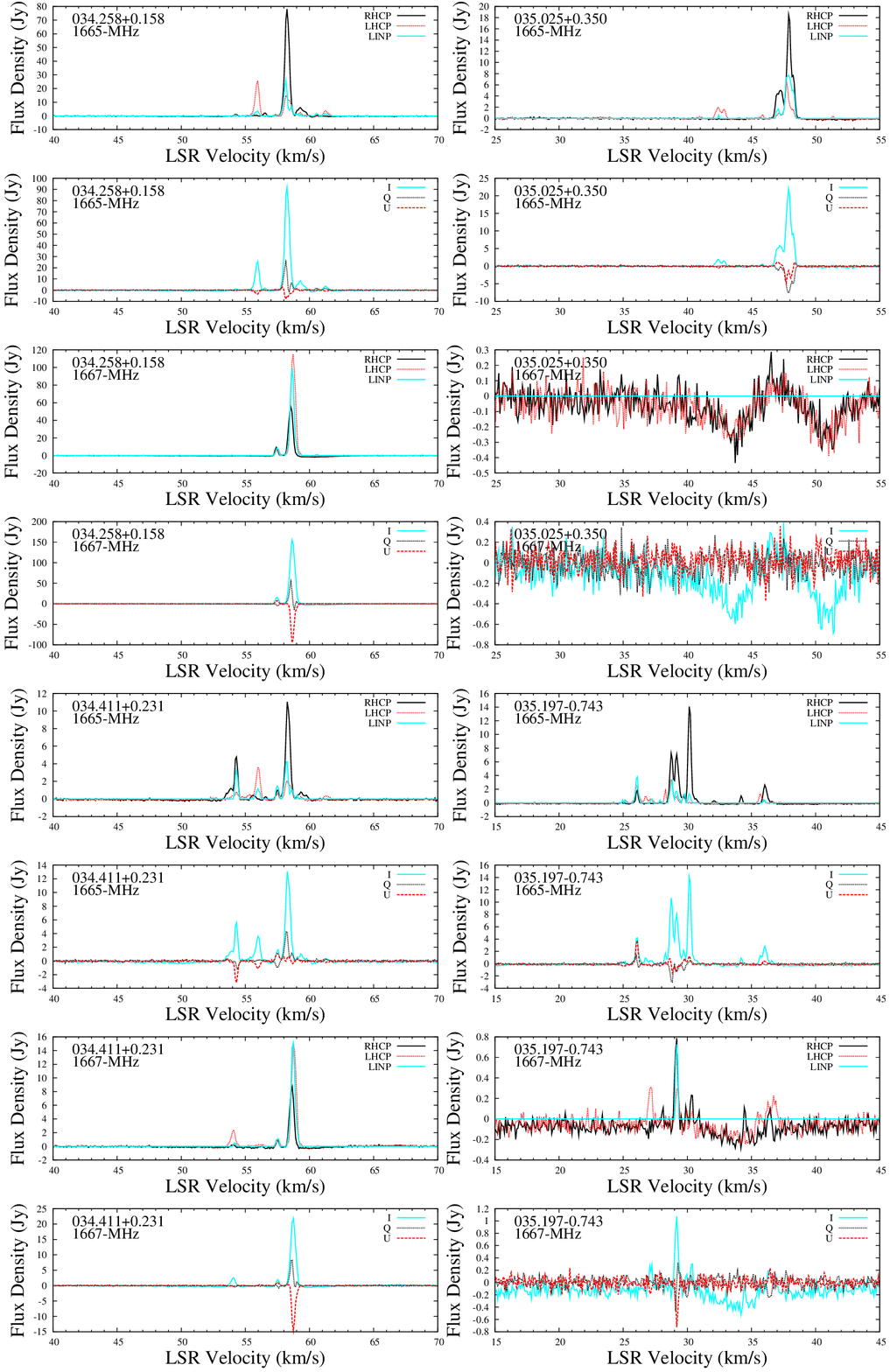}

\caption{\textit{- continued p22 of 23}}

\label{fig1p22} 

\end{figure*}

\begin{figure*}
 \centering

\addtocounter{figure}{-1}

\includegraphics[width=15.5cm]{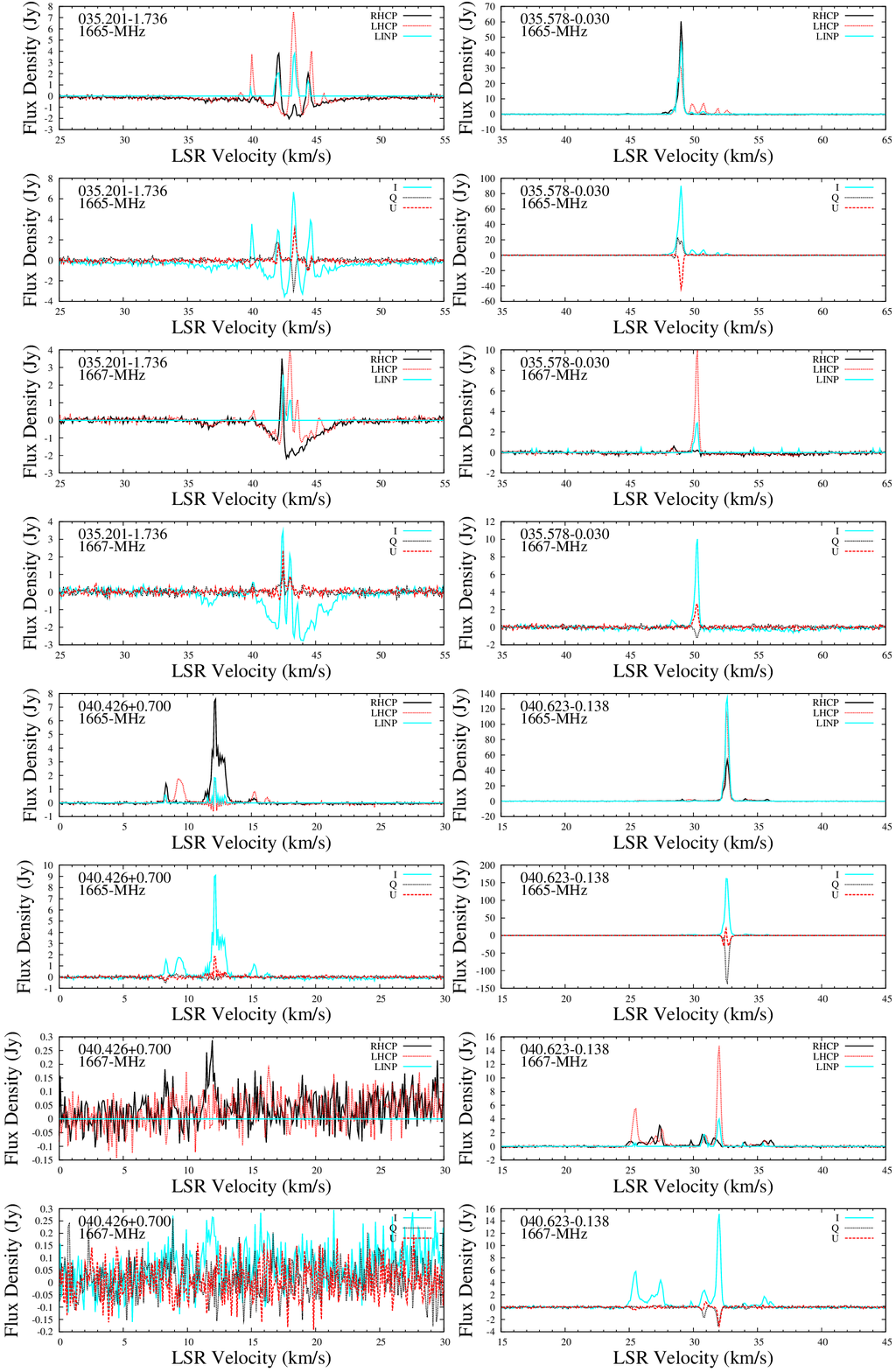}

\caption{\textit{- continued p23 of 23}}

\label{fig1p23} 

\end{figure*}

\label{lastpage}

\end{document}